\documentclass[12pt]{elsarticle}
\usepackage{slashed}
\usepackage{epsfig,graphics}
\usepackage{hyperref}
\usepackage{bbm}
\usepackage{mathtools}
\usepackage{amssymb}
\usepackage{mathrsfs}
\usepackage{amsmath}
\usepackage{natbib}

\newcommand{\bra}[1]{\langle #1|}
\newcommand{\ket}[1]{|#1\rangle}

\newcommand{\roundbra}[1]{(#1|}
\newcommand{\roundket}[1]{|#1)}

\def\Dslash{D \hspace{-2.7mm}/ \;}
\def\quarknumberoperator{{\mathbbm 1}\,}

\newcommand{\mm}[1]{m^{\!\!\!\!\!^o\;\,2}_{#1}}

\newlength{\llslash}

\begin{document}

\begin{frontmatter}
\title{On the convergence of the chiral expansion for the baryon ground-state masses}
\author[GSI,TUD]{M.F.M. Lutz}
\author[GSI]{Yonggoo Heo}
\author[GSI]{and Xiao-Yu Guo}
\address[GSI]{GSI Helmholtzzentrum f\"ur Schwerionenforschung GmbH,\\
Planck Str. 1, 64291 Darmstadt, Germany}
\address[TUD]{Technische Universit\"at Darmstadt, D-64289 Darmstadt, Germany}
\begin{abstract}
We study the chiral expansion of the baryon octet and decuplet masses in the isospin limit. It is illustrated that a chiral expansion 
of the one-loop contributions is rapidly converging up to quark masses that generously encompasses the mass of the physical strange quark.
We express the successive orders in terms of physical meson and baryon masses. In addition, 
owing to specific correlations amongst the chiral moments, we suggest a reordering of terms that make the convergence 
properties more manifest. Explicit expressions up to chiral order five are derived for all baryon masses at the one-loop level. The baryon masses obtained do not depend on the 
renormalization scale. Our scheme is tested against 
QCD lattice data, where the low-energy parameters are systematically correlated by large-$N_c$ sum rules.
A reproduction of the baryon masses from PACS-CS, LHPC, HSC, NPLQCD, QCDSF-UKQCD and ETMC is achieved for ensembles with pion and kaon masses smaller than 600 MeV. 
Predictions for baryon masses on ensembles from CLS as well as all low-energy constants that enter the baryon masses at N$^3$LO are made.
\end{abstract}
\end{frontmatter}
\newpage 

\tableofcontents

\newpage

\section{Introduction}

By now there is large set of QCD lattice data on the lowest baryon masses composed of up, down and strange quarks 
and with $J^P = \frac{1}{2}^+$ and $\frac{3}{2}^+$  \cite{MILC2004,LHPC2008,PACS-CS2008,HSC2008,BMW2008,Alexandrou:2009qu,Durr:2011mp,WalkerLoud:2011ab,NPLQCD:2011,Andersen:2017una}.
The recent and unexpected success of a quantitative description of this data set based on the chiral 
three-flavour Lagrangian \cite{Semke:2012gs,Lutz:2013kq,Lutz:2014oxa,Ren:2013dzt,Ren:2013oaa} makes it paramount to unravel further 
the chiral convergence domain of this sector of QCD. It is an important question how small the quark masses have to be 
chosen as to render a chiral expansion framework meaningful in the flavour $SU(3)$ case. Based on two flavour studies of the 
nucleon mass the power-counting domain (PCD) was estimated in various studies to be as 
low at $m_\pi < $ 200 MeV$-$300 MeV \cite{Young:2002ib,Leinweber:2003dg,Beane:2004ks,Leinweber:2005xz,McGovern:2006fm,Djukanovic:2006xc,Schindler:2007dr,Hall:2010ai}. 
Due to the important role played by the baryon decuplet fields it is not quite straight forward to 
conduct such studies in the three flavour case. In the two-flavour case there are different power counting schemes 
as to incorporate the spin three half field in a consistent manner \cite{Banerjee:1994bk,Banerjee:1995wz,Fettes:2001cr,Pascalutsa:2002pi,Long:2009wq}. 
It is an open challenge how to adapt such schemes to the flavour $SU(3)$ case.   

A strict chiral expansion for the baryon masses following the rules of heavy-baryon $\chi$-PT
appears futile. Applications to current QCD lattice data do not seem 
promising \cite{MILC2004,LHPC2008,PACS-CS2008,HSC2008,BMW2008,Alexandrou:2009qu,Durr:2011mp,WalkerLoud:2011ab}. In oder to make progress we have to leave the safe haven of conventional $\chi$PT. 
State of the art chiral extrapolations of the lattice data set are based on the relativistic chiral Lagrangian with baryon 
octet and decuplet fields \cite{Semke:2011ez,Semke:2012gs,Lutz:2012mq,Ren:2012aj,Ren:2013oaa,Geng:2013xn,Lutz:2014oxa}. A quantitative 
reproduction of the lattice data requires a summation scheme and the consideration of counter terms that 
turn relevant at next-to-next-to-next-to-leading order (N$^3$LO). Any summation scheme, be it a phenomenological form factor, relativistic kinematics or the use of 
physical masses in the loop functions, is set up in a manner that the results of a strict chiral expansion are recovered 
to some given order upon the neglect of formally higher order terms. One may rightfully argue that possibly some model 
dependence is encountered once the quark masses reach their physical values. 
While in \cite{Semke:2011ez,Semke:2012gs,Lutz:2012mq} the $\chi-$MS scheme of \cite{Semke2005,Semke:2006hd} was used,
the EOMS of \cite{Gegelia:1999gf} was used in \cite{Ren:2013dzt,Ren:2013oaa}. Both schemes protect the analytic
structure of the one-loop contributions in the baryon self energies as requested by micro causality. In contrast,
the IR scheme of \cite{Becher:1999he} was applied in \cite{Bruns:2012eh}.

The target of our present study is  to scrutinize and further improve the most comprehensive analysis of the QCD lattice data set in the isospin limit \cite{Lutz:2014oxa}. 
A quantitative description of more than 220 lattice data points  for the baryon octet and decuplet states of five different lattice groups, PACS-CS, LHPC, HSC, NPLQCD 
and QCDSF-UKQCD \cite{PACS-CS2008,LHPC2008,HSC2008,NPLQCD:2011,Bietenholz:2011qq}, was 
achieved in terms of a 12 parameter fit. The number of parameters was reduced significantly by large-$N_c$ sum 
rules \cite{Lutz:2010se,Lutz:2011fe,Lutz:2014jja}. More recent lattice simulation data from the ETM collaboration \cite{Alexandrou:2014sha} for the 
baryon decuplet and octet masses were predicted by this approach. The analysis was based on the relativistic chiral Lagrangian 
and the use of physical meson and baryon masses in the one-loop expressions that include effects up to N$^3$LO. 
A conventional counting, where $M_\Delta - M_N \sim m_K $ was taken as the guide for the construction of the chiral extrapolation 
formulae. While such a counting is justified for larger meson masses, in the limit of very small meson masses like 
$m_\pi \ll  M_\Delta - M_N$ some chiral constraints are realized only in an approximate manner. It is the purpose of the 
present work to establish an approximation strategy that is efficient uniformly well from the small up and down quark masses up to  
a large strange quark mass. 

The work is organized as follows. In the second chapter we review all large-$N_c$ sum rules that are relevant for the chiral extrapolation of 
the baryon masses. So far unknown subleading terms in the $1/N_c$ expansion will be derived for the first time.
It follows Chapter 3 in which the previous approach \cite{Lutz:2014oxa} is further developed 
to restore chiral constraints close to the chiral limit at $m_\pi \ll  M_\Delta - M_N$.  
A novel chiral expansion scheme, in which the various moments are expressed in terms of the physical meson 
and baryon masses is presented in Chapter 4. The convergence properties of such an expansion are examined. 

In Chapter 5 we report on a global fit to the baryon masses from PACS-CS, LHPC, HSC, NPLQCD, QCDSF and ETMC  \cite{PACS-CS2008,LHPC2008,HSC2008,NPLQCD:2011,Bietenholz:2011qq,Alexandrou:2013joa}.
Based on parameter sets obtained predictions for the baryon sigma terms at physical quark masses and baryon masses 
for the ensembles of the CLS collaboration are made \cite{Bruno:2014jqa,Bruno:2016plf}. Since we do not consider yet lattice discretization effects  a systematic error analysis 
of our results is outside the scope of this work.  

\newpage

\section{Primer on large-$N_c$ sum rules}

\noindent
Before we turn to the chiral extrapolation of the baryon masses we recall and further develop the large-$N_c$ sum rules 
previously applied in the studies \cite{Semke:2011ez,Semke:2012gs,Lutz:2014oxa}. They are of crucial importance in a chiral extrapolation study of 
the baryon masses since they provide a significant parameter reduction at N$^3$LO. The sum rules are a consequence of a large-$N_c$ analysis of 
baryon matrix elements of axial-vector, vector and pseudo-scalar, scalar  
quark currents of QCD, which are
\begin{eqnarray}
&& A_\mu^{(a)}(x)= \bar q(x) \,\gamma_\mu \,\gamma_5 \,\frac{\lambda^{(a)}}{2} \,q(x)\,, \qquad  V_\mu^{(a)}(x)= \bar q(x) \,\gamma_\mu  \,\frac{\lambda^{(a)}}{2} \,q(x)\,, 
\nonumber\\
&& P^{(a)}(x)= \bar q(x) \, \gamma_5\,\frac{\lambda^{(a)}}{2} \,q(x)\,,\qquad \quad \, S^{(a)}(x)= \bar q(x) \, \frac{\lambda^{(a)}}{2} \,q(x)\,.
\end{eqnarray}
On the one hand one may compute such matrix elements from a given truncated chiral Lagrangian. On the other hand they may be analyzed
systematically in a $1/N_c$ expansion. A matching of the two results leads to a correlation of the low-energy parameters of the chiral Lagrangian 
formulated with the baryon octet and decuplet fields. We use here the conventions for the various low-energy parameters as introduced previously in 
\cite{Semke:2011ez}.

For the reader's convenience all terms relevant 
for our study are recalled explicitly in this chapter. The applied conventions are illustrated at the hand of the kinetic terms
\begin{eqnarray}
&& \mathcal{L}^{(1)} =
\mathrm{tr}\, \Big\{ \bar B\, (i\, \Dslash\, - M_{[8]})\, B \Big\}  +
 \mathrm{tr}\, \Big\{ \bar B_\mu \cdot \big((i\,\Dslash\, - M_{[10]})\,g^{\mu\nu} 
\nonumber\\ 
&& \qquad \qquad \qquad  -\,i\,(\gamma^\mu D^\nu + \gamma^\nu D^\mu) + \gamma^\mu(i\,\Dslash + M_{[10]})\gamma^\nu \big)\, B_\nu \Big\}\,,
\nonumber\\
&&\Gamma_\mu ={\textstyle{1\over 2}}\,u^\dagger \,\big[\partial_\mu -i\,(v_\mu + a_\mu) \big] \,u
+{\textstyle{1\over 2}}\, u \,
\big[\partial_\mu -i\,(v_\mu - a_\mu)\big] \,u^\dagger\,,  
\nonumber\\
&&  D_\mu \, B \;\,= \partial_\mu B +  \Gamma_{\mu}\, B -
B\,\Gamma_{\mu} \,, \qquad \qquad \qquad
 u = e^{i\,\frac{\Phi}{2\,f}}  \,,
\label{def-L1}
\end{eqnarray}
for the baryon spin 1/2 and 3/2 fields $B$ and $B_\mu$. Here we encounter the covariant derivative $D_\mu$ which involves the vector and axial vector
source fields $v_\mu$ and $a_\mu$. The SU(3) matrix field of the Goldstone bosons is $\Phi$. Throughout this chapter we apply the convenient flavour '$\cdot $' product notation for terms involving 
the symmetric baryon fields $B_\mu$ and $\bar B_\mu$ (see  \cite{Lutz:2001yb,Semke:2011ez}). For instance the product of the two flavour decuplet fields  $\bar B_\mu \cdot B_\nu $  is constructed
to transform as a flavour octet.

\subsection{Matrix elements of an axial-vector  current }
\vskip0.3cm

We recall the well known operator analysis for the axial-vector current, which correlate the axial-vector coupling constants, $F,D,C,H$ 
of the baryon octet and decuplet states \cite{Dashen1994,Dashen1995}. 
The matrix elements at zero momentum transfer,
\begin{eqnarray}
 &&\bra{\bar p\,,\,\bar\chi\,}\,A^{(a)}_i(0)\,
\ket{\,p\,,\chi } = \roundbra{\bar\chi\,}\,  \hat g_1\,G^{a}_i + \hat g_2\,\big\{ J_i\,,\,T^a\big\}\,
\roundket{\chi } + {\mathcal O}\left(\frac{1}{N_c} \right)\,,
\label{1-axial-opt}
\end{eqnarray}
are parameterized in terms of $\hat g_1$ and $\hat g_2$ and the spin, $J_i$, flavour, $T^a$ and spin-flavour, $G_i^a$ operators \cite{LutzSemke2010}. 
It suffices to compute matrix elements at $N_c=3$ only for which the effective spin-flavour states $\ket{\,p\,,\chi }$ are introduced. As emphasized in \cite{Lutz:2001yb,Semke:2011ez,Semke:2012gs}
the hierarchy of the large-$N_c$ sum rules can efficiently be derived once a complete compilation of the action of the three operators $J_i$, $T^a$ and $G_i^a$ on the spin-flavour states $\ket{\,p\,,\chi }$ is available. 
Such a compilation is provided in \cite{Lutz:2001yb,Semke:2011ez} for the first time. 
The $1/N_c$ expansion is implied by a proper selection of spin-flavour operators in the truncation (\ref{1-axial-opt}) (see \cite{tHooft74,Witten1979,Luty1994,Dai1995} for more technical details). 
We recall the relevant terms in the chiral Lagrangian
\begin{eqnarray}
&& \mathcal{L}^{(1)}= F\, \mathrm{tr} \Big\{ \bar{B}\, \gamma^\mu \gamma_5\, [i\,U_\mu,B]\, \Big\} + D\, \mathrm{tr}\Big\{ \bar{B}\, \gamma^\mu \gamma_5\, \{i\,U_\mu,\,B\}\, \Big\}
\nonumber \\
&& \qquad +\, C\left( \mathrm{tr} \Big\{ (\bar{B}_\mu \cdot i\, U^\mu)\, B\Big\} + \mathrm{h.c.} \right)
+ H\, \mathrm{tr} \Big\{ (\bar{B}^\mu\cdot \gamma_5\,\gamma_\nu   B_\mu)\, i\,U^\nu \Big\}\,,
\nonumber\\
&& U_\mu = {\textstyle{1\over 2}}\,u^\dagger \, \big(
\partial_\mu \,e^{i\,\frac{\Phi}{f}} \big)\, u^\dagger
-{\textstyle{i\over 2}}\,u^\dagger \,(v_\mu+ a_\mu)\, u
+{\textstyle{i\over 2}}\,u \,(v_\mu-a_\mu)\, u^\dagger\;, 
\label{def-FDCH}
\end{eqnarray}
where the object $(\bar B_\mu \cdot U^\mu)$ transforms as a flavour octet by construction. 
For the  particular case (\ref{1-axial-opt}) the well known hierarchy of sum rules 
\begin{eqnarray}
&& F= \frac{2}{3}\, D \,,\qquad C=2\,D \,,\qquad H = 3\,D\,, \qquad {\rm for } \qquad \hat g_2 =0  \,,
\nonumber\\
&&  C=2\,D\,,\qquad H= 9\,F-3\,D \,,\qquad\qquad \qquad \! {\rm for } \qquad \hat g_2 \neq 0  \,,
\label{res-FDCHs}
\end{eqnarray}
is recovered.  At leading order $( N_c)$ there are three sum rules and  at subleading order $(N_c^0)$ there remain two sum rules only. 

\subsection{Matrix elements of  a scalar current }
\vskip0.3cm

Less well studied are matrix elements of the scalar current
\begin{eqnarray}
&&\bra{\bar p\,,\,\bar\chi\,}\,S^{(a)}(0)\,
\ket{\,p\,,\chi } = \roundbra{\bar\chi\,}\, \delta_{a0}\,\big(\hat b_1\,{\mathbbm 1} + \hat b_2\,J^2 \big) 
\nonumber\\
&& \qquad \qquad +\, \hat b_3\,T^{a} + \hat b_4\,\big\{ J_i\,,\,G_i^a\big\}\,
\roundket{\chi } + {\mathcal O}\left(\frac{1}{N^2_c} \right)\,,
\label{1-scal-opt}
\end{eqnarray}
where the spin-flavour operators are supplemented by their flavour 
singlet counter parts
\begin{eqnarray}
 T^0 = \sqrt{ \frac 1 6}\, {\mathbbm 1} \,, \qquad G_i^0 =\sqrt{ \frac 1 6}\,J_i \,.
\end{eqnarray}
We recall the relevant terms in the chiral Lagrangian
\begin{eqnarray}
&& \mathcal{L}^{(2)}_\chi = 2\, b_0 \,\mathrm{tr} \left(\bar B \,B\right) \mathrm{tr}\left(\chi_+\right) + 2 \,b_D\,\mathrm{tr}\left(\bar{B}\,\{\chi_+,\,B\}\right) + 2\, b_F\,\mathrm{tr}\left(\bar{B}\,[\chi_+,B]\right) \nonumber \\
&& \qquad  -\, 2\, d_0\, \mathrm{tr}\left(\bar B_\mu \cdot B^\mu\right) \mathrm{tr}(\chi_+) - 2\, d_D\, \mathrm{tr} \left( \left(\bar{B}_\mu \cdot B^\mu\right) \chi_+\right)\,,
\nonumber\\
&& \chi_+ = {\textstyle{1\over 2}}\, \big( u \,\chi_0 \,u + u^\dagger \,\chi_0 \,u^\dagger \big) = 2\,B_0\, s + \cdots \,,
\label{def-b-d}
\end{eqnarray}
where we encounter the scalar source field $s$. For the  particular case (\ref{1-scal-opt}) the  hierarchy of sum rules 
\begin{eqnarray}
&& b_D = 0\,,\qquad  d_0 - b_0 = - \frac13\,d_D\,,\qquad d_D = 3 \,b_F\, \qquad {\rm for} \qquad \hat b_{2,4} = 0 \,,
\nonumber\\
&& b_F + b_D = \frac13\,d_D\,, \qquad  d_0+ \frac13\,d_D- b_0 = 2\,b_D\,, \; \qquad {\rm for} \qquad \hat b_{2} = 0 \,,
\nonumber\\
&& b_F + b_D = \frac13\,d_D\,, \qquad \qquad \qquad \qquad \qquad \qquad \qquad \;{\rm for} \qquad \hat b_{2,4} \neq 0 \,,
\label{res-bds}
\end{eqnarray}
is predicted. The relations (\ref{res-bds}) are a straight forward consequence of the technical results of \cite{Dashen1994,Dashen1995,LutzSemke2010}. 
At leading order $( N_c)$ there are three sum rules, at subleading order $(N_c^0)$ there are two sum rules and at the 
accuracy level $(1/N_c)$ there remains one sum rule only. The leading order symmetry breaking parameters of the chiral 
Lagrangian $b_0,b_F, b_D$ and $d_0, d_D$ as used in \cite{Semke:2011ez,Semke:2012gs,Lutz:2014oxa} are correlated. 

Such relations (\ref{res-bds}) were obtained before in \cite{Jenkins1995}, however, with a method  
slightly distinct to our approach. In  \cite{Jenkins1995} the contribution of the counter terms to the baryon masses 
was analyzed and expanded in powers of the flavour breaking parameter 
\begin{eqnarray}
\varepsilon \sim m_s- \frac{1}{2}\,(m_u +m_d)\,,
\end{eqnarray}
with the current quark masses $m_{u,d,s}$ of QCD. 
Accordingly the large-$N_c$ operator expansion was sorted according to powers of 
the $T_8$ and $G_8^i$ operators. At subleading order in $\varepsilon$, the following operators were considered
\begin{eqnarray}\label{lbl:Nc_baryonmass_epsilon2}
&& M_B = \roundbra{\bar \chi}\, c_1^{(0,1)}\mathbbm 1 + c_2^{(0,1)} J_i\, J_i + \varepsilon\, c_1^{(0,8)} T^8 
\nonumber\\
&& \qquad \qquad +\, \varepsilon\,c_2^{(0,8)} \{J_i, G^{8}_i\} + \varepsilon^2\, c_2^{(0,27)} \{T^8, T^8\}\, \roundket{\chi}\,,
\quad 
\label{old-ces}
\end{eqnarray}
where for the matching of the symmetry breaking parameters $b_0,b_F, b_D$ and 
$d_0, d_D$ the first four operators are relevant only. Keeping the first four operators in (\ref{old-ces}) 
the last sum rule in (\ref{res-bds}) is obtained. 

\subsection{Product of a scalar and a vector current}
\vskip0.3cm

A further set of symmetry breaking low-energy constants of the chiral Lagrangian can be large-$N_c$ correlated upon a study of 
the current-current correlation with one scalar and one vector current
\begin{eqnarray}
&& O^{ab}_\mu (q) = i\,\int d^4 x \,e^{-i\,q\cdot x}  \,{\mathcal T}\,S^{(a)} (x)\,V^{(b)}_\mu(0) \,,
\label{def:scalar-scalar}
\end{eqnarray}
in the baryon states. Terms in the chiral Lagrangian that involve a covariant derivative and a scalar source term will contribute to this correlation function. 
In \cite{Semke:2012gs} such terms were constructed systematically in terms of the five parameters $\zeta_0,\zeta_D, \zeta_F$ and $\xi_0, \xi_D$, where the 
flavour structure resembles the one of the $Q^2$ parameters $b_0,b_D, b_F$ and $d_0, d_D$ studied in the previous section. The wave-function renormalization terms 
of the chiral Lagrangian are recalled  with
\begin{eqnarray}
&& \mathcal{L}^{(3)}_\chi = \zeta_0\, \mathrm{tr} \big(\bar{B}\, (i\,\Dslash -M_{[8]})\, B\big)\, \mathrm{tr}(\chi_+) + \zeta_D\, \mathrm{tr} \big(\bar{B}\, (i\,\Dslash -M_{[8]})\, [B, \chi_+]\big) \nonumber\\
&& \qquad + \,\zeta_F\, \mathrm{tr} \big(\bar{B}\, (i\,\Dslash -M_{[8]})\, \{B, \chi_+\} \big) 
- \xi^{}_0\, \mathrm{tr} \big(\bar B_\mu\,\cdot  (i\,\Dslash -M_{[10]})\, B^\mu \big)\, \mathrm{tr}(\chi_+) 
\nonumber \\
&& \qquad -\, \xi^{}_D\, \mathrm{tr} \big( \big(\bar B_\mu\,\cdot (i\,\Dslash -M_{[10]})\,B^\mu \big)\, \chi_+ \big) \,.
\label{def-zeta-xi}
\end{eqnarray}
We introduce the leading orders operator ansatz for the correlator as
\begin{eqnarray}
&&\bra{\bar p, \, \bar \chi }\, O^{ab}_i\, \ket{p,\, \chi} = (\bar p+p)_i\,\roundbra{\bar \chi}\,
+ \delta_{a0}\,\hat \zeta_1\, T^{b}
\nonumber\\
&& \qquad \qquad+\, \hat \zeta_2\,\{T^{a} ,\,T^{b}\} + \hat \zeta_3\,\{J_i,\,\{T^b,\, G^a_i\}\} \,\roundket{\chi} + \cdots \,,
\label{def-SV-expansion}
\end{eqnarray}
which leads to the following two sum rules
\begin{eqnarray}
&& 3\,\zeta_F + 3\,\zeta_D= \xi_D \,,\qquad \qquad  
\xi_0+ \frac{1}{3}\,\xi_D=\zeta_0+ 2\,\zeta_D\,.
\label{def-combinations}
\end{eqnarray}
If the subleading operator with $\hat \zeta_3$ in (\ref{def-SV-expansion}) would be dropped the additional relation $\zeta_D=0$ arises. Such results are in full analogy to the 
relations obtained for the symmetry breaking $Q^2$ parameters in (\ref{res-bds}). 

\subsection{Product of two scalar currents}

\vskip0.3cm

We turn to matrix elements of the product of two scalar currents. They probe the symmetry breaking counter terms $c_i$ and $e_i$ of 
the chiral Lagrangian that are proportional to the square of the current quark masses. We recall their particular form
\begin{eqnarray}
\mathcal{L}^{(4)}_\chi &=& c_0\, \mathrm{tr}\left(\bar B\, B\right) \mathrm{tr} \left(\chi_+^2\right) + c_1\, \mathrm{tr} \left(\bar B \,\chi_+\right) \mathrm{tr}\left(\chi_+ B\right) \nonumber \\
&+& c_2\, \mathrm{tr} \left( \bar B\, \{\chi_+^2,\, B\} \right)  + c_3\,\mathrm{tr} \left( \bar B\, [\chi_+^2, \,B] \right)
\nonumber \\
&+& c_4\, \mathrm{tr} \left(\bar B\, \{\chi_+,\,B\} \right) \mathrm{tr} (\chi_+) + c_5\, \mathrm{tr}\left(\bar B\, [\chi_+,\,B]\right) \mathrm{tr} (\chi_+) \nonumber \\
&+& c_6\, \mathrm{tr} \left(\bar B \,B\right) \left(\mathrm{tr}(\chi_+)\right)^2
\nonumber \\
&-& e_0\, \mathrm{tr}\left(\bar B_\mu \cdot B^\mu \right) \mathrm{tr}\left(\chi_+^2\right) - e_1\, \mathrm{tr}\left( \left(\bar{B}_\mu \cdot \chi_+\right) \left(\chi_+ \cdot B^\mu\right) \right)
\nonumber \\
&-& e_2\, \mathrm{tr}\left( \left(\bar B_\mu \cdot B^\mu\right)\cdot \chi_+^2\right) - e_3\, \mathrm{tr}\left( \left(\bar B_\mu \cdot B^\mu\right)\cdot \chi_+\right) \mathrm{tr}(\chi_+) \nonumber \\
&-& e_4\, \mathrm{tr}\left(\bar B_\mu \cdot B^\mu\right) \left(\mathrm{tr}(\chi_+)\right)^2\,.
\label{def-c-e}
\end{eqnarray}
Sum rules are derived by a study of the following 
matrix elements
\begin{eqnarray}
&&i\, \int  d^4 x \,e^{-i\,q\cdot x}  \,\bra{\bar p\,,\,\bar\chi\,}\, {\mathcal T}\,S^{(a)} (x)\,S^{(b)}(0) \ket{\,p\,,\chi }
=  \roundbra{\bar\chi\,}\,\hat c_1\,\delta_{a0}\,\delta_{b0}\,{\mathbbm 1}
\nonumber\\
&& \quad \!+ \,\hat c_2\,\delta_{ab}\,{\mathbbm 1}
+ \hat c_3\,\big( T^a\,\delta_{b0} + \delta_{a0}\,T^b \big) 
+ \hat c_4\,d_{abc}\,T^c +\hat c_5\,\{ T^a,T^b\}
\nonumber\\
&&  \quad \!+\,  \hat c_6\,d_{abc}\,\{ J_i,G_i^c\}
 + \hat c_7\,\Big( \big\{ J_i\,,\,G_i^a \big\}\,\delta_{b0}
+ \delta_{a0}\,\big\{ J_i\,,\,G_i^b \big\}\Big)
\nonumber\\
&&  \quad \!+\, \hat c_8 \,\Big( \big\{ J_i\,,\big\{T^a,\,\,G_i^b \big\}\big\} +\big\{ J_i\,,\big\{T^b,\,\,G_i^a \big\}\big\} \Big)
\roundket{\chi } 
+ \cdots\,,
\label{SS-expanded}
\end{eqnarray}
where we consider leading and subleading order operators only. Note that the two sums over $c$   in (\ref{SS-expanded})  are from $c=0$ to $c=8$.
At leading order with $\hat c_6 =\hat c_7 =\hat c_8 =0$
a matching with the tree-level expression derived from the chiral Lagrangian leads to the following 7 sum rules
\begin{eqnarray}
&& 2\,c_2=-3\,c_1\,,\qquad 2\,c_0 = c_1 +2\,(c_3+e_0)\,, \qquad 3\,c_1 =e_1\,,\qquad c_4=c_1\,,
\nonumber\\
&& 3\,e_1+2\,e_2 =6\,c_3 \,, \qquad e_3=3\,(c_4+c_5)\,, \qquad c_6 = c_5+e_4 \,.
\label{res-ces}
\end{eqnarray}
At subleading order there remain 4 sum rules only
\begin{eqnarray}
&& c_0 = 2\,c_3 + e_0 - {\textstyle{ 1\over 6}}\,e_1 - {\textstyle{ 1\over 3}}\,e_2 - {\textstyle{ 1\over 2}}\,c_1\,, \qquad 
c_1 = {\textstyle{ 1\over 3}}\,(e_1 + e_2) - c_2 - c_3\,,
\nonumber\\
&& e_3 = 3\,(c_4 + c_5)\,, \qquad
 e_4 = c_0 + c_2 + c_4 + c_6 - c_3 - c_5 - e_0\,.
\label{ces-subleading}
\end{eqnarray}

It is interesting to observe that sum rules (\ref{res-ces}) differ from the sum rules applied previously 
in \cite{Semke:2011ez,Semke:2012gs,Lutz:2014oxa}
 \begin{eqnarray}
&&c_1 = 2\,c_0, \qquad c_2=-3\,c_0,\qquad  c_3=0 \,, \quad
\nonumber\\
&& e_0 = 0\,, \quad e_1 = 6\, c_0, \qquad  e_2 = -9\,  c_0\,, \qquad  e_3 = 3\, ( c_4 + c_5)\,.
\label{result:large-Nc-chi}
\end{eqnarray} 
The latter were derived in application of the $\varepsilon$ expansion (\ref{old-ces}), where the contribution of 
the counter terms to the baryon masses was considered. Here the fifths term proportional to $\varepsilon^2$ turns relevant. 
The 12 symmetry breaking counter terms are expressed in terms of the five coefficients in (\ref{old-ces}). This leads to the 
7 sum rules in (\ref{result:large-Nc-chi}). 

\subsection{Product of two axial-vector currents}

\vskip0.3cm

There remain the sum rules derived from the product of two axial-vector currents, studied previously in \cite{LutzSemke2010}. They correlate
two-body meson-baryon counter terms. A complete list of chiral symmetry conserving $Q^2$ counter terms was given in \cite{Lutz2002a,LutzSemke2010}.
Here we recall all terms relevant for the calculation of the N$^3$LO baryon masses. The terms are grouped according to 
their Dirac structure $ \mathcal{L}^{(2)}=\mathcal{L}^{(S)} + \mathcal{L}^{(V)} $ with
\allowdisplaybreaks[1]
\begin{eqnarray}
&& \mathcal{L}^{(S)} =- \frac{1}{2}\,g_0^{(S)}\,\mathrm{tr} \,\Big\{\bar{B}\,B \Big\}\, \mathrm{tr}\Big\{ U_\mu\, U^\mu \Big\} - \frac{1}{2}\,g_1^{(S)}\,\mathrm{tr} \,\Big\{ \bar{B}\, U^\mu \Big\}\, \mathrm{tr}\,\Big\{U_\mu\, B \Big\}
\nonumber \\
&&\qquad -\,\frac{1}{4}\,g_D^{(S)} \mathrm{tr}\,\Big\{\bar{B}\left\{\left\{U_\mu, U^\mu\right\}, B\right\}\Big\}
-\frac{1}{4}\,g_F^{(S)}\mathrm{tr}\,\Big\{ \bar{B}\left[\left\{U_\mu, U^\mu\right\}, B\right]\Big\}
\nonumber\\
&&\qquad + \,\frac{1}{2}\,h_1^{(S)}\,\mathrm{tr}\,\Big\{ \bar{B}_\mu \cdot B^\mu \Big\}\, \mathrm{tr}\,\Big\{U_\nu\; U^\nu\Big\} +
\frac{1}{2}\,h_2^{(S)}\,\mathrm{tr}\,\Big\{\bar{B}_\mu \cdot B^\nu \Big\}\, \mathrm{tr}\,\Big\{U^\mu\, U_\nu\Big\}
\nonumber \\
&& \qquad + \,h_3^{(S)}\,\mathrm{tr}\,\Big\{\Big(\bar{B}_\mu \cdot B^\mu\Big)\, \Big(U^\nu\, U_\nu\Big) \Big\} + \frac{1}{2}\,h_4^{(S)}\,\mathrm{tr}\,\Big\{ \Big(\bar{B}_\mu \cdot B^\nu\Big)\, \{U^\mu,\, U_\nu \} \Big\}
\nonumber \\
&&\qquad  +\, h_5^{(S)}\, \mathrm{tr}\, \Big\{ \Big( \bar{B}_\mu \cdot U_\nu\Big)\, \Big(U^\nu\cdot B^\mu \Big) \Big\}
\nonumber \\
&& \qquad +\, \frac{1}{2}\,h_6^{(S)}\, \mathrm{tr} \Big\{ \Big( \bar{B}_\mu \cdot U^\mu\Big)\, \Big(U^\nu\cdot B_\nu \Big)
+\Big( \bar{B}_\mu \cdot U^\nu\Big)\, \Big(U^\mu\cdot B_\nu \Big) \Big\} \, ,
\nonumber\\
&&\mathcal{L}^{(V)} = -\frac{1}{4}\,g_0^{(V)}\, \Big( \mathrm{tr}\,\Big\{\bar{B}\, i\,\gamma^\mu\, D^\nu B\Big\} \,
\mathrm{tr}\,\Big\{ U_\nu\, U_\mu \Big\}\Big)
\nonumber \\
&& \qquad -\, \frac{1}{8}\,g_1^{(V)} \,\Big( \mathrm{tr}\,\Big\{\bar{B}\,U_\mu \Big\} \,i\,\gamma^\mu \, \mathrm{tr}\,\Big\{U_\nu\, D^\nu B\Big\} 
+ \mathrm{tr}\,\Big\{\bar{B}\,U_\nu \Big\} \,i\,\gamma^\mu \, \mathrm{tr}\,\Big\{U_\mu\, D^\nu B\Big\} \Big)
\nonumber \\
&& \qquad -\, \frac{1}{8}\,g_D^{(V)}\, \Big(\mathrm{tr}\,\Big\{\bar{B}\, i\,\gamma^\mu \left\{\left\{U_\mu,\, U_\nu\right\}, D^\nu B\right\}\Big\} \Big)
\nonumber\\
&& \qquad -\, \frac{1}{8}\,g_F^{(V)}\,\Big( \mathrm{tr}\,\Big\{ \bar{B}\, i\,\gamma^\mu\, \left[\left\{U_\mu,\, U_\nu\right\},\, D^\nu B \right]\Big\}  \Big)
\nonumber \\
&& \qquad +\, \frac{1}{4}\,h_1^{(V)}\,\Big(\mathrm{tr}\,\Big\{ \bar{B}_\lambda \cdot i\,\gamma^\mu\, D^\nu B^\lambda\Big\} \,\mathrm{tr}\,\Big\{U_\mu\, U_\nu\Big\}\Big)
\nonumber \\
&& \qquad +\, \frac{1}{4}\,h_2^{(V)}\,\Big(\mathrm{tr}\,\Big\{ \left(\bar{B}_\lambda \cdot i\,\gamma^\mu\, D^\nu B^\lambda \right) \{U_\mu,\, U_\nu\}\Big\} \Big)
\nonumber \\
&& \qquad +\, \frac{1}{4}\,h_3^{(V)}\, \Big( \mathrm{tr}\, \Big\{ \left( \bar{B}_\lambda \cdot U_\mu\right) i\,\gamma^\mu \left(U_\nu\cdot D^\nu B^\lambda \right)
\nonumber\\
&& \qquad \qquad \qquad \qquad +\, \left( \bar{B}_\lambda \cdot U_\nu\right) i\,\gamma^\mu \left(U_\mu\cdot D^\nu B^\lambda \right) \Big\} \Big)
 + \mathrm{h.c.} \, .
\label{def-Q2-terms}
\end{eqnarray}
We extend the previous large-$N_c$ analysis  \cite{LutzSemke2010} and construct the subleading contributions.  Altogether we find the 
relevance of the following operators
\begin{eqnarray}
&&i\,\int d^4 x \,e^{-i\,q\cdot x}  \,\bra{\bar p\,,\,\bar\chi\,}\, {\mathcal T}\,A^{(a)}_i (x)\,A^{(b)}_j(0) \ket{\,p\,,\chi }
\nonumber\\
& & = ( \bar \chi | - \delta_{ij}\,\Big[
\, \hat g_1\,\delta_{ab}\,\quarknumberoperator+\hat g_2\,
d_{abc}\,T^c  + \hat g_3\,\{T^a,\,T^b\}  + \hat g_4\,d_{abc}\, \{J_k,\,G_k^c\} 
\nonumber\\
& & \quad + \,\hat g_5\,\Big( \{ J_k,\, \{T^{a},\,G^{b}_k\} \} + \{ J_k,\,\{G^{a}_k,\,T^{b} \} \Big)   \Big]
+ \hat g_6\,\Big( \{G^{a}_i,\,G^{b}_j\}  +\{G^{a}_j,\,G^{b}_i \} \Big)
\nonumber\\
& &  + \,(\bar p+p)_i \,(\bar p+p)_j \,\Big[
 \,\hat g_7\,\delta_{ab}\,\quarknumberoperator+ \hat g_8\,
d_{abc}\,T^c + \hat g_9\,\{T^a,\,T^b \} + \hat g_{10}\,d_{abc}\, \{J_k,\,G_k^c\} 
\nonumber\\
& & \quad 
+ \,\hat g_{11}\,\Big( \{ J_k,\, \{T^{a},\,G^{b}_k\} \} + \{ J_k,\,\{G^{a}_k,\,T^{b} \} \Big) \Big]
 \, | \chi ) + \cdots \,.
\label{QCD-identity-AA}
\end{eqnarray}
A matching  with the terms form (\ref{def-Q2-terms}) leads to the following six sum rules
\begin{eqnarray}
&& \bar g^{(S)}_F =  \bar g^{(S)}_0 + \frac{3}{2}\, \bar g^{(S)}_1 +  \bar g^{(S)}_D -  \bar h^{(S)}_1 - \frac{1}{3}\, \bar h^{(S)}_5 + \frac{2}{9}\, \bar h^{(S)}_6\,,
\nonumber\\
&&  \bar h^{(S)}_2 = 0\,, \qquad  \bar h^{(S)}_4 = -  \bar h^{(S)}_6\,,
\nonumber\\
&&  \bar h^{(S)}_3 = \frac{3}{2}\,\bar g^{(S)}_0 + \frac{15}{4}\, \bar g^{(S)}_1 + 3\, \bar g^{(S)}_D - \frac{3}{2}\, \bar h^{(S)}_1 
- \frac{3}{2}\, \bar h^{(S)}_5 + \frac{1}{3}\, \bar h^{(S)}_6\,, \qquad 
\nonumber\\ \nonumber\\      
&& \bar g^{(V)}_F =  \bar g^{(V)}_0 + \frac{3}{2}\, \bar g^{(V)}_1 +  \bar g^{(V)}_D -  \bar h^{(V)}_1 - \frac{1}{3}\, \bar h^{(V)}_3\,, \qquad
\nonumber\\      
&&   \bar h^{(V)}_2 = \frac{3}{2}\, \bar g^{(V)}_0 + \frac{15}{4}\, \bar g^{(V)}_1 + 3\, \bar g^{(V)}_D - \frac{3}{2}\, \bar h^{(V)}_1 - \frac{3}{2}\, \bar h^{(V)}_3\,,
\label{Q4-subleading}
\end{eqnarray}
with $\bar g = g$ and $\bar h = h$.
The leading order operators of \cite{LutzSemke2010} are recovered with $3\,\hat g_1 = \hat g_2$, $3\,\hat g_7 = \hat g_8$ and $\hat g_4 = \hat g_5 = 0 =
\hat g_{10} = \hat g_{11} =0 $.  This leads to an additional six leading order sum rules
\begin{eqnarray}
&&  \bar h^{(S)}_1 = 0\,,\qquad 
 \bar h^{(S)}_5 =  \bar g^{(S)}_D + 3\, \bar g^{(S)}_1\,\qquad   \bar h^{(S)}_6 = -3\,\Big( \bar g^{(S)}_D + \frac{3}{2}\,\bar g^{(S)}_1 \Big)\,,  
\nonumber\\
&& \bar  h^{(V)}_1= 0 \,, \qquad  \bar g^{(V)}_D = -\frac{3}{2}\,\bar g^{(V)}_1\,,\qquad \quad     \bar h^{(V)}_3 = \frac{3}{2}\, \bar g^{(V)}_1\,.
\label{Q4-leading}
\end{eqnarray}
Combining the sum rules of (\ref{Q4-subleading}, \ref{Q4-leading}) confirms the leading order sum rules derived first in \cite{LutzSemke2010}
\footnote{We correct a misprint in \cite{LutzSemke2010}. The coupling constants $g^{(V)}_{0,1,F,D}$ should be 
multiplied by the factor $1/2$ in equations (32).}. 

As a further cross check we computed the contributions from the s- and u-channel baryon 
exchange diagrams. They may modify the sum rules obtained previously and lead to a renormalization of the low-energy constants with $g \to \bar g$ and $h \to \bar h$. 

Here we need to be more specific about the kinematics 
of the correlation function. We assume on-shell baryons with $p_0= \sqrt{M^2+\vec p\,^2} = \bar p_0$. The energy of the current $A^{(a)}_i$ in (\ref{QCD-identity-AA}) is $q_0 $. 
For simplicity we assume the frame where $\vec p+\vec q= 0$, i.e. we have only the two three vectors of $ p$ and $\bar p$ around. 
In an analysis of the correlation function there are additional terms that are singular in $q_0$, which are a consequence of the s- 
and u-channel exchange diagrams.  The sum rules (\ref{Q4-subleading}, \ref{Q4-leading}) with $\bar g = g$ and $ \bar  h =  h$ follow from the analysis of the 
contributions that are regular in the limits $q_0 \to 0$ and $q_0 \to \pm\, 2\,M$.  However, significant results can only be expected if in addition the poles 
at $q_0 \to \pm \,2\,M$ are expanded for small $q_0$. This leads to a renormalization of the sum rules. The result of such an analysis  is 
\begin{eqnarray}
&& \bar g_0^{(S)} = g_0^{(S)} - \frac{4}{3}\, g^{(S)}_C \,,  \qquad \quad 
 \bar g_1^{(S)} = g_1^{(S)}  +\frac{1}{3}\,g^{(S)}_C \,,  
\nonumber\\
&& \bar g_D^{(S)} =   g_D^{(S)} + g^{(S)}_C\,,   \qquad \qquad 
 \bar g_F^{(S)} =g_F^{(S)}  - g^{(S)}_C \,,\qquad g^{(S)}_C =  \frac{2\,C^2}{3\,M}\,\alpha_5 \,,
\nonumber\\ 
&& \bar g_0^{(V)} =  g_0^{(V)}  - \frac{4}{3}\, g^{(V)}_C \,, \qquad 
\bar  g_1^{(V)} =  g_1^{(V)}   + \frac{1}{3}\,g^{(V)}_C\,,
\nonumber\\
&& \bar g_D^{(V)} =   g_D^{(V)}  + g^{(V)}_C\,,\qquad \quad 
\bar g_F^{(V)} =  g_F^{(V)} - g^{(V)}_C\,,\qquad g^{(V)}_C = \frac{C^2}{3\,M^2}\,\alpha_{6}  \,,
\nonumber\\
&&\bar  h_1^{(V)} =  h_1^{(V)}   \,, \qquad \qquad \qquad 
 \bar h_2^{(V)} =  h_2^{(V)}  - \frac{2}{9}\,\frac{H^2}{(M+ \Delta)^2}\,, 
\nonumber \\
&& \bar h_3^{(V)} = h_3^{(V)}  + \frac{4}{27}\,\frac{H^2}{(M+ \Delta)^2} -  \frac{1}{6}\,\frac{\beta_{6}\,C^2}{M\,(M+\Delta)}\,,
\nonumber\\ \nonumber\\
&& \bar h_n^{(S)} = h_n^{(S)} \quad {\rm for} \qquad n \neq 5 \qquad {\rm and }\qquad \bar h_5^{(S)} = h_5^{(S)} +  \frac{C^2}{3\,M}\,\beta_5\,,
\label{Q4-renormalization}
\end{eqnarray}
where the coefficients $\alpha_{5,6}$ and $\beta_{5,6}$ depend on the ratio $\Delta/M$ only and approach $ 1$ in the limit with $\Delta \to 0$ (see Appendix A and B). 

An analysis of the singular terms may be used to correlate the axial-vector coupling constants, $F,D,C, H$ of the baryon octet 
and decuplet states. It is more convenient, however, to derive such sum rules in terms of matrix elements of a single axial-vector 
current (\ref{1-axial-opt}).

\newpage

\section{Chiral extrapolation of baryon masses}

We consider the chiral extrapolation of the baryon masses to unphysical quark masses. Assuming exact isospin symmetry, 
the hadron masses are functions of  $m_u=m_d\equiv m$ and $m_s$. The ultimate goal is to establish a decomposition of the baryon masses 
into their power-counting moments
\begin{eqnarray}
M_B = \sum_{n=0}^\infty M^{(n)}_B \,,
\end{eqnarray}
where there is a significant controversy in the community \cite{Bernard:1993nj,Banerjee:1994bk,Banerjee:1995wz,Lehnhart2004,Semke2005} 
as to whether any significant results can be obtained in the flavour SU(3) case.

At leading order the octet and decuplet masses are determined by the tree-level mass parameters of the chiral Lagrangian. 
At next-to-leading order the chiral symmetry breaking counter 
terms $b_0,b_D, b_F$ and $d_0, d_D$ in (\ref{res-bds}) turn relevant
\begin{eqnarray}
&& M^{(2)}_{N}=  -2\,\mm{\pi}\,(b_0+2\,b_F) - 4\,\mm{K}\, (b_0+b_D-b_F)\,,
\nonumber\\
&& M^{(2)}_{\Sigma} - M^{(2)}_{\Lambda}={\textstyle{16\over 3}}\, b_D\, (\mm{K}-\mm{\pi})\,, \quad
 M^{(2)}_{\Xi} - M^{(2)}_{N}=- 8\, b_F \,(\mm{K}-\mm{\pi})\,,
\nonumber\\
&& M^{(2)}_\Xi- M^{(2)}_\Sigma = - 4\,(b_D+b_F)\,(\mm{K}-\mm{\pi})
\,,
\nonumber\\ \nonumber\\
&&  M^{(2)}_{\Delta}
=  - 2\,(d_0+d_D)\,\mm{\pi}-4\, d_0\,\mm{K}\,,
\nonumber\\
&& M^{(2)}_\Sigma-  M^{(2)}_\Delta
= - {\textstyle{4\over 3}}\,d_D \,(\mm{K}-\mm{\pi}) \,, \quad 
 M^{(2)}_{\Xi}-  M^{(2)}_\Sigma
= -  {\textstyle{4\over 3}}\,d_D\,(\mm{K}-\mm{\pi}) \,,
\nonumber \\
&& M^{(2)}_{\Omega}- M^{(2)}_\Xi
= -  {\textstyle{4\over 3}}\,d_D\,(\mm{K}-\mm{\pi}) \,,
\label{res-Q2}
\end{eqnarray}
where the meson masses are to be taken at leading order with e.g. $\mm{\pi} = 2\,B_0\,m$. 
Based on (\ref{res-Q2}) the parameters $b_D, b_F$ and $d_D$ may be adjusted to the mass differences of the
physical baryon states:
\begin{eqnarray}
&& b_D \simeq  0.07 \,{\rm GeV}^{-1} \,, \quad 
b_F \simeq - 0.21 \,{\rm GeV}^{-1} \,, \quad
 d_D \simeq - 0.49  \,{\rm GeV}^{-1} \,,
\label{parameters-Q2A} 
\end{eqnarray}
where the uncertainties as implied by the use of distinct mass differences is rather small. 
The estimate of the parameter $b_0$ and $d_0 $ requires additional input. If we insist on the 
leading chiral moments of the baryon masses as deduced recently from a comprehensive analysis 
of the available lattice data \cite{Lutz:2014oxa} we obtain the estimates
\begin{eqnarray}
&& \qquad  M= M^{(0)}_N \simeq 800 \,{\rm MeV}  \qquad \qquad M+ \Delta = M^{(0)}_\Delta \simeq 1100 \,{\rm MeV} \,,
\nonumber\\
\rightarrow &&\qquad  b_0 \simeq -0.39\,{\rm GeV}^{-1} \,, \qquad \qquad \quad d_0 \simeq -0.11\,{\rm GeV}^{-1} \,.
\label{parameters-Q2B} 
\end{eqnarray}
All together at this order the physical baryon masses can be reproduced with an uncertainty of 3.1 MeV and 3.5 MeV for the octet and decuplet 
states respectively.  

Yet, it is well known that the loop contributions to the baryon masses that arise in a strict chiral expansion are very 
large - much too large as to suggest a convincing expansion pattern convergent at physical quark masses \cite{Bernard:1993nj,Banerjee:1994bk,Banerjee:1995wz,Lehnhart2004,Semke2005}. 
Thus the above parameter estimate, despite its deceiving success, may not to be very reliable. Nevertheless, such 
parameters play an important role in low-energy QCD. In principle they can be extracted from 3-flavour QCD lattice data at 
sufficiently small up, down and strange quark masses. While at present such simulations are not available one may explore 
that chiral power counting domain of QCD by analyzing extrapolation studies of the current lattice data set.

\subsection{The bubble-loop contributions to the baryon masses }
\vskip0.3cm

We take the most comprehensive analysis \cite{Lutz:2014oxa} that is based on relativistic 
kinematics and the use of physical masses in the loop expressions. This will serve as our starting point 
to reconsider the convergence properties of the chiral expansion for the baryon octet and decuplet masses with three 
light flavours. We will scrutinize further the particular summation scheme that is implied by the use of physical baryon and meson masses inside the 
one-loop expressions. In a more conventional scheme for such masses approximate values are assumed. For a rapidly 
converging system either of the two approaches is fine. In contrast, for a slowly converging system 
a summation scheme can be of advantage even though this may imply some model dependence. The target of this and the following sections is to decompose the 
one-loop expressions for the baryon self energies into chiral moments, which may depend on physical meson and baryon masses.  

The one-loop self energy of a baryon of type $B$, can be written as a sum of contributions characterized by $Q$ and $R$, where $Q$ and $R$ indicate 
the presence of an intermediate meson or baryon state of mass, $m_Q$ or $M_R$, respectively.  The previous computations \cite{Semke2005,Semke:2011ez,Semke:2012gs,Lutz:2012mq,Lutz:2014oxa} are 
based on the Passarino-Veltman reduction scheme. Results that are consistent with the expected chiral power as deduced by conventional dimensional counting rules are 
obtained by a minimal subtraction of the scalar loop diagrams that define the Passarino-Veltman reduction \cite{Semke2005}. At the one-loop level expressions can always be 
cast into a form where the scale dependence of a given diagram is exclusively 
determined by the coefficient in front of the scalar tadpole term $\bar I_Q$, a structure already encountered here in (\ref{def-tadpole-integral}). 
It was argued in \cite{Semke2005} that tadpole terms involving a 'heavy' particle mass, i.e. $M_R$ in our case, 
can be dropped consistently without violating chiral Ward identities. Scale independent results are obtained once suitable counter terms of chiral order $Q^4$ and higher 
are activated. 
The leading chiral order of the one-loop diagrams starts at chiral order $Q^3$, where one does not expect a scale dependence. Thus at this order  all terms proportional to 
$m_Q^{2\,n}\,\bar I_Q$ with $n \geq 1$ should be dropped at least. This is an immediate consequence 
of the chiral counting rule
\begin{eqnarray}
 \frac{m_Q^2}{M_R^2} \sim \frac{m_Q^2}{M_B^2} \sim Q^2\,.
 \label{def-counting}
\end{eqnarray}
Using the previous results derived in 
 \cite{Semke2005,Semke:2011ez,Semke:2012gs,Lutz:2012mq,Lutz:2014oxa} it is straight forward to verify the according expressions for the baryon octet and decuplet states with
\allowdisplaybreaks[1]
\begin{eqnarray}
&&\Sigma^{\rm bubble}_{B \in [8]} = \sum_{Q\in [8], R\in [8]}
\left(\frac{G_{QR}^{(B)}}{2\,f} \right)^2  \Bigg\{ \frac{M^2_R-M^2_B}{2\,M_B}\, \bar I_Q
\nonumber\\
&& \qquad \qquad \qquad  
- \,\frac{(M_B+M_R)^2}{E_R+M_R}\, p^2_{QR}\,\bar I_{QR} 
\Bigg\}
\nonumber \\
&& \qquad  \;\,\,\,+\sum_{Q\in [8], R\in [10]}
\left(\frac{G_{QR}^{(B)}}{2\,f} \right)^2 \, \Bigg\{
 \frac{(M_R+M_B)^2}{12\,M_B\,M^2_R}\,\Big(M^2_R-M^2_B\Big)\,\bar I_Q 
\nonumber\\
&& \qquad \qquad \qquad  
 -\, \frac{2}{3}\,\frac{M_B^2}{M_R^2}\,\big(E_R+M_R\big)\,p_{QR}^{\,2}\,
\bar I_{QR} + \frac{4}{3}\,\alpha^{(B)}_{QR}
   \Bigg\}\,,
\label{result-loop-8} \\ \nonumber\\
&& p_{Q R}^2 =
\frac{M_B^2}{4}-\frac{M_R^2+m_Q^2}{2}+\frac{(M_R^2-m_Q^2)^2}{4\,M_B^2} \,,\qquad \qquad 
E_R^2=M_R^2+p_{QR}^2 \,,
\nonumber\\ \nonumber\\
&&\Sigma^{\rm bubble}_{B\in [10]} = \sum_{Q\in [8], R\in [8]}
\left(\frac{G_{QR}^{(B)}}{2\,f} \right)^2  \Bigg\{ 
 \frac{(M_R+M_B)^2}{24\,M^3_B}\,\Big(M^2_R-M^2_B\Big)\,\bar I_Q 
\nonumber\\
&& \qquad \qquad \qquad 
-\,\frac{1}{3}\,\big( E_R +M_R\big)\,p_{QR}^{\,2}\,
\bar I_{QR}+ \frac{2}{3}\,\alpha^{(B)}_{QR}
\Bigg\}
\nonumber\\
&& \qquad \;\,\,\,+\sum_{Q\in [8], R\in [10]}
\left(\frac{G_{QR}^{(B)}}{2\,f} \right)^2 \, \Bigg\{
- \frac{ M_B^2 + M_R^2}{18\,M^2_B\,M_R}\,\Big(M^2_R-M^2_B\Big)\,\bar I_Q 
\nonumber\\
&& \qquad \qquad \qquad 
+\,\frac{M_R^4+M_B^4  + 12\,M_R^2\,M_B^2 }{36\,M^3_B\,M_R^2}\,\Big(M^2_R-M^2_B\Big)\,\bar I_Q 
\nonumber\\
&& \qquad \qquad \qquad 
-\,\frac{(M_B+M_R)^2}{9\,M_R^2}\,\frac{2\,E_R\,(E_R-M_R)+5\,M_R^2}{E_R+M_R}\,
p_{QR}^{\,2}\,\bar I_{QR} 
\Bigg\}\,,
\label{result-loop-10}
\end{eqnarray}
where the sums in (\ref{result-loop-8}, \ref{result-loop-10}) extend over the intermediate Goldstone bosons and  the baryon
octet and decuplet states. The notations of \cite{Semke2005,Semke:2011ez} are applied throughout this work. 
All Clebsch coefficients are listed therein. The coupling constants $G_{QR}^{(B)}$  are determined by the axial-vector 
coupling constants $F,D,C,H$ of the baryon states in (\ref{def-FDCH}). The renormalized scalar bubble loop 
integral $\bar I_{Q R}$ and the additional subtraction terms $\alpha^{(B)}_{QR}$ will be discussed in more detail below. 
It is emphasized that with (\ref{result-loop-8}, \ref{result-loop-10}) we have yet an intermediate result only. A further decomposition into chiral moments, in particular of the 
scalar bubble loop $\bar I_{Q R}$ is required. Such a decomposition will rely crucially on how to power count the mass differences $M_B-M_R$, 
a central issue of the following development.  

There is a subtle issue owing to the manner the scalar tadpole integral  $\bar I_Q$  appears in 
(\ref{result-loop-8}, \ref{result-loop-10}). 
The renormalized tadpole integral 
\begin{eqnarray}
&& \bar I_Q =\frac{m_Q^2}{(4\,\pi)^2}\,
\log \left( \frac{m_Q^2}{\mu^2}\right)\,,
\label{def-tadpole-integral}
\end{eqnarray}
depends on the renormalization scale $\mu$. Unlike a term proportional to 
$m_Q^2\,\bar I_Q$, which causes a renormalization of the symmetry preserving $g$ and $h$ coupling constants in (\ref{def-Q2-terms}), a term proportional to $(M_R -M_B)\,\bar I_Q$ cannot be 
dropped without jeopardizing the chiral Ward identities. Moreover such a term implied a particular renormalization scale dependence of the low-energy parameters $c_i$ and $e_i$ in (\ref{def-c-e}). 
From the results of Appendix A and B we extract the leading behavior in the $1/N_c$ expansion
\begin{eqnarray}
 \mu\, \frac{d}{d\,\mu} \,c^{\rm ano}_i \sim \left(\frac{\hat g_1}{4\,\pi f} \right)^2 \,\hat b_{1,3} \sim N^2_c\,,\quad \quad 
 \mu\, \frac{d}{d\,\mu} \,e^{\rm ano}_i \sim \left(\frac{\hat g_1}{4\,\pi f} \right)^2 \,\hat b_{1,3} \sim N^2_c\,,
 \label{def-c-e-ano}
\end{eqnarray}
which is determined by the axial coupling constant $\hat g_1 \sim N_c$ of (\ref{1-axial-opt}) and the 
scalar coupling constants $\hat b_1 \sim N_c $ and $\hat b_3 \sim N_c$ of (\ref{1-scal-opt}). We conclude the leading scaling behavior $\sim N_c^2$ is in conflict with the maximal scaling behavior $\sim N_c$ set by QCD.
This signals further terms not considered here that are expected to mend this anomalous behavior. A power-counting respecting remedy to this problem is provided by   
the following simple rewrite
\begin{eqnarray}
 \big(M_R -M_B\big)\,\bar I_Q = \underbrace{\frac{M_R-M_B}{(4\pi)^2}\,m_Q^2 \,\log \frac{m_Q^2}{M_R^2}}_{\equiv \,(M_R-M_B)\,I_Q^R} + \underbrace{\frac{M_R-M_B}{M_R^2}\,m_Q^2\,\bar I_R}_{\to \;{\rm counter \;terms}}\,,
 \label{eliminate-mu}
\end{eqnarray}
which suggests the scale dependent part to be systematically absorbed into counter terms of chiral order $Q^2, Q^4$ and higher. 
Owing to our renormalization prescription that drops all baryon tadpole contributions like $\bar I_R$
the second term in (\ref{eliminate-mu}) should be moved into counter terms in any case. Since we are not in the position to follow up all possible terms proportional to a baryonic tadpole $\bar I_R$
this offers an easy way to get rid of the anomalous $ N_c^2$-terms. After dropping the $\bar I_R$ terms all scale dependent terms relevant for the running of the $c_i$ and $e_i$ parameters 
scale with at most $\sim N_c $. In view of this prescription the results 
 (\ref{result-loop-8}, \ref{result-loop-10}) can be considered scale invariant and therefore used to study the convergence properties of 
the chiral decomposition we are after. This is what we will do in the following. The corresponding baryon self energy we denote with $\bar \Sigma_B^{\rm bubble}$.

We turn to the subtracted scalar bubble loop integral $\bar I_{Q R}= \bar I_{QR}(p^2\!=\!M_B^2)$ which a priori is a function of the 4-momentum $p$ of the considered 
baryon of type $B$. It is finite and does not depend on the renormalization scale. A dispersion-integral representation of the following form holds for the unsubtracted 
bubble function
\begin{eqnarray}
&& I_{Q R}(p^2)  =  \frac{\bar I_Q-\bar I_R}{M_R^2-m_Q^2} + \int_{(m_Q+M_R)^2}^\infty \frac{d s}{8\,\pi^2}
\,\frac{p^2}{s^{3/2}}\,\frac{p_{Q R}(s)}{s-p^2}\,,
\nonumber\\
&& p_{Q R}^2(s) = \frac{s}{4}-\frac{M_R^2+m_Q^2}{2}+\frac{(M_R^2-m_Q^2)^2}{4\,s} \,.
\label{disp-integral}
\end{eqnarray}
In the previous works  \cite{Semke2005,Semke:2011ez,Semke:2012gs,Lutz:2012mq,Lutz:2014oxa}
the scalar bubble was subtracted such that upon a conventional chiral expansion the leading moment is of chiral order $Q$ as expected from 
dimensional counting rules. In more detail it was argued that the scale invariant combination
\begin{eqnarray}
 \bar I_{QR}(p^2 ) = I_{QR}(p^2) - \frac{\bar I_Q-\bar I_R}{M_R^2-m_Q^2}    -\frac{1}{16\,\pi^2}\,,
 \label{def-previous}
\end{eqnarray}
is of chiral order $Q$ once the baryon tadpole contribution $\bar I_R $ is dropped. While this procedure can be implemented unambiguously at the one-loop level it introduces an artificial pole  
at $m_Q = M_R$ once the renormalization prescription  $\bar I_R \to 0$ is applied. Since the singularity occurs at outrageously large masses $m_Q$ there is a priori no conceptual problem with it. 
However, if possible one should remedy this issue. 
In this work we further improve the renormalization scheme by the following prescription. We insist on a rewrite analogously to (\ref{eliminate-mu})  introducing an 
updated scale invariant bubble function
\begin{eqnarray}
&& \bar I_{QR}(p^2) = I_{QR}(p^2) + \frac{\bar I_R}{M_R^2}    -\frac{1 - \gamma^R_{B}}{16\,\pi^2}\,,
\nonumber\\
{\rm with } \qquad && \gamma^R_{B} = -  \lim_{m, m_s\to 0}\,\frac{M_R^2-M_B^2}{M_B^2}\,\log \left|\frac{M_R^2-M_B^2}{M_R^2}\right| \,,
\label{def-IQR-new}
\end{eqnarray}
where again the renormalization prescription  $\bar I_R \to 0$ is applied in the following. 
Altogether at $p^2=M_B^2$ our subtracted bubble loop takes the form
\begin{eqnarray}
&& \bar I_{Q R}=\frac{1}{16\,\pi^2}
\left\{ \gamma^R_{B} -   \left(\frac{1}{2} +\frac{m_Q^2-M_R^2}{2\,M_B^2}
\right)
\,\log \left( \frac{m_Q^2}{M_R^2}\right)
\right.
\nonumber\\
&& +\left.
\frac{p_{Q R}}{M_B}\,
\left( \log \left(1-\frac{M_B^2-2\,p_{Q R}\,M_B}{m_Q^2+M_R^2} \right)
-\log \left(1-\frac{M_B^2+2\,p_{Q R}\,M_B}{m_Q^2+M_R^2} \right)\right)
\right\}\;,
\nonumber\\
&& p_{Q R}^2 =
\frac{M_B^2}{4}-\frac{M_R^2+m_Q^2}{2}+\frac{(M_R^2-m_Q^2)^2}{4\,M_B^2} \,,\qquad
E_R^2=M_R^2+p_{QR}^2 \,.
\label{def-master-loop}
\end{eqnarray}
According to the Passarino-Veltman reduction scheme 
all one-loop integrals can be unambiguously  decomposed into the updated renormalized bubble $\bar I_{QR}(p^2= M_B^2)$, the tadpole terms $\bar I_Q$ or $\bar I_R$ and 
an infinite hierarchy of finite, scale invariant and power-counting respecting set of scalar loop integrals \cite{Semke2005}.  
Like the prescription (\ref{def-previous}) the new function $\bar I_{QR}(p^2)$ introduced in (\ref{def-IQR-new}) is consistent with the chiral power expected for the bubble function 
from dimensional counting rules.

In contrast to the previous works \cite{Semke2005,Semke:2011ez,Semke:2012gs,Lutz:2012mq,Lutz:2014oxa} the finite subtraction term $\gamma^R_{B}$  in (\ref{def-master-loop}), 
discriminates the case $R\in [8]$ from $R\in [10]$. Note that the dimension less 
$\gamma^R_{B}\neq 0 $ is active only for the off-diagonal cases with neither $B,R \in[8]$ nor $B,R \in [10]$.  It depends on the chiral limit values, $M$ and $M+ \Delta$ of 
the baryon octet and decuplet masses only. 
Such subtractions are useful in a study of the chiral regime where $m_Q \ll \Delta$, which we will recapitulate briefly in the following.

In the chiral regime all meson and baryon masses are expanded strictly in powers of the quark masses with $m_Q \ll \Delta$. 
While the loop expressions (\ref{result-loop-8}, \ref{result-loop-10}) can be consistently 
expanded according to the counting rule $\Delta \sim m_Q$, a further renormalization may be required in the chiral regime.
Such a need is nicely illustrated by terms proportional to $m_Q^3\,\Delta^2 $ that arise from an expansion valid for  $m_Q \ll \Delta$. 
Within the counting world $\Delta \sim m_Q$ such terms are of order $Q^5$, which are beyond the accuracy of the one-loop level. Two loop effects are 
expected to modify such terms and therefore such terms are not fully controlled. They would be part of a summation scheme. If they are 
numerically small they do not cause a problem for physical meson masses. However, in the strict chiral limit at unphysically small 
quark masses a conceptual issue arises.  This is so since all terms proportional to $m_Q^3$ are protected by a chiral theorem that has to be recovered 
in the chiral regime. Rather than computing explicitly such contributions from two-loop integral it suffices to construct a suitable subtraction scheme. 
This is achieved by the terms $\gamma^R_{B}$ and $\alpha^{(B)}_{QR}$. 

Our $\gamma^R_{B}$ and $\alpha^{(B)}_{QR}$ subtractions are effective  in contributions to the baryon self energies only 
that are proportional to $C^2$. In particular the scalar bubble-loop function 
$ \bar I_{Q R}$ vanishes in the chiral limit with $ \bar I_{Q R} \sim m_Q^2/\Delta$ irrespective of the particular behavior of 
$M_R \neq M_B $ in that limit. This is convenient since this allows for an efficient integrating-out of the decuplet degrees of freedom 
in the chiral limit region where $m_Q \ll \Delta$. We will return to this issue below. Here we detail the specific form of the further 
subtraction constant $\alpha^{(B)}_{QR}$ with 
\begin{eqnarray}
&& \alpha^{(B\in\,[8])}_{QR}\, = \frac{\alpha_1\,\Delta^2}{(4\,\pi)^2} \Bigg\{ 
- \Big( M_B - M \Big)\, \Big( \frac{\Delta\,\partial}{\partial\,\Delta} -\frac{\Delta\,\partial}{\partial\,M} 
+ \frac{M+ \Delta}{M} \Big)
\nonumber\\
&& \qquad  +\, \Big( M_R - M -\Delta  \Big)\, \Big( \frac{\Delta\,\partial}{\partial\,\Delta} 
+ 1 \Big) \, \Bigg\}\,\gamma_1 
 +  \frac{\Delta\, m_Q^2}{(4\,\pi)^2}\,\alpha_1\,\gamma_2 \,, 
\nonumber\\
&& \alpha^{(B\in[10])}_{QR} = \frac{\beta_1\,\Delta^2}{(4\,\pi)^2} \Bigg\{ 
+\Big( M_B - M - \Delta\Big)\, \Big( \frac{\Delta\,\partial}{\partial\,\Delta} + 1\Big)
\nonumber\\
&& \qquad  -\,\Big( M_R - M \Big)\, \Big( \frac{\Delta\,\partial}{\partial\,\Delta} - \frac{\Delta\,\partial}{\partial\,M} 
+ \frac{M+ \Delta}{M}\Big)
\Bigg\}\,\delta_1 
 + \frac{\Delta\,m_Q^2}{(4\,\pi)^2}\,\beta_1\,\delta_2 \,, 
\label{def-alphaBR}
\end{eqnarray}
where the dimension less parameters $\alpha_n, \beta_n$ and $\gamma_n, \delta_n$ depend on the 
ratio $\Delta/M$ only. They are detailed in Appendix A and Appendix B. We note that while the $\alpha_n$ and $\beta_n$ characterize the 
chiral expansion of the coefficients in front of $\bar I_{QR}$ and $\bar I_Q$ 
in  (\ref{result-loop-8}, \ref{result-loop-10}), the $\gamma_n$ and $\delta_n$ follow from a chiral 
expansion of $\bar I_{QR}$  at $m_Q < \Delta$. In the limit $\Delta \to 0 $ it holds $\alpha_n \to 1$ and 
$\beta_n \to 1$. In contrast the coefficients $\gamma_i$ and $\delta_i$ show a 
log divergence in this limit. For instance we find $\gamma_1 \to + 2\, \log (2\,\Delta/M) $ and 
$\delta_1 \to -   2\, \log (2\,\Delta/M) $. 

As a consequence of the subtraction term $\gamma^R_{B}$ in (\ref{def-master-loop}) the loop functions (\ref{result-loop-8}, \ref{result-loop-10})
do not affect the baryon masses in the chiral limit. The additional subtraction term $\alpha^{(B)}_{QR}$ has various effects. 
The combination of all terms in (\ref{def-alphaBR}) prevent a renormalization of the counter terms $b_0, b_D, b_F $ and $d_0, d_D$ in (\ref{res-bds}). 
This is in contrast to the previous works  \cite{Semke2005,Semke:2011ez,Semke:2012gs,Lutz:2012mq,Lutz:2014oxa} where the chiral limit masses 
as well as the parameters $b_0, b_D, b_F $ and $d_0, d_D$ are renormalized by loop effects. 

\subsection{Renormalization scale invariance: meson masses }
\vskip0.3cm

The main target of our work is an attempt to reformulate conventional $\chi$PT, which is constructed in terms of bare masses rather than the physical meson and baryon masses. 
An immediate concern arises as to whether this can be done keeping the renormalization scale independence of the effective field theory. 
We discuss this issue at hand of the meson masses first and then turn to the more complicated baryon masses in the next section. 

Within a conventional $\chi$PT approach loop correction terms to the meson masses would impact the baryon masses at N$^4$LO, which is beyond the target of our work. However, 
since we wish to formulate our chiral expansion scheme in terms of physical meson masses a reliable and quantitative approximation of the 
meson masses should be used, particularly in any chiral extrapolation attempt of QCD lattice data. This requires the consideration of chiral correction terms to the meson masses. 
Let us consider the NLO expressions for the pion, kaon and eta  masses of \cite{Gasser:1984gg} where we keep the physical meson masses wherever they occur in the evaluation of the 
relevant diagrams. The expressions
\allowdisplaybreaks[1]
\begin{eqnarray}
&& m_\pi^2 \,=2\,B_0\,m - \frac{1}{3\,f^2}\,\Big\{ \underbrace{5\,B_0 \, m - 4\,m_\pi^2}_{\to \,\frac{1}{6}\,(-10\, m_\pi^2 + \,4\,m_K^2- 3\,m_\eta^2) }\Big\}\,\bar I_\pi 
\nonumber\\
&& \quad \; \;\; -\,\frac{1}{3\,f^2}\,\Big\{\underbrace{B_0 \, (3\,m+m_s) - m_\pi^2-m_K^2}_{\to\, 0}\Big\}\,\bar I_K  - \frac{1}{6\,f^2}\,\underbrace{2\,B_0 \, m}_{\to \,m_\pi^2}\,\bar I_\eta 
\nonumber\\
&& \quad \; \;\; +\,\frac{1}{f^2}\,\underbrace{32\,B^2_0\,m\, (2\, m+m_s)}_{\to \,8\,m_\pi^2\,(m_\pi^2 + \,2\,m_K^2)}\,(2\,L_6-L_4)
\nonumber\\
&& \quad \;\; \;+\,\frac{1}{f^2}\, \underbrace{32\,B^2_0\,m^2}_{\to \,8\,m_\pi^4}\, (2\,L_8-L_5)\,,
\nonumber\\
&& m_K^2 = B_0\,(m+m_s) -\frac{1}{4\,f^2}\,\Big\{ \underbrace{B_0 \,(3\,m +m_s) - m_\pi^2 -m_K^2}_{\to \,0} \Big\}\,\bar I_\pi  
\nonumber\\
&& \quad \; \;\; -\,\frac{1}{f^2}\,\Big\{ \underbrace{B_0 \, (m +m_s) -m_K^2}_{\to \,\frac{1}{6}\,( m_\pi^2 -\,4\, m_K^2 +\,3\,m_\eta^2 )}\Big\}\,\bar I_K
\nonumber\\
&& \quad \; \;\;-\,\frac{1}{12\,f^2}\,\Big\{ \underbrace{B_0 \, (m+3\,m_s) -3\,m_K^2-3\,m_\eta^2}_{\to\,-4\,m_K^2}\Big\}\,\bar I_\eta
\nonumber\\
&& \quad \;\; \;+\,\frac{1}{f^2}\,\underbrace{16 \,B^2_0\,(m+ m_s)\,(2\,m+m_s)}_{\to \,12\,m_K^2\,(m_\pi^2+\,m_\eta^2)}\,(2\,L_6-L_4)
\nonumber\\
&& \quad \;\; \;+\,\frac{1}{f^2}\, \underbrace{8\,B^2_0\,( m+m_s)^2}_{\to\, 8\,m_K^4}\,(2\,L_8-L_5)\,,
\nonumber\\
&& m_\eta^2 \,= \frac{2}{3}\,B_0\,(m+2\,m_s)- \frac{1}{2\,f^2}\,\Big\{ \underbrace{2\,B_0\,m}_{\to\,m_\pi^2} \Big\} \,\bar I_\pi  
- \frac{1}{9\,f^2}\,\Big\{\underbrace{B_0\,(m+8\,m_s)}_{\to \,\frac{3}{2}\,(7\,m_\eta^2-\,4\,m_K^2)} \Big\} \bar I_\eta 
\nonumber\\
&& \quad \;\; \;-\,\frac{1}{3\,f^2}\,\Big\{ \underbrace{B_0\,(m+ 3\,m_s)-3\,m_K^2-3\,m_\eta^2}_{\to \, -4\,m_K^2} \Big\} \,\bar I_K
\nonumber\\
&& \quad \;\;\;  +\, \frac{8}{3\,f^2}\,\underbrace{4\,B^2_0\,(m + 2\,m_s)\,(2\, m+m_s)}_{\to \,9\,m_\eta^2\,( 2\,m_K^2-\,m_\eta^2)}\,(2\,L_6-L_4) 
\nonumber\\
&& \quad \;\; \;+\,\frac{8}{9\,f^2}\,\underbrace{4\, B^2_0\,(m + 2\,m_s)^2}_{\to \,9\,m_\eta^4}\, (2\,L_8-L_5) 
\nonumber\\
&& \quad \;\; \;+\,\frac{128}{9\,f^2}\,\underbrace{B_0\,( m_s-m )^2}_{\to\, \frac{9}{40}\,(3\,m_\pi^4-\,8\,m_K^4-\,8\,m_\eta^2\,m_K^2+\,13\,m_\eta^4)} \,(3\,L_7+L_8) \,,
\label{meson-masses-q4}
\end{eqnarray} 
involve  a set of low-energy parameters $L_i$ and 
the renormalized mesonic tadpole integral $\bar I_Q$  already recalled in (\ref{def-tadpole-integral}).

The parameter $f$ labels the pion-decay constant in the flavour $SU(3)$ limit with $m=m_s =0$. We recall that the tadpole contributions in (\ref{meson-masses-q4}) have two distinct sources. The 
terms proportional to the quark masses are a consequence of the symmetry breaking counter terms that give rise to the leading order Gell-Mann-Oakes-Renner relations. 
The remaining terms are implied by the symmetry conserving Weinberg-Tomozawa interaction terms.

The renormalization scale $\mu$ can be absorbed into the low-energy constants $L_i$ if all meson masses on the r.h.s. of (\ref{meson-masses-q4}), particularly in the tadpole terms $\bar I_Q$,  
are used at leading chiral order as given by the Gell-Mann-Oakes-Renner relations. We assure, that alternatively, scale invariant results are implied also 
if the replacement rules as indicated in (\ref{meson-masses-q4}) are applied. In this case the physical meson masses can be used in the tadpole contributions. 
Such rules can be viewed as a solution to a suitable renormalization group equation that generates specific higher order terms 
that are needed to arrive at renormalization scale independent results. 
In this work we will use the latter procedure. For a given set of quark masses this requires the numerical solution of 
a set of three coupled and non-linear equations. By construction this set of non-linear equations recovers the conventional NLO result of $\chi$PT if expanded in powers of the quark masses. 

\subsection{Renormalization scale invariance: baryon masses}
\vskip0.3cm

We turn to the N$^3$LO effects in the baryon masses. In a conventional $\chi$PT approach this is the minimal oder at which renormalization scale dependent 
counter terms turn relevant. There are three types of contributions to a fourth order approximation:
terms from symmetry breaking and conserving counter terms and the fourth order moment of the one-loop 
expressions (\ref{result-loop-8}, \ref{result-loop-10}). 

Let us begin  with a discussion of the contributions from the symmetry breaking and conserving counter terms. 
Using the notations of \cite{Semke:2011ez} we recall such terms
\begin{eqnarray}
&& \Sigma^{(4-\rm ct )}_{B} = \frac{1}{(2\,f)^2}\sum_{Q\in [8]} \Big(  G^{(\chi )}_{BQ} 
- m_Q^2\,G^{(S)}_{BQ} - \frac 14 \, m_Q^2\,M^{(0)}_B \,G^{(V)}_{BQ}\Big)\, \bar I_Q
\nonumber\\
&& \qquad \quad \,+\,\Sigma^{(4-\chi)}_B \,.
\label{def-tadpole} 
\end{eqnarray}
While the coefficients $G_{QR}^{(\chi)}$ probe the  symmetry 
breaking parameters $b_0,b_D, b_F$ and $d_0,d_D$ already encountered in (\ref{def-b-d}, \ref{res-Q2}), the scalar and vector coupling constants $G^{(S)}_{BQ}$ and $G^{(V)}_{BQ}$ 
encode the symmetry conserving set of parameters $ g^{(S)}_{\dots}, h^{(S)}_{\dots}$  and $g^{(V)}_{\dots}, h^{(V)}_{\dots}$ respectively (see (\ref{def-Q2-terms})). 

Here it is important to remember that the parts of the one-loop contribution that are proportional to $m_Q^2\,\bar I_Q $ were excluded in (\ref{result-loop-8}, \ref{result-loop-10}) and therefore renormalized 
values  have to be taken in (\ref{def-tadpole}). The bare coupling constants $g, h$ are to be replaced by renormalized ones $\bar g, \bar h$. 
In application of the previous results \cite{Semke2005,Semke:2011ez,Semke:2012gs,Lutz:2012mq,Lutz:2014oxa} we derive their specific form for the octet and decuplet baryons. 
It should not be surprsing that we recover identically the expressions  (\ref{Q4-renormalization}) of  Section 2.5 obtained in an analysis of the s- and u-channel baryon exchange 
contributions to the correlation function of two axial-vector currents in the baryon gound states. Note that our large-$N_c$ sum rules (\ref{Q4-subleading}) and (\ref{Q4-leading}) hold for such renormalized 
low-energy constants (\ref{Q4-renormalization}).

We recall from \cite{Lutz:2014oxa} that 
the scalar and vector terms can be discriminated by their distinct behavior in the finite box variant of our approach. 
In this case  two different types of tadpole integrals occur. While the scalar terms $g^{(S)}_ {\cdots }$ come with $\bar I_Q$, 
the vector terms $g^{(V)}_{\cdots}$ come with $\bar I^{(2)}_Q$. The two structures are redundant only in the 
infinite volume limit with 
$
4\,\bar I^{(2)}_Q \to m_Q^2\,\bar I_Q\,.
$
Here we wish to correct an  error in \cite{Lutz:2014oxa} where the scalar coupling constants $h_n^{(S)}$ were not treated properly in the finite volume case. 
It was overlooked that the latter contribute to $G^{(S)}_{BQ}$ {\it and} $G^{(V)}_{BQ}$. 
Our remedy is readily implemented by using  
$\tilde h^{(S)}_n$ {\it and } $\tilde h^{(V)}_n$ parameters with 
\begin{eqnarray}
&&  \tilde h_1^{(S)} = \bar h_1^{(S)} + \frac{1}{3}\, \bar h_2^{(S)}\,, \qquad \qquad \tilde  h_1^{(V)} =  \bar h_1^{(V)} -  \frac{\bar h_2^{(S)}\,}{3\,(M + \Delta)} \,
\nonumber\\
&&  \tilde h_2^{(S)} = \bar h_3^{(S)} + \frac{1}{3}\,\bar h_4^{(S)} \,, \qquad  \qquad\tilde  h_2^{(V)} =  \bar h_2^{(V)} -  \frac{\bar h_4^{(S)}\,}{3\,(M + \Delta)}\,,\quad 
\nonumber\\
&& \tilde h_3^{(S)} =\bar h_5^{(S)} +\frac{1}{3}\, \bar h_6^{(S)} \, ,\qquad \qquad \tilde  h_3^{(V)} =  \bar h_3^{(V)} -  \frac{\bar h_6^{(S)}\,}{3\,(M + \Delta)} \,,
\label{def-tildeh}
\end{eqnarray}
in (\ref{def-tadpole}) where the Clebsch are to be taken from Tab. I of \cite{Semke:2011ez}.  Note that the factors $1/4$ rather than the $1/3 $ in (\ref{def-tildeh}) were claimed in 
the identification of $\tilde h_n^{(S)}$ coupling constants.

\begin{table}[t]
\setlength{\tabcolsep}{1.5mm}
\renewcommand{\arraystretch}{1.2}
\begin{center}
\begin{tabular}{c|cc}\hline
$\Sigma_B^{(4-\chi)}$  & $B=N$\,                                                                        & $B=\Lambda$                                                              \\  \hline \hline
$m_\pi^4$              & $ 63 \,\tilde c_1 + 9 \,\tilde c_2 + 9 \,\tilde c_3 - 3 \,\tilde c_6$          & $42 \,\tilde c_1 + 6 \,\tilde c_3 - 3 \,\tilde c_6 $                     \\
$m_K^4$                & $ 80 \,\tilde c_1 - 6 \,\tilde c_2 + 18 \,\tilde c_3 - 4 \,\tilde c_6$         & $ 4\,(35 \,\tilde c_1 + 5 \,\tilde c_3 - \,\tilde c_6)$                  \\
$m_\eta^4$             & $35 \,\tilde c_1 - 3 \,\tilde c_2 + 5 \,\tilde c_3 - \,\tilde c_6 $            & $ -4 \,\tilde c_1 + 6 \,\tilde c_3 - \,\tilde c_6$                       \\ \hline
$B_0\,m\,m_\pi^2$      & $27 \,\tilde c_4 - 9 \,\tilde c_5 $                                            & $ 18 \,\tilde c_4$                                                       \\ 
$B_0\,(m+m_s)\,m_K^2$  & $3 \,(9 \,\tilde c_4 + \,\tilde c_5) $                                         & $30 \,\tilde c_4 $                                                       \\
$B_0\,m\,m_\eta^2$     & $3 \,\tilde c_4 - \,\tilde c_5 $                                               & $2 \,\tilde c_4 $                                                        \\ 
$B_0\,m_s\,m_\eta^2$   & $ 4 \,(3 \,\tilde c_4 + \,\tilde c_5)$                                         & $ 16 \,\tilde c_4$                                                       \\ \hline
$B_0^2\,\tilde c_0\,m\,m_s$   & $-943/r  - 946 - 829 \,r         $                                             & $ -6 \,(71/r + 198 + 184\,r)$                                            \\  \\

$\Sigma_B^{(4-\chi)}$  & $B=\Sigma$\,                                                                   & $B=\Xi$                                                                   \\  \hline \hline
$m_\pi^4$              & $80 \,\tilde c_1 + 18 \,\tilde c_3 - 3 \,\tilde c_6 $                          &  $63 \,\tilde c_1 - 9 \,\tilde c_2 + 9 \,\tilde c_3 - 3 \,\tilde c_6 $    \\
$m_K^4$                & $84 \,\tilde c_1 + 12 \,\tilde c_3 - 4 \,\tilde c_6 $                          &  $80 \,\tilde c_1 + 6 \,\tilde c_2 + 18 \,\tilde c_3 - 4 \,\tilde c_6 $   \\
$m_\eta^4$             & $14 \,\tilde c_1 + 2 \,\tilde c_3 - \,\tilde c_6 $                             &  $35 \,\tilde c_1 + 3 \,\tilde c_2 + 5 \,\tilde c_3 - \,\tilde c_6 $      \\ \hline
$B_0\,m\,m_\pi^2$      & $54 \,\tilde c_4 $                                                             &  $ 9 \,(3 \,\tilde c_4 + \,\tilde c_5)$                                   \\ 
$B_0\,(m+m_s)\,m_K^2$  & $18 \,\tilde c_4 $                                                             &  $ 27 \,\tilde c_4 - 3 \,\tilde c_5$                                      \\
$B_0\,m\,m_\eta^2$     & $ 6 \,\tilde c_4$                                                              &  $ 3 \,\tilde c_4 + \,\tilde c_5$                                         \\ 
$B_0\,m_s\,m_\eta^2$   & $0 $                                                                           &  $12 \,\tilde c_4 - 4 \,\tilde c_5 $                                      \\ \hline
$B_0^2\,\tilde c_0\,m\,m_s$   & $-2 \,(1247/r + 110 + 2 \,r) $                                                  &  $- 943/r - 946  - 829 \,r $                                              \\ \hline
\end{tabular}
\caption{A rewrite of (\ref{result-counter-terms-octet}) with $r =m_s/m$ and (\ref{def-tilde-c}). Here the wave function parameters $\zeta_0 = \zeta_F = \zeta_D =0$
are put to zero. }
\label{tab:zz1}
\end{center}
\end{table}

\begin{table}[t]
\setlength{\tabcolsep}{1.5mm}
\renewcommand{\arraystretch}{1.2}
\begin{center}
\begin{tabular}{c|cc}\hline
$\Sigma_B^{(4-\chi)}$  & $B=\Delta$\,                                                    & $B=\Sigma^*$                                                              \\  \hline \hline
$m_\pi^4$              & $ 51 \,\tilde e_1 + 9 \,(\tilde e_2 - 6 \,\tilde e_4)  $        &  $ 11 \,\tilde e_1 + 6 \,(\tilde e_2 - 9 \,\tilde e_4) $   \\
$m_K^4$                & $ 6 \,(-2 \,\tilde e_1 + \tilde e_2 - 12 \,\tilde e_4)  $       &  $ 76 \,\tilde e_1 + 8 \,(\tilde e_2 - 9 \,\tilde e_4) $   \\
$m_\eta^4$             & $ 21 \,\tilde e_1 + \,\tilde e_2 - 18 \,\tilde e_4  $           &  $ -27 \,\tilde e_1 + 2 \,(\tilde e_2 - 9 \,\tilde e_4) $   \\ \hline
$B_0\,m\,m_\pi^2$      & $  -27 \,\tilde e_3 $                                           &  $-18 \,\tilde e_3  $   \\ 
$B_0\,(m+m_s)\,m_K^2$  & $ -9 \,\tilde e_3  $                                            &  $ -12 \,\tilde e_3 $   \\
$B_0\,m\,m_\eta^2$     & $  -3 \,\tilde e_3 $                                            &  $ -2 \,\tilde e_3 $   \\ 
$B_0\,m_s\,m_\eta^2$   & $ 0  $                                                          &  $ -4 \,\tilde e_3 $   \\ \hline
$B_0^2\,\tilde e_0\,m\,m_s$   & $  11/r + 14 + 2\,r $                                           &  $11/r + 14 + 2\,r  $   \\  \\

$\Sigma_B^{(4-\chi)}$  & $ B=\Xi^*$                                                      &  $  B=\Omega $    \\  \hline \hline
$m_\pi^4$              & $ -6 \,\tilde e_1 + 3 \,\tilde e_2 - 54 \,\tilde e_4  $         &  $ -54 \,\tilde e_4 $   \\
$m_K^4$                & $  72 \,\tilde e_1 + 10 \,\tilde e_2 - 72 \,\tilde e_4 $        &  $ -12 \,(2 \,\tilde e_1 - \,\tilde e_2 + 6 \,\tilde e_4) $   \\
$m_\eta^4$             & $  3 \,(-2 \,\tilde e_1 + \,\tilde e_2 - 6 \,\tilde e_4) $      &  $ 84 \,\tilde e_1 + 4 \,\tilde e_2 - 18 \,\tilde e_4 $   \\ \hline
$B_0\,m\,m_\pi^2$      & $  -9 \,\tilde e_3 $                                            &  $ 0 $   \\ 
$B_0\,(m+m_s)\,m_K^2$  & $  -15 \,\tilde e_3 $                                           &  $ -18 \,\tilde e_3 $   \\
$B_0\,m\,m_\eta^2$     & $  -\,\tilde e_3 $                                              &  $ 0 $   \\ 
$B_0\,m_s\,m_\eta^2$   & $  -8 \,\tilde e_3 $                                            &  $ -12 \,\tilde e_3 $   \\ \hline
$B_0^2\,\tilde e_0\,m\,m_s$   & $ 11/r + 14 + 2\,r  $                                           &  $11/r + 14  + 2\,r  $   \\ \hline
\end{tabular}
\caption{A rewrite of (\ref{result-counter-terms-decuplet}) with $r =m_s/m$ and (\ref{def-tilde-e}). Here the wave function parameters $\xi_0 = \xi_D =0$
are put to zero.}
\label{tab:zz2}
\end{center}
\end{table}

We continue with the symmetry breaking counter terms, $\Sigma^{(4-\chi)}_B$.  The specific form of their contributions to the 
baryon octet and decuplet self energies is recalled in Appendix A and B, where also more details on the effect of the renormalization as implied by (\ref{eliminate-mu}) is provided. 
There are two classes of contributions. There are terms that contribute to the baryon wave function renormalization and 
terms that are proportional to the product of two quark masses (see (\ref{def-c-e})).
The latter contributions are driven by the parameters  $c_i$ and  $e_i$, for which in Chapter 2 we derived 
their large-$N_c$ sum rules in (\ref{res-ces}, \ref{ces-subleading}). They are the only renormalization scale dependent low-energy parameters relevant in our work. 
Altogether the term $\Sigma^{(4-\rm ct )}_B$ should not depend on the renormalization scale $\mu$. This can indeed be achieved by either expanding the meson masses in $\bar I_Q$  and $m_Q^2\,\bar I_Q$
to leading order in the quark masses or similarly to (\ref{meson-masses-q4}), by reinterpreting the contributions proportional to  $c_i$ and  $e_i$ in terms of physical 
meson masses. In Tab. \ref{tab:zz1} and Tab. \ref{tab:zz2} the details of such a rewrite are provided for the baryon octet and baryon decuplet terms respectively. 
For that purpose it is convenient to identify linear combinations that go together with specific combinations of quark and meson masses. In the octet sector 
\begin{eqnarray}
&&  \tilde c_0 = \frac{2}{2277}\,\Big(33\,c_0 + 6\,c_1 + 22\,c_2 - 30\,c_4 - 45\,c_6 \Big) \,,\qquad \tilde c_1 = \frac{3}{23}\,c_1\,,\qquad 
\nonumber\\
&& \tilde c_2 = \frac{1}{46}\,\Big( 22\,c_3 - 3\,c_5\Big) \,,\qquad 
 \tilde c_3 = \frac{1}{46}\,\Big( 22\,c_2 - 3\,c_4\Big) \,,\qquad 
\nonumber\\
&& \tilde c_4 = \frac{1}{23}\,\Big(22\,c_0 + 6\,c_2 - 23\,c_4 - 30\,c_6 \Big) \,,\qquad 
 \tilde c_5 = \frac{1}{23}\,\Big( 26\,c_3 + 9\,c_5\Big) \,,\qquad 
\nonumber\\ 
 && \tilde c_6 = \frac{1}{253}\,\Big( 924\,c_0 + 621\,c_1 + 528\,c_2 - 828\,c_4 - 846\,c_6\Big) \,,\qquad 
 \label{def-tilde-c}
 \end{eqnarray}
we identify the parameter combinations 
$\tilde c_1, \tilde c_2, \tilde c_3$ and $\tilde c_6$ that probe the fourth power of some meson mass.  As can be seen from Tab. \ref{tab:zz1} only the particular term $\tilde c_0$ 
keeps the original structure being a product of two quark masses. The remaining parameters $\tilde c_4$ and $\tilde c_5$ select the terms involving the product of a quark mass with the square of some meson mass. 
Analogous combinations in the decuplet sector are:
\begin{eqnarray}
 &&  \tilde e_0 = \frac{4}{207}\,\Big(33\,e_0 + 12\,e_1 + 11\,e_2 - 15\,e_3 - 45\,e_4 \Big)\,,\qquad  
  \tilde e_1 = \frac{1}{23}\,e_1\,,\qquad  
\nonumber\\  
 &&  \tilde e_2 = \frac{1}{46}\,\Big(22\,e_2 - 3\,e_3 \Big)\,,\qquad  
   \tilde e_3 = \frac{1}{69}\,\Big(12\,e_1 + 26\,e_2 + 9\,e_3 \Big)\,,\qquad  
\nonumber\\
 &&  \tilde e_4 = \frac{1}{138}\,\Big(7\,e_0 +\,e_2 - 3\,e_3 + 3\,e_4 \Big)\,.\qquad  
 \label{def-tilde-e}
\end{eqnarray}
The expressions in Tab. \ref{tab:zz1} and (\ref{result-counter-terms-octet}) and also Tab. \ref{tab:zz2} and (\ref{result-counter-terms-decuplet}) agree identically if the Gell-Mann-Oakes Renner relations for the meson masses are used. We recall that the merit of 
Tab. \ref{tab:zz1} and Tab. \ref{tab:zz2}  lie in their property of making the fourth order contribution (\ref{def-tadpole}) independent on the renormalization scale $\mu$ even if the physical masses for the pion, kaon and eta meson are used.
We note a particularity: at leading order the effects of $b_0$ and $ d_0$ in $G^{(\chi )}_{BQ}$ cannot be discriminated from $g_0^{(S)}$ and $h_1^{(S)}$ in $G^{(S)}_{BQ}$. Scale invariance requires to consider 
the particular combinations 
\begin{eqnarray}
 g_0^{(S)} - 8\,b_0\,,\qquad \qquad \qquad h_1^{(S)} - 8\,d_0\,,
 \label{scale-particular}
\end{eqnarray}
in $G^{(S)}_{BQ}$ and in turn use $b_0 = 0 = d_0$ in $G^{(\chi)}_{BQ}$.
In the following we will continue with such  scale invariant representation of the term $\Sigma^{(4-\rm ct )}_B$.

\subsection{Effects from the wave-function renormalization}
\vskip0.3cm

The bubble-loop contributions to the baryon self energy implies a renormalization of the baryon wave-function $Z_B$ of the form
\begin{eqnarray}
Z_B - 1= \frac{\partial }{\partial M_B}\,\bar \Sigma^{\rm bubble}_B\,,
\label{def-ZB}
\end{eqnarray}
where we insist on (\ref{result-loop-8}) and (\ref{result-loop-10}) to be renormalized expressions already. In the meson-baryon 
coupling constant $G_{QR}^{(B)}$ the wave-function factor $\sqrt{Z_R\,Z_B}$ is already incorporated. The renormalized 
coupling constants $G_{QR}^{(B)}$ are approximated by the renormalized leading order parameters $F,D,C,H$ as introduced in (\ref{def-FDCH}) at tree-level. 

We emphasize that as a consequence of  $\alpha^{(B)}_{QR}$ in (\ref{result-loop-8}, \ref{result-loop-10}) the wave-function factors are identical to one in the chiral limit. 
This is implied by the terms proportional to $M_B -M$  in (\ref{def-alphaBR}). Recall that  
the terms involving $M_B -M$ {\it and} $M_R -M$ are indispensable to cancel terms proportional to $m_Q^3\,\Delta^2$. The need of such a cancellation 
was discussed below (\ref{def-master-loop}).

The Dyson equation for the baryon propagator determines the physical baryon masses $M_B$. A set of coupled equations is obtained since the 
renormalized loop functions depend themselves on the physical masses of the baryons. We find  
\begin{eqnarray}
 M_B - M^{(0)}_B - M_B^{(2)}  - \bar \Sigma^{\rm bubble}_B / Z_B = 0\,,
\label{gap-equation-A}
\end{eqnarray}
where $  M^{(0)}_B =M$ and $  M^{(0)}_B = M + \Delta$ for the octet and decuplet cases respectively.
The second order terms $ M_B^{(2)}$ are the tree-level second order contributions (\ref{res-Q2}) written in terms of the parameters 
$b_0, b_D, b_F $ and $d_0, d_D$. It should be stressed that the  wave-function renormalization $Z_B$ has a quark-mass dependence which 
cannot be fully moved into the counter terms of the chiral Lagrangian. Therefore it is best to work with the bare parameters first and derive 
any possible renormalization effect explicitly. 

\begin{table}[t]
\setlength{\tabcolsep}{3.5mm}
\renewcommand{\arraystretch}{1.2}
\begin{center}
\begin{tabular}{c|rr||c|rr }\hline
$B$        & $Z_B$\,       & $\bar \Sigma^{\rm bubble }_B$     &  $B$ & $Z_B$\, & $ \bar \Sigma^{\rm bubble }_B$   \\ \hline \hline
$N$        & 1.118         &  -303.9                    &  $\Delta$      & 1.570  &   -313.7   \\
$\Lambda$  & 2.064         &  -458.9                    &  $\Sigma^*$    & 1.915  &   -324.9  \\
$\Sigma$   & 2.507         &  -653.6                    &  $\Xi^*$       & 2.438  &   -340.2  \\
$\Xi$      & 3.423         &  -764.8                    &  $\Omega$      & 3.064  &   -371.4   \\ \hline
\end{tabular}
\caption{The axial coupling constants are $F=0.45$ and $D =0.80$ together with $C=2\,D$ and $H= 9\,F-3\,D$.  The self energies 
$\Sigma^{\rm bubble }_B$ [MeV] are evaluated with physical meson and baryon masses properly averaged over isospin states. }
\label{tab:1}
\end{center}
\end{table}


We compute the numerical values of the wave-function terms assuming that all masses can take their physical values. The results for 
the size of the loop and wave-function contributions are collected in Tab. \ref{tab:1}.
The presence of the subtraction terms $\gamma^R_B$ and $\alpha^{(B)}_{QR}$ changes the size of the wave-function factor and the loop function 
significantly, in particular for the octet states. For instance at $\gamma^R_{B} = 0 = \alpha^{(B)}_{QR}$ we would obtain $Z_\Xi \simeq 1.60$ and 
$\Sigma^{\rm bubble}_{\Xi}\simeq -1291$ MeV as compared 
to $Z_\Xi \simeq 3.42$ and $\Sigma^{\rm bubble}_{\Xi}\simeq -765$ MeV for $\gamma^R_{B} \neq  0 \neq \alpha^{(B)}_{QR}$. Moreover, with $\gamma^R_B =0$ and $\alpha^{(B)}_{QR}= 0$  
it followed $Z_{[8]} \simeq -0.19$  in the chiral limit. This is striking since 
\begin{eqnarray}
Z_{[8]} - 1= \underbrace{\frac{10}{3}\,\left(\frac{C\,\Delta}{4\,\pi\,f} \right)^2 
\Bigg( 1+ 3\,\log \frac{2\,\Delta}{M}\Bigg)}_{\simeq \,0.09 } + \,{\mathcal O} \left( \Delta^3\right)\,,
\label{def-Z8-expanded}
\end{eqnarray}
the change of the wave-function factor is suppressed formally by a factor $Q^2 \sim (\Delta/M)^2$ in the conventional 
counting \cite{Banerjee:1994bk,Banerjee:1995wz}. The leading order result $Z_{[8]} - 1 \simeq 0.09$ of (\ref{def-Z8-expanded}) 
is to be compared with the exact unexpanded value $Z_{[8]} - 1 \simeq  -1.19$. This indicates that any expansion in powers 
of $\Delta/M \sim Q$ is converging rather slowly. 
 
The source of such a slow convergence is readily uncovered. Factors like $(M+ \Delta)^{-n}$ with $n> 1$  arise typically from relativistic 
kinematics (see Appendix A and B). A formal expansion that is truncated at low orders is not able to provide a reliable estimate always. 
Consider the expansion 
\begin{eqnarray}
&& \frac{1}{(M+ \Delta)^n} = \frac{1}{M^n} \,\Big( 1- n\,\frac{\Delta}{M} 
+ \frac{1}{2}\,n\,(n+1)\,\frac{\Delta^2}{M^2} 
\nonumber\\
&& \qquad \qquad \qquad \qquad -\,\frac{1}{6}\,n\,(2+3\,n +n^2)\,\frac{\Delta^3}{M^3} + \cdots  \Big)\,,
\label{example-inefficient}
\end{eqnarray}
which is convergent for $\Delta < M$ but requires more and more terms in the alternating expansion as $n$ gets 
larger. In particular the first two terms in the expansion have opposite signs and may be of almost equal size. This 
may cause trouble making the conventional expansion ineffective. Thus the counting should be modified towards $\Delta/M \sim Q^{1/2}$ or 
even more extremely $\Delta/M \sim Q^{0}$. Since we are primarily interested in the quark mass 
dependence of the baryon masses we may easily avoid this issue by expanding the baryon self energy in the quark 
masses only, where the ratio $\Delta/M$ is kept fixed. This is what we do in the following.

\begin{table}[t]
\setlength{\tabcolsep}{3.5mm}
\renewcommand{\arraystretch}{1.2}
\begin{center}
\begin{tabular}{|l||rr|r|}\hline
                             & $Z_B\neq 1$ &  $Z_B=1$  & tree-level \\ \hline \hline
  $ b_0\, \hfill \mathrm{[GeV^{-1}]}$        & -0.617   & -0.994  & -0.370\\
  $ b_D\, \hfill \mathrm{[GeV^{-1}]}$        &  0.087   &  0.206  &  0.063\\
  $ b_F\, \hfill \mathrm{[GeV^{-1}]}$        & -0.172   & -0.438  & -0.198\\
  $ d_0\, \hfill \mathrm{[GeV^{-1}]}$        & -0.297   & -0.396  & -0.110\\
  $  d_D\, \hfill \mathrm{[GeV^{-1}]}$       & -0.377   & -0.519  & -0.460\\
\hline
\end{tabular}
\caption{In application of (\ref{gap-equation-A}) the parameters at N$^2$LO (1-loop level) and NLO (tree-level) are adjusted to the physical baryon masses with $M =800$ MeV 
and $\Delta = 300$ MeV. The large-$N_c$ sum rules $C= 2\,D$ and $H = 9\,F-3\,D$ together with $f= 92.4$ MeV, $F = 0.45$, $D=0.80$   are used. 
}
\label{tab:FitParameters:N2LO}
\end{center}
\end{table}

In order to illustrate the resumed N$^2$LO approximation we adjust the values of the low-energy parameters $M, \Delta$ and 
$ b_0, b_D, b_F $ and $ d_0, d_D$ to the physical baryon masses. 
We use $ \mm{\pi} = m_\pi^2$, $\mm{K} = m_K^2$ and $\mm{\eta} = m_\eta^2$ in (\ref{res-Q2}) and  
take the chiral limit values of the octet and decuplet masses as assumed at NLO in (\ref{parameters-Q2A}, \ref{parameters-Q2B}). 
The remaining parameters are fitted to the empirical baryon masses. We perform two types of fits. First we assume the wave-function 
factors, $Z_B$,
of Tab. \ref{tab:1} and second we insist on $Z_B=1$. The resulting parameters are shown in Tab. \ref{tab:FitParameters:N2LO}. In both cases the 
isospin averaged baryon masses are recovered quite accurately. The averaged error in the octet and decuplet masses is 4.3 MeV and 0.8 MeV only 
for the case with $Z_B \neq 1$. Assuming $Z_B=1$ a slightly worse description with an error of 9.5 MeV and 1.8 MeV is obtained instead. 
In both cases the size of the error is similarly good as the description at NLO, which is characterized by a typical error 3 MeV. 
Note, that the typical isospin splittings in the baryon masses, which is not considered in this work, is about 3 MeV also.  

At sufficiently small quark masses a linear dependence of the baryon masses is expected as recalled in (\ref{res-Q2}).
The associated slope parameters $b_0, b_D, b_F$ and $d_0,d_D$ are scale independent. We find remarkable that the values 
of the parameters in the first column  in Tab. \ref{tab:FitParameters:N2LO} determined  at the one-loop level are quite 
compatible with the tree-level estimate (\ref{parameters-Q2A}, \ref{parameters-Q2B}). 

We close this section with a discussion of how to generalize the third order ansatz (\ref{gap-equation-A}) to the fourth order case where 
the effect of $\Sigma^{(4-{\rm ct})}_{B } $ should be considered. Here the low-energy parameter $\zeta_{0,D,F}$ and $\xi_0, \xi_D$ have an additional 
impact on the wave-function renormalization of the baryons  (see (\ref{def-zeta-xi})). The set of Dyson equations that determine the physical baryon masses should take the form
\begin{eqnarray}
&&  M_B - M^{(0)}_B - M_B^{(2)}  - \Sigma^{(4-{\rm ct})}_{B } -  \bar \Sigma^{{\rm bubble} }_B / Z_B = 0\,,
\nonumber\\
&&  Z_B = \Big(1 + \frac{\partial }{\partial M_B}\,\bar \Sigma^{\rm bubble}_B \Big)/ \Big( 1- \frac{\partial }{\partial M_B}\,\Sigma^{(4-{\rm ct})}_B\Big)\,,
\label{gap-equation-B}
\end{eqnarray}
with the updated wave-function renormalization $Z_B$. We emphasize the 
importance of the  wave function factor $Z_B$ in (\ref{gap-equation-B}). Only in the presence of this factor it is justified to take tree-level estimates for the coupling constants $F, D , C$ and $H$ from 
the empirical axial-vector coupling constants of the baryon octet and decuplet states.

\subsection{Large-$N_c$ sum rules and loop effects}
\vskip0.3cm

We close this chapter with a discussion of the role of possible loop corrections to the large-$N_c$ sum rules. 
One may expect that loop contributions are suppressed as compared to tree-level contributions in the $1/N_c$ expansion. 
Thus leading-order sum rules should not be renormalized. On the other hand sum rules that are derived at subleading orders 
in the $1/N_c$ expansion may have to be renormalized to sustain the claimed higher accuracy level. In the target application of 
this work, the chiral extrapolation of baryon masses at N$^3$LO, the axial-coupling constants are not considered at the accuracy 
level where loop corrections would contribute. Thus it is justified to  use the subleading sum rules (\ref{res-FDCHs}) without 
a further renormalization. A similar argument holds for the sum rules (\ref{Q4-subleading}). 

In contrast, the sum rules (\ref{res-bds}) for the symmetry breaking parameters $b_0,b_D, b_F$ 
 and $d_0,d_D$ are affected by the one-loop diagrams considered in this work. 
There is a correction to the matrix elements of the scalar currents proportional to
\begin{eqnarray}
 \frac{C^2\,\Delta}{(4\,\pi f)^2} \sim N_c^0  \quad \qquad {\rm with} \quad \qquad 
\Delta = \lim_{m_u,m_d,m_s\to 0} (M_\Delta - M_N) \,,
\end{eqnarray}
that needed to be considered to keep the accuracy of the relations in the second and third lines of  (\ref{res-bds}). In our work this effect are taken care of by the 
suitable subtraction scheme that avoids a renormalization of the parameters $b$ and $d$ such that the predictions (\ref{res-bds}) can be scrutinized in our work directly.

How about the sum rules derived from the study of the product of two scalar currents? Here the low-energy parameters develop 
a renormalization-scale dependence 
\begin{eqnarray}
 \mu\, \frac{d}{d\,\mu} \,c_i = - \frac{1}{2}\,\frac{\Gamma_{c_i}}{(4\,\pi f)^2}  \,,\qquad \qquad 
 \mu\, \frac{d}{d\,\mu} \,e_i = - \frac{1}{2}\,\frac{\Gamma_{e_i}}{(4\,\pi f)^2}  \,,
 \label{c-e-running}
\end{eqnarray}
which specific form is worked out in (\ref{res-Gamma-ci}) and (\ref{res-Gamma-ei}) of Appendix A and B. The coefficients $\Gamma_{c_n}$ and $\Gamma_{e_n}$ depend on the symmetry breaking parameters $b, d$ and 
symmetry preserving parameters $\bar g, \bar h$. Insisting on the leading order identities (\ref{res-ces}) the corresponding seven relations 
\begin{eqnarray}
&& 2\,\Gamma_{c_2}=-3\,\Gamma_{c_1}\,,\qquad 2\,\Gamma_{c_0} = \Gamma_{c_1} +2\,(\Gamma_{c_3}+\Gamma_{e_0})\,, \qquad 3\,\Gamma_{c_1} =\Gamma_{e_1}\,,
\nonumber\\
&& 3\,\Gamma_{e_1}+2\,\Gamma_{e_2} =6\,\Gamma_{c_3} \,, \qquad \Gamma_{e_3}=3\,(\Gamma_{c_4}+\Gamma_{c_5})\,, \qquad \Gamma_{c_4}=\Gamma_{c_1}\,,\qquad 
\nonumber\\
&& \Gamma_{c_6} = \Gamma_{c_5}+\Gamma_{e_4}\,,
\label{res-ces-B}
\end{eqnarray}
control the scale invariance of the correlation function (\ref{SS-expanded}). It is interesting to analyze the impact of (\ref{res-ces-B}) on the low-energy parameters 
$b, d$ and $\bar g, \bar h$. This can readily be done in application of the detailed expressions provided in Appendix A and B. Initially,  
the set of equations is examined insisting on the leading order large-$N_c$ sum rules for the $b, d$ and $\bar g, \bar h$ parameters. Using the  first line of  (\ref{res-bds}) together 
with (\ref{Q4-subleading}, \ref{Q4-leading}) scale invariance of (\ref{SS-expanded}) is observed if and only if 
\begin{eqnarray}
 M\,\bar g_1^{(V)} = -4\,\bar g_1^{(S)}\,, \qquad \qquad \bar g_1^{(S)} = -\frac{2}{3}\,\bar g_D^{(S)} = \frac{2}{9}\,\Big( 3\,\bar h_5^{(S)} + \bar h_6^{(S)}\Big)\,,
 \label{res-large-Nc-mu-A}
\end{eqnarray}
holds. This is a remarkable result: the seven scale equations (\ref{res-ces-B}) are largely compatible with the leading order large-$N_c$ sum rules. 
Only two additional scale-invariance relations arise\footnote{We note that 
insisting on the set of equations (\ref{result:large-Nc-chi}) obtained within the $\varepsilon$-expansion instead, would lead to significant inconsistencies with (\ref{res-bds})  
and (\ref{Q4-subleading}, \ref{Q4-leading}). }. 

What is the true nature of the two additional constraint equations (\ref{res-large-Nc-mu-A}) discovered by the requirement of a renormalization scale invariant correlation function?
We argue, that in fact they are a direct consequence of large-$N_c$ QCD. This can be seen by a study of the large-$N_c$ scaling behavior of (\ref{c-e-running}). 
At leading order the coefficients $ \Gamma_{c_n},\Gamma_{e_n}\sim N_c$ scale linear in $N_c$. Therefore, from a large-$N_c$ point of view the set of 
equations (\ref{c-e-running}) can be significant only, if the sum rules for the $c_n$ and  $e_n$ are imposed to subleading order. This implies a set of four equations (\ref{ces-subleading}) only. 
From the requirement of scale independence of the correlation function (\ref{SS-expanded}) the following  set of corresponding large-$N_c$ sum rules is obtained 
\begin{eqnarray}
&& \Gamma_{c_0} = 2\,\Gamma_{c_3} + \Gamma_{e_0} - {\textstyle{ 1\over 6}}\,\Gamma_{e_1} - {\textstyle{ 1\over 3}}\,\Gamma_{e_2} - {\textstyle{ 1\over 2}}\,\Gamma_{c_1}\,, \qquad 
\nonumber\\
&& \Gamma_{c_1} = {\textstyle{ 1\over 3}}\,(\Gamma_{e_1} + \Gamma_{e_2}) - \Gamma_{c_2} - \Gamma_{c_3}\,, \qquad \qquad \Gamma_{e_3} = 3\,(\Gamma_{c_4} + \Gamma_{c_5})\,, \qquad
\nonumber\\
&&  \Gamma_{e_4} = \Gamma_{c_0} + \Gamma_{c_2} + \Gamma_{c_4} + \Gamma_{c_6} - \Gamma_{c_3} - \Gamma_{c_5} - \Gamma_{e_0}\,,
\label{ces-subleading-Gamma}
\end{eqnarray}
valid for the leading large-$N_c$ moments of the $\Gamma_{c_n}$ and $\Gamma_{e_n}$. This is an interesting result because (\ref{ces-subleading-Gamma}) provides  additional 
leading order large-$N_c$ constraints on the symmetry preserving parameters $\bar g , \bar h$ in (\ref{Q4-subleading}, \ref{Q4-leading}). 
With this we rediscover the two scale relations (\ref{res-large-Nc-mu-A}), i.e.
\begin{eqnarray}
 M\,\bar g_1^{(V)} = -4\,\bar g_1^{(S)}\,, \qquad \qquad \bar g_1^{(S)} = -\frac{2}{3}\,\bar g_D^{(S)} = \frac{2}{9}\,\Big( 3\,\bar h_5^{(S)} + \bar h_6^{(S)}\Big)\,,
 \label{res-large-Nc-mu}
\end{eqnarray}
which now are shown to be valid at leading order in the $1/N_c$ expansion. In the derivation of (\ref{res-large-Nc-mu}) we used the leading order relations, i.e. the first line of  (\ref{res-bds}) together with 
the two set of equations 
(\ref{Q4-subleading}, \ref{Q4-leading}). We find remarkable that the result (\ref{res-large-Nc-mu}) does not depend on any of the symmetry breaking parameters $\hat b_{1,3}$ and 
also that the four equations in (\ref{ces-subleading-Gamma}) provide only two additional constraints as given in (\ref{res-large-Nc-mu}). 

We may analyze the type of relations (\ref{ces-subleading-Gamma}) at the next order in the $1/N_c$ expansion for the $\Gamma_{c_n}$ and $\Gamma_{e_n}$. Note, however, that including into (\ref{SS-expanded}) all 
operators relevant at order $1/N_c$ reduces the number of sum rules further, in fact there would be no obvious sum rule constraint left at this accuracy level. However, it should be recalled that given our framework 
we cannot exclude the existence of some residual relations valid at this accuracy level. This phenomenon is illustrated by the newly discovered leading order large-$N_c$ sum rules (\ref{res-large-Nc-mu}). 
Therefore we feel that it is reasonable to extend our initial analysis to the subleading order level.

In the following we will insist on  (\ref{ces-subleading}) together with (\ref{Q4-subleading}, \ref{ces-subleading-Gamma}) as parametric relations.  
The symmetry conserving parameters $\bar g$ and $\bar h$ relevant at subleading order in the $1/N_c$ expansion are invoked. With the
second line of  (\ref{res-bds}) and (\ref{Q4-subleading})  it follows from (\ref{ces-subleading-Gamma}) the four relations:
\begin{eqnarray}
&&  M\, \bar  g_1^{(V)} = - 4\, \bar  g_1^{(S)} + \frac{8}{3}\, \bar  h_5^{(S)} + \frac{2}{3}\, \bar  h_6^{(S)} -\frac{2}{3}\, (M +\Delta )\, \bar  h_2^{(V)}\, ,\qquad   \bar h_3^{(V) } = - \bar h_2^{(V)} \,,
\nonumber\\
&&  \bar h_3^{(S)}  =   \frac{3}{2}\,\bar  g_0^{(S)} + \frac{183}{652}\, \bar  g_1^{(S)} -  h_5^{(S)}  + \frac{3}{8}\,M\, \bar  g_0^{(V)} + 
      \frac{183}{2608 }\,M\, \bar  g_1^{(V)} 
\nonumber\\
&& \qquad \qquad   -\, \frac{3}{8}\,( M+\Delta)\, \bar  h_1^{(V)}-\frac{3}{2}\, \bar   h_1^{(S)}
+24\,b_D\,\frac{111}{ 163}\,,
\label{res-large-Nc-mu-B}\\
&&  \bar  h_2^{(V)} =  \frac{1467 \,\Delta \,\bar  h_1^{(V)} + 3780\, \bar  h_5^{(S)} + 2738\, \bar  h_6^{(S)}}{945\,M +1434\, \Delta }
-  24\, b_D\,\frac{96 }{315\, M + 478 \,\Delta }\,.
\nonumber
\end{eqnarray}
Our result (\ref{res-large-Nc-mu-B}) is surprising to the extent that it suggests the existence of large-$N_c$ sum rules that correlate the low-energy parameters 
$b, d$ and $\bar g, \bar h$. We are not in a position to add any more to this at this stage, but can only repeat that the relations (\ref{res-large-Nc-mu-B}) are mandatory identities 
to protect the renormalization scale invariance of the correlation function (\ref{SS-expanded}) in a scenario where its loop corrections are evaluated with  
large-$N_c$ sum rules for the $b, d$ and $\bar g, \bar h$ low-energy parameters accurate to subleading order.

We summarize the two scenarios I) and II) scrutinized so far. Both cases will lead to renormalization scale invariant results for the baryon masses. 
\begin{itemize}
 \item[I)] We insist of the leading order sum rules uniformly. That leaves the  parameters $b_0,b_F$ together with $e_0, e_1, e_2, e_3, c_6$ and 
 $g_0^{(S)}, h_5^{(S)}, g_0^{(V)}$. Altogether with $M, \Delta$ we count 12 independent parameters. 
 \item[II)] We insist on subleading order sum rules uniformly.  That leaves the  parameters $b_0,b_D,b_F$ together with $e_0, e_1, e_2, e_3, c_2, c_3, c_4 ,c_6$ and 
 $g_0^{(S)},h_1^{ (S)},\- h_5^{(S)}, h_6^{(S) }, g_0^{(V)}, g_1^{(V)}, h_1^{ (V)}$. Altogether with $M, \Delta$ we count 20 independent parameters. 
\end{itemize}
Our parameter count reveals the relevance of 12 and 20 parameters for the two scenarios
considered. Those values are reasonably small for an attempt to interpolate
the quite large set of QCD lattice simulation data on the baryon masses at
various choices of the quark masses and lattice volumes of about 300 points
altogether. Using the isospin averages of the empirical baryon masses as
further strict constraints reduces the number of fit parameters down to 4 and 12 respectively.

There is a subtle issue to be discussed that is related to  the symmetry breaking parameters $b$ and $d$. They  enter the computation of the baryon self energy at different chiral orders. On the one hand 
they determine the strength of the contributions linear in the quark masses, but there are also one-loop tadpole contributions that are proportional to those parameters. 
Clearly, only the latter have impact on the $\Gamma_{c_n}$ and $\Gamma_{e_n}$. While we worked out a strategy how to deal with the tadpole terms, what to do with the chirally more important tree-level 
terms? We see two distinct strategies how to proceed. First we may simply use slightly different values for the $b$ and $d$ parameters depending on the chiral accuracy level they enter. For instance the tree-level 
parameters may be left unconstrained by large-$N_c$ sum rules or only related by the subsubleading order sum rules in (\ref{res-bds}). The price to pay is 
a breaking of chiral constraints, which however, should be suppressed in  a large-$N_c$ world. The second path is to insist on universal $b$ and $d$ parameters but simply update the scale relations (\ref{res-large-Nc-mu-B}) 
accordingly. This is readily achieved. 
Giving up on the leading order relations (\ref{res-bds}) our result (\ref{res-large-Nc-mu-B}) receives further terms proportional to the $b$ and $d$ parameters. In this case it holds for instance 
\begin{eqnarray}
&& M\, \bar  g_1^{(V)} = - 4\, \bar  g_1^{(S)} + \frac{8}{3}\, \bar  h_5^{(S)} + \frac{2}{3}\, \bar  h_6^{(S)} -\frac{2}{3}\, (M +\Delta )\, \bar  h_2^{(V)}\, 
\nonumber\\
&& \qquad \qquad +\, \frac{416}{21}\,\Big(M - \frac{10}{13}\,\Delta\Big)\,\frac{d_D- 3\,(b_D+b_F)}{\Delta}\,,
\nonumber\\
&&  \bar h_3^{(V) } = - \bar h_2^{(V)} + \frac{208}{7}\,\frac{d_D- 3\,(b_D+b_F)}{\Delta}\,,
\label{res-large-Nc-mu-C}
\end{eqnarray}
where the corresponding expressions for $\bar h_3^{(S)}$ and $\bar h_2^{(V)}$ are considerably more complicated and involve the singlet parameters $b_0$ and $d_0$ in addition. 
We have two options here: either  insist on the subsubleading order relation $d_D = 3\,(b_D+b_F)$ (scenario III) or keep the parameters $b$ and $d$ fully unconstrained (scenario IV).
The two cases lead to a total number of fit parameters of 13 and 14 respectively, a  minor increase for the number of 12 fit parameters in our second scenario.

\clearpage

\section{A power-counting decomposition of the bubble loop}
\vskip0.3cm

In the previous chapter we presented the contributions of the set of counter terms together with the tadpole and bubble-loop contributions 
as they are implied by the chiral Lagrangian as recalled in Chapter 2. While one may well justify the use of the expressions of Chapter 3 at a phenomenological level, 
it can a priori not be linked to a systematic power-counting expansion as it is requested in any effective field theory approach. 
In particular, given the counting rules of strict $\chi$PT the bubble-loop 
contribution must be truncated if one claims to work at the level N$^3$LO. As we will illustrate in the 
following and as it is well known from many previous studies, any conventional chiral expansion attempt of the bubble loop function does not appear to 
converge sufficiently fast as to arrive at any significant result. How could one justify its application to the flavour SU(3) case nevertheless? We would argue that the only way out of this misery is to 
modify the power-counting rules as to make them more effective. Could this be the case once the counting rules 
are formulated in terms of physical meson and baryon masses?

The purpose of the following sections is to decompose the loop function $\bar \Sigma_B^{\rm bubble}$ into power counting moments
\begin{eqnarray}
 \bar \Sigma_B^{\rm bubble} = \bar \Sigma_B^{{\rm bubble}-3} + \bar \Sigma_B^{{\rm bubble}-4}+ \bar \Sigma_B^{{\rm bubble}-5} + \cdots \,,
\end{eqnarray}
where we will derive explicit expressions for the first three moments. A useful decomposition will rely on 
a novel counting scheme formulated in terms of physical masses. In particular it avoids an expansion in powers of $\Delta/M$.

\subsection{Convergence studies for the scalar bubble-loop function}

\vskip0.3cm

In this section we will try to shed further light on the convergence properties of a chiral expansion for the baryon masses. 
Is it possible to further extend the convergence domain beyond the chiral regime with 
\begin{eqnarray}
m_Q \ll \Delta = M_\Delta^{(0)}-M_N^{(0)} \,?
\label{def-chiral-regime}
\end{eqnarray}
This may be the case upon the summation of terms of the form $(m_Q/\Delta)^n$. Since the physical kaon mass is significantly larger than 
$\Delta$  a successful chiral expansion in a flavour SU(3) context must rest necessarily on some summation scheme. 
Is it possible to identify the convergence domain of any such approach? To what extent is it required to impose the use of physical baryon 
masses inside the multi-loop contributions as suggested repeatedly by the first author?

Based on the general one-loop expressions (\ref{result-loop-8}) and (\ref{result-loop-10}) expressed 
in terms of the physical meson and baryon masses the chiral expansion of the baryon masses can be 
scrutinized \cite{Young:2002ib,Hall:2012iw,Geng:2013xn}. At the center of any convergence study for the baryon masses are the 
properties of the scalar bubble integral $\bar I_{QR} = \bar I_{QR}(M_B,m_Q,M_R)$ as recalled in  (\ref{def-master-loop}). How to 
reliably expand this integral into its chiral moments? 

\begin{figure}[t]
\begin{center}
\includegraphics[width=14cm,clip=true]{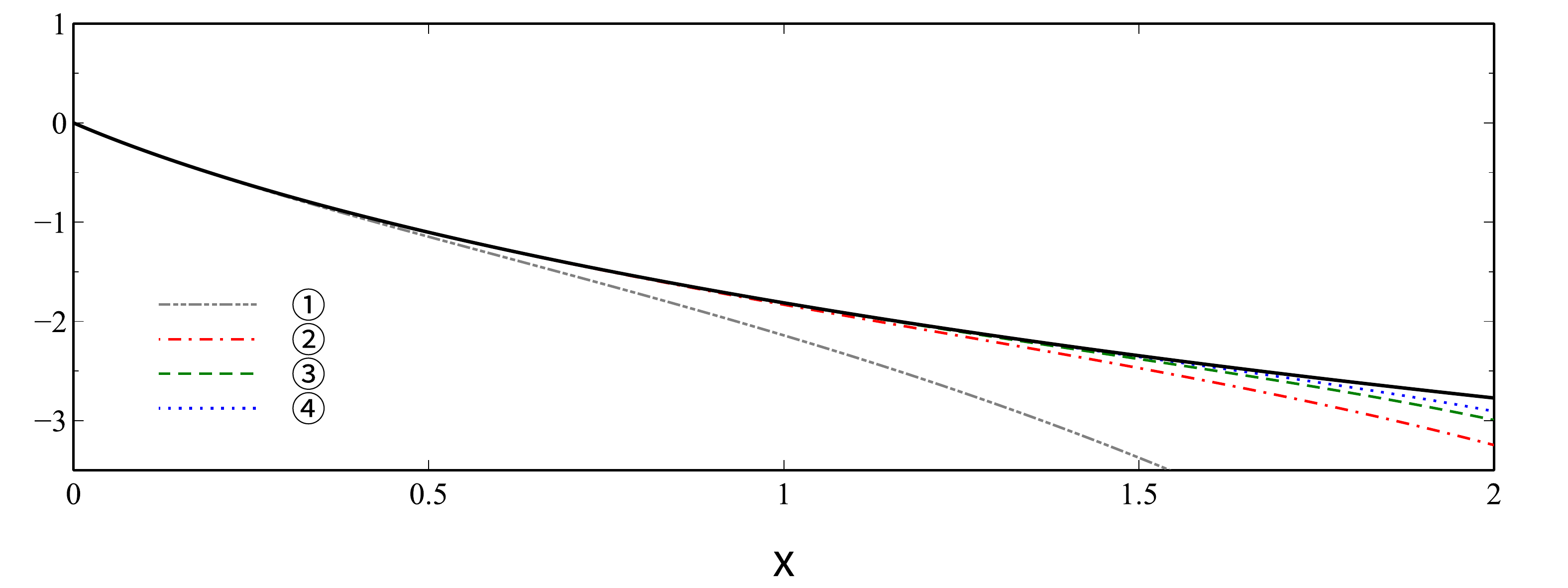}
\end{center}
\caption{The loop function $(4\pi)^2\,\bar I_{QR}$ is plotted as a function of $x=m_Q/M_B$ at $M_B=M_R$ (solid line). The different broken lines correspond to 
the correlated truncation of the three functions $f_n(x)$ in (\ref{IQR-x}, \ref{fn-exp}), where 1, 2, 3, or 4 terms in  (\ref{fn-exp}) are kept. }
\label{fig:IQRa}
\end{figure}

Let us first recapitulate some properties of the scalar 
bubble at $M_R=M_B$ with either $R,B\in[8]$ or $R,B\in[10]$. 
In this case it is a function of $x = m_Q/M_B$ only, which we may want to expand in powers of $x$ following the conventional power 
counting rule $x\sim m_Q \sim Q$. The function in (\ref{IQR-x}) is analytic in $x$ with branch points at $x = 0$ and $x =\pm\, 2$ only. Thus 
an expansion around $x=0$ involves necessarily some terms that are non-analytic at $x=0$. The function may be decomposed as follows 
\begin{eqnarray}
(4\pi)^2\,\bar I_{QR} = - \pi\,\sqrt{x^2}\,f_1(x^2)  + x^2\,f_2(x^2) - \frac{1}{2}\,x^2\,f_3 (x^2)\,\log x^2 \,,
\label{IQR-x}
\end{eqnarray}
where the functions  $f_n(x^2)$ are analytic in the complex plane with the exception of a branch point at $x^2 = 4$. Therefore the 
functions $f_n(x^2)$ with $f_n(0) =1 $ may be Taylor expanded around $x^2 =0$ with the convergence domain of $|x|< 2$.  We observe  a significant 
cancellation amongst the three terms in (\ref{fn-exp}) at small values of $x$ already. It is emphasized that such a cancellation is not a consequence of 
fine-tuned low-energy parameters, rather a general consequence of the analytic structure of the bubble loop. In turn it is justified and efficient 
to expand the three functions $f_n(x^2)$ uniformly, i.e. the first order term is defined by 
$f_n(x^2) = 1 $, the second order terms are implied by  $f_n(x^2) = 1 + x^2\,f_n'(0)$ etc. 
This leads to the following approximation hierarchy
\begin{eqnarray}
&& (4\pi)^2\,\bar I_{QR} = -\Big\{ 1 - \frac{1}{8}\,x^2 - \frac{1}{128}\, x^4 - \frac{1}{1024} \,x^6
+ {\mathcal O} (x^8)\Big\} \,\pi \,\sqrt{x^2}
\nonumber\\
&& \qquad \qquad \;\;\;\;\, +\, \Big\{ 1 - \frac{1}{12}\,x^2 -\frac{1}{120}\, x^4  - \frac{1}{840}\,x^6 + {\mathcal O} (x^8) \Big\} \,x^2
\nonumber\\
&& \qquad \qquad  \;\;\;\;\,
-  \,\frac{1}{2}\,x^2\,\log x^2    \,,
\label{fn-exp}
\end{eqnarray}
where in fact it holds $f_3(x^2) = 1$. 
In  Fig. \ref{fig:IQRa} it is illustrated that indeed such an expansion converges rapidly up to the convergence 
limit $|x| < 2$. This would imply a surprisingly  large convergence radius for a chiral expansion bounded by $m_Q < 2\,M_B$.

We continue with a study of the bubble loop $ \bar I_{QR}$ this time evaluated at  
$M_R \neq  M_B $ with $B\in [8]$ and $R\in[10]$ for instance. It is a function of two variables 
\begin{eqnarray}
x = \frac{m_Q}{M_B} \,,\qquad \qquad d = \frac{M_R}{M_B} - 1\,,
\label{def-x-d}
\end{eqnarray}
only, which we may want to expand in $x$ at fixed value of $d$. For $x \ll d$ an expansion in powers of $x$ with
\begin{eqnarray}
&& (4\pi)^2\,\bar I_{QR} = \gamma_B^R + d\,\gamma_1(d) 
\nonumber\\
&& \qquad \qquad \;+\, d\,\sum_{n=1}^\infty \Big[ \gamma_{2\,n}(d) + 2\,\gamma_{2\,n+1}(d)\, \log \frac{x}{1+d}\Big]\left(\frac{x}{d}\right)^{2\,n}\,,
\label{def-gammas}
\end{eqnarray}
can be justified in terms of specific functions, $\gamma_n(d)$, that are regular and nonzero in the limit $d \to 0$. For values $x \sim d$ a reorganization of the expansion 
(\ref{def-gammas}) is required. The function in (\ref{IQR-x-delta}) is analytic in $x$ with branch points 
at $x = 0$, $x =  -d$ and $x =\pm\, (2 + d)$. Thus an expansion around $x=0$ involves necessarily some terms that are non-analytic 
at $x=0$ and at $x = -d$, if the expansion is expected to be effective at $x^2 > d^2$. The function may be decomposed into four terms
\begin{eqnarray}
&& (4\pi)^2\,\bar I_{QR} = \gamma_B^R + \underbrace{x_d\,\Big( \log (d+ x_d)  - \log (d- x_d)}_{\equiv f^{(d)}_0 (x^2)}\Big)\,f^{(d)}_1(x^2)  
\nonumber\\
&& \qquad \qquad \;+\, \Big[x^2-d^2\Big]\,f^{(d)}_2(x^2)  - \frac{1}{2}\,\Big[x^2-d\,(2+d) \Big]\,\log\frac{x^2}{(1+d)^2} \,,
\nonumber\\
&&  x_d = \big( d^2-x^2 \big)^{1/2} \,,
\label{IQR-x-delta}
\end{eqnarray}
where the functions  $f^{(d)}_{n\neq 0}(x^2)$ are  analytic in the complex plane with the exception of a branch point at $x^2 = (2+d)^2$. Therefore the 
functions $f^{(d)}_{n\neq 0}(x^2)$  may be  Taylor expanded around $x^2 =0$ with the convergence domain of $|x|< 2+d$.

\begin{figure}[t]
\begin{center}
\includegraphics[width=14cm,clip=true]{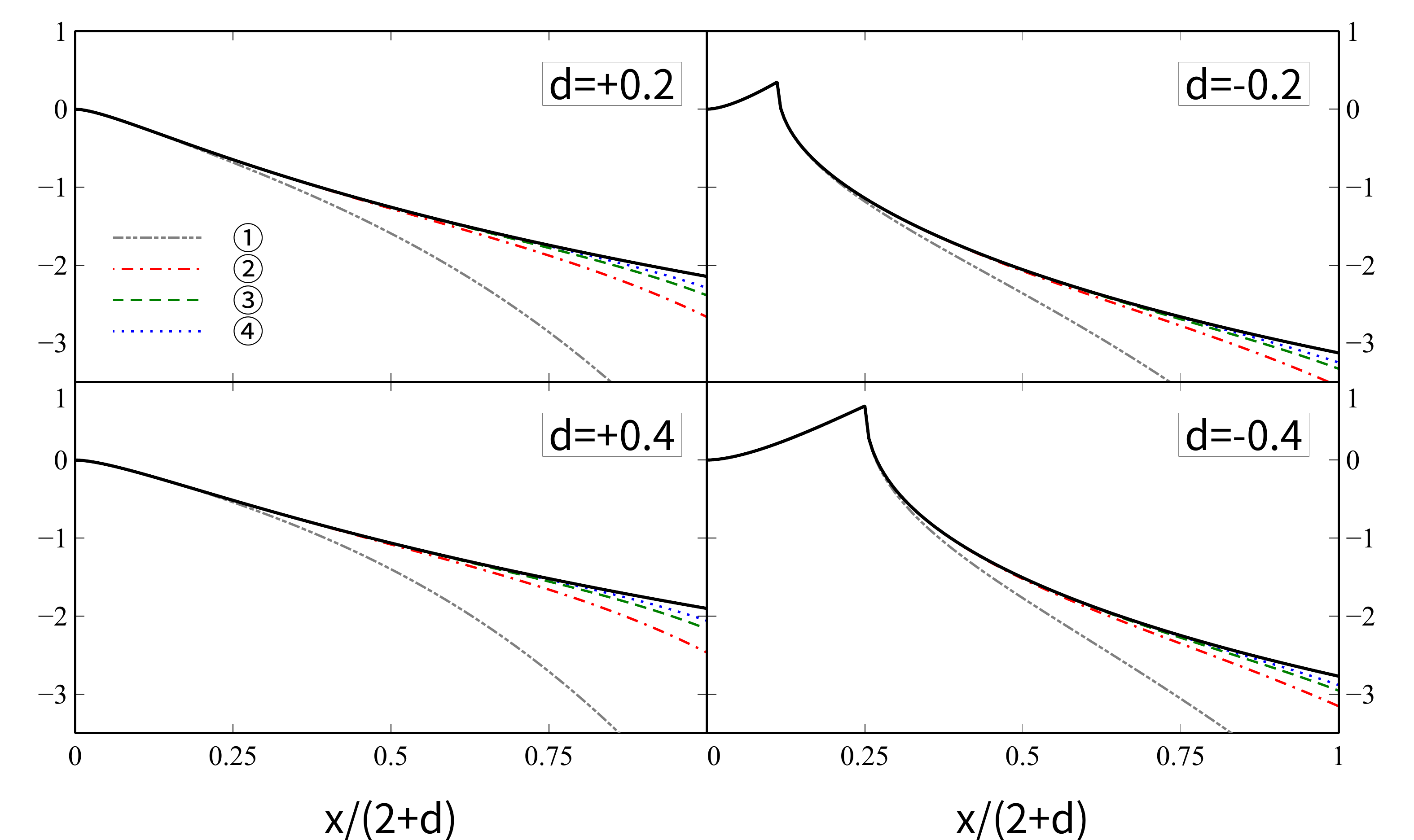}
\end{center}
\caption{The loop function $(4\pi)^2\,\bar I_{QR}$ is plotted as a function of $x=m_Q/M_B$ at fixed value of $d =M_R/M_B -1$ (solid lines). 
The different broken lines correspond to the correlated truncations of the three functions $f^{(d)}_n(x)$ 
in (\ref{IQR-x-delta}, \ref{fndelta-exp}).}
\label{fig:IQRb}
\end{figure}
We chose a normalization with $f^{(d)}_{n\neq 0}(0) =1 $ at $d=0$ that matches the convention used already in (\ref{IQR-x}). 
The leading order coefficients are derived 
\allowdisplaybreaks[1]
\begin{eqnarray}
&& f^{(d)}_1(x^2 ) = 1+ {\textstyle{1 \over 2}}\,d  -  \textstyle{ \frac{1}{4}\, \frac{x^2}{(2+d)}   } 
 - \textstyle{ \frac{1}{16}\, \frac{x^4}{(2+d)^3}   } 
 - \textstyle{ \frac{1}{32}\, \frac{x^6}{(2+d)^5}   } + {\mathcal O} (x^8) \,,
\label{fndelta-exp} \\ \nonumber\\
&& f^{(d)}_2(x^2 ) = \textstyle{\frac{2+d}{d}\,\log\frac{2\,(1+d)}{2+d} }
 -\,\Big( \textstyle{ \frac{4+d}{4\,d^2\,(2+d)} - \frac{2\,(1+d)}{d^3\,(2+d)}\,\log\frac{2\,(1+d)}{2+d} }\Big)\,x^2 
\nonumber\\
&& \qquad \quad \;\; -\,\Big(\textstyle{ \frac{(4 + 3 \,d)\, (24 + d \,(12 + d))}{32\,d^4\,(2+d)^3}
- \frac{2\,(1+d)\,(3+ d\,(3+d))}{d^5\,(2+d)^3}\,\log\frac{2\,(1+d)}{2+d} } \Big)\,x^4 
\nonumber\\
&& \qquad \quad \;\;-\,\Big(\textstyle{ \frac{ (960 + d \,(2160 + d\, (1856 + d \,(750 + d \,(126 + 5\, d))))) }{96\,d^6\,(2+d)^5} }
\nonumber\\
&& \qquad \qquad \qquad \quad\;\;   -\,\textstyle{\frac{2\,(1+d)\,(10 + d\, (20 + d\, (16 + d\, (6 + d ))))}{d^7\,(2+d)^5}\,
\log \frac{2\,(1+d)}{2+d} }  \Big)\,x^6
+ {\mathcal O} (x^8) \,,
\nonumber
\end{eqnarray} 
where it is emphasized that all coefficients in front of the $(x^{2})^{n} $ terms are analytic at $d=0$ and therefore can be expanded in a Taylor 
series around that point. This is not immediate from (\ref{fndelta-exp}), but follows from the general property 
\begin{eqnarray}
f^{(d=0)}_{n\neq 0}(x^2 ) = f_n(x^2 ) \,,
\end{eqnarray}
where we remind the reader of the functions $f_n(x^2 )$ Taylor expanded already in (\ref{fn-exp}) around $x=0$.
While the convergence domain of an expansion around $d=0$ is readily identified with $|d| <1$, we observe a rather slow and 
inefficient convergence behavior. This again reflects our previous  observation (\ref{example-inefficient}). 

There are two ways how to use the expansion (\ref{IQR-x-delta}, \ref{fndelta-exp}). Within a chiral expansion one would 
identify $M_B = M, M_R = M+\Delta$ or $M_R = M, M_B = M+\Delta$. This implies
\begin{eqnarray}
&& d= \frac{M+ \Delta}{M} -1 = \frac{\Delta}{M}\,\qquad {\rm or} \qquad    d= \frac{M}{M+\Delta} -1 = -\frac{\Delta}{M+\Delta}\,,
\nonumber\\
&& \qquad {\rm with }\qquad \gamma_B^R = -  d\,(2+d)\,\log \Bigg|\frac{d\,(2+d)}{(1+d)^2}\Bigg|\,,
\end{eqnarray}
and that the function $\bar I_{QR} =\bar I_{QR}(x,d)$ vanishes at $x=0$ for any choice of $d$. This is the case illustrated in Fig. \ref{fig:IQRb}.
While on the l.h.p. of the figure the two cases with $d=+0.2$ and $d=+0.4$,  
on the r.h.p. of the figure the two cases with $d=-0.2$ and $d=-0.4$ are shown. 
Like for the limiting case $d=0$ in Fig. \ref{fig:IQRa} a simultaneous truncation of the three functions $f^{(d)}_n(x^2)$ with increasing 
and correlated order in $x^2$ leads to  a rapidly converging approximation to the bubble function $\bar I_{QR}(x,d)$. 
The figure illustrates that such an expansion converges uniformly and rapidly within the convergence domain $|x/(2+d)| < 1$. 

Yet there is a potential source for a small convergence radius of the chiral expansion. This is linked to 
the case where one keeps the quark-mass dependence of $M_B$ and $M_R$ in (\ref{def-x-d}). That asks for a further expansion of the system 
(\ref{IQR-x-delta}, \ref{fndelta-exp}). In any conventional counting 
\begin{eqnarray}
 d = d(x)= d_0 + \# \,x^2 + \cdots
\end{eqnarray}
the term proportional to $x^2$ is at least one power down as compared  to the first term $d_0$ with 
$d_0=\Delta/M$ or $d_0 = -\Delta/(M+\Delta)$. Thus to further scrutinize the convergence properties of the chiral expansion 
requires a study of the analytic properties of  $\bar I_{QR}= \bar I_{QR}(x,d)$ in the variable $d$.

For the two functions $f_{n\neq 0}^{(d)}(x^2)$ in (\ref{IQR-x-delta}) the convergence domain in this variable $d$ is readily identified  with 
\begin{eqnarray}
|d-d_0 | < 1 + d_0 \,,
\label{d-condition-A}
\end{eqnarray}
which is implied by the presence of the structure $\log (1+d)$ in the expansion coefficients (\ref{fndelta-exp}). 
For the baryon octet and decuplet the conditions 
\begin{eqnarray}
&& \Bigg| \frac{M_R}{M_B} -  \frac{M+ \Delta}{M}\Bigg| <  \frac{M+ \Delta}{M} \qquad {\rm for}\qquad B\in[8] \;\,\,\quad \&\quad R\in [10] \,,
\nonumber\\
&& \Bigg|\frac{M_R}{M_B} - \frac{M}{M+ \Delta}  \Bigg| <  \frac{M}{M+ \Delta}  \qquad  {\rm for}\qquad B\in[10] \, \quad \& \quad R\in [8] \,,
\label{def-ineq-1}
\end{eqnarray}
arise. For physical quark masses we are far from violating these conditions. Any accessible pair of physical masses $M_B, M_R$ satisfies 
the inequalities in (\ref{def-ineq-1}) comfortably for reasonable choices of $M$ and $\Delta$. For $\Delta =0.3$ GeV and $M= 0.8$ GeV the ratio of left-hand 
to right-hand sides in (\ref{def-ineq-1}) is always smaller than $\simeq 0.25$ and $\simeq 0.35$ for the octet and decuplet baryons. 

How about the remaining so far not considered structure in (\ref{IQR-x-delta})? It is readily expanded 
\allowdisplaybreaks[1]
\begin{eqnarray}
&& f^{(d)}_0(x^2)= x_d\,\Big( \log (d+ x_d)  - \log (d- x_d)\Big)\,,
\end{eqnarray}
with
\begin{eqnarray}
&&   f^{(d)}_0(x^2) =\, f_0  + \Big( 2+ \frac{d_0}{x_0^2}\, f_0\Big)\,(d-d_0) 
+ \Big( \frac{d_0}{x_0^2} - \frac{x^2}{2\,x_0^4}\,  f_0\Big)\, (d-d_0) ^2  
\nonumber\\
&&  \qquad \quad \;\;-\,\Big( \frac{d^2_0 +2\,x^2}{3\,x_0^4} - \frac{d_0\,x^2}{2\,x_0^6}\,  f_0\Big)\, (d-d_0)^3
\nonumber\\
&&  \qquad \quad \;\;+\,\Big( \frac{(2\,d^2_0 +13\,x^2)\,d_0}{12\,x_0^6} - \frac{(4\,d^2_0 +x^2)\,x^2}{8\,x_0^8}\,  f_0\Big)\, (d-d_0)^4
+ {\mathcal O}\left( d-d_0\right)^5\!,
\nonumber\\ \nonumber\\
&& f_0 = x_0\,\Big( \log (d_0+ x_0)  - \log (d_0- x_0)\Big) \,,\qquad  x_0 = \sqrt{d_0^2-x^2}\,,
\label{f0-exp}
\end{eqnarray}
around $d=d_0$, where we note that all expansion coefficients in (\ref{f0-exp}) are regular at $x_0 =0$.   
The determination of the analytic structure of $f^{(d)}_0(x^2)$ considered as a function of $d$ at fixed 
value of $x$ may not be obvious. We identify one branch point at $d =-\sqrt{x^2}$ rather than two branch points at $d = \pm \,\sqrt{x^2}$ as one may 
naively but erroneously expect. Our claim is readily confirmed by plotting the function $f^{(d)}_0(x^2)$ in the complex plane. 
As a consequence we obtain the  convergence condition 
\begin{eqnarray}
 \big|d - d_0 \big| < |x+d_0 |\,.
 \label{d-condition-B}
\end{eqnarray}
Though the result (\ref{d-condition-B}) does not directly determine the convergence domain, it does provide a necessary condition 
that should hold if there is convergence.  We need to discriminate two cases with either $d_0> 0$ or $d_0< 0$. 

\begin{figure}[t]
\begin{center}
\includegraphics[width=14cm,clip=true]{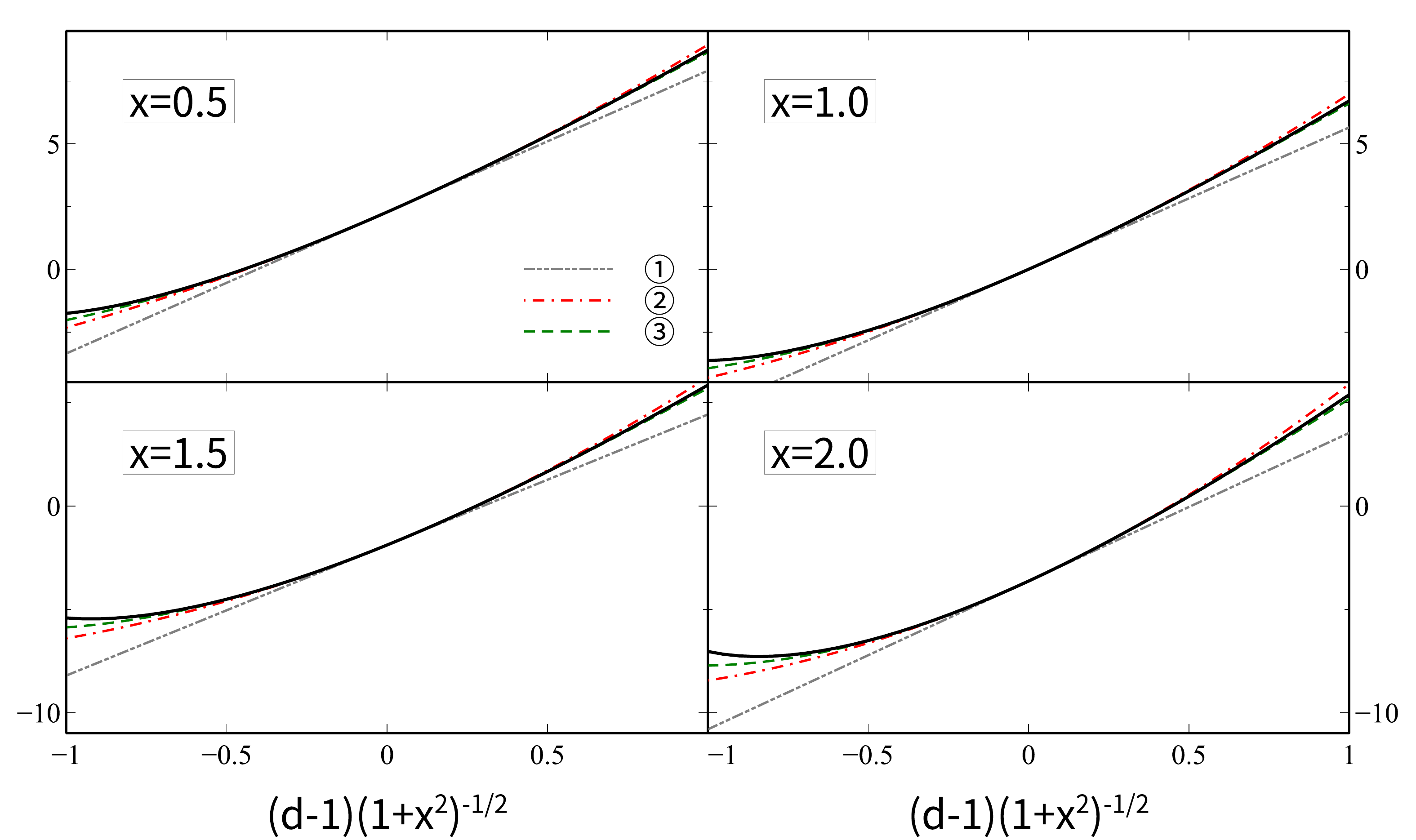}
\end{center}
\caption{The multi-valued function $f_0^{(d)}(x^2)$ is plotted in the variable $(d-d_0)/|d_0 +x|$ at fixed values of $x$ and 
$d_0=1$. 
The different broken lines correspond to the first few  truncations in (\ref{f0-exp}). The range in $d$ is shown where the 
expansion (\ref{f0-exp}) is expected to convergence.}
\label{fig:IQRc}
\end{figure}

First we assume $d_0 >0$ as 
implied by the baryon octet states. We illustrate the convergence behavior in Fig. \ref{fig:IQRc} where the function $f^{(d)}_0(x^2)$
is plotted in the variable $d$ at various choices of $x$. Note that only the ratios $d/x$ and $d_0/x$ are relevant for our convergence study. 
Therefore it suffice to set $d_0=1$ for convenience and select a few representative values for x with $x \in\{0.5,1.0,1.5,2.0\}$. 
According to the convergence condition (\ref{d-condition-B}) we expect the expansion in (\ref{f0-exp}) to be faithful for $1-(1+x) <d< 1+(1+x)$. 
This is indeed confirmed by the figure, which verifies the approximation hierarchy within the expected convergence domain. 

Similar results are obtained for $d_0 <0$ as implied by the baryon decuplet states. It is emphasized however, that in this case convergence is expected only 
in the significantly smaller interval $-1-|1-x| <d< -1+|1-x|$. The convergence domain is quite unfavorable, which ultimately 
is a consequence of open decay channels in the decuplet states.

Let us scrutinize the condition (\ref{d-condition-B}) for physical quark masses, for which all meson and baryon masses are known 
accurately: 
\begin{eqnarray}
&&\Bigg| \frac{M_R}{M_B} -  \frac{M+ \Delta}{M}\Bigg| \,\theta^{(B)}_{QR}\,<  \Bigg|\frac{m_Q }{M_B} +\frac{\Delta}{M}  \Bigg| 
\qquad \quad \quad {\rm for}\quad B\in[8] \;\,\,\& \;R\in [10] \,,
\nonumber\\
&& \Bigg|\frac{M_R}{M_B} - \frac{M}{M+ \Delta}  \Bigg|\,\theta^{(B)}_{QR}\, <  \Bigg|\frac{m_Q }{M_B} - \frac{\Delta}{M+\Delta} \Bigg|
\qquad  {\rm for}\quad B\in[10] \;\& \;R\in [8] \,,
\nonumber\\
&& \qquad \,{\rm with} \qquad \theta^{(B)}_{QR} = \left\{ 
\begin{array}{ll}
1 \qquad & {\rm if } \qquad G^{(B)}_{QR}  \neq 0 \\
0 \qquad & {\rm else}
 \end{array}
 \right. \,.
\label{res-ratios} 
\end{eqnarray}
Taking the ratio of the left-hand sides over the right-hand side in (\ref{res-ratios}) provides a convenient measure 
for the convergence property of the expansion in (\ref{f0-exp}). The ratios depend somewhat on the choice of $\Delta$ and $M$. For the particular 
choice $\Delta =0.3$ GeV and $0.8$ GeV $ < M< 0.9$ GeV we find the range $0.30-0.44 $ for the maximum of all ratios for the baryon octet. Note that for 
the nucleon this value is significantly smaller with $ 0.12-0.16$. We conclude that for the baryon octet states the function $f_0^{(d)}(x^2)$  may be expanded 
around $d=d_0$ successfully.  This is distinct for the baryon decuplet states for which the maximum of all ratios
is larger than one with the range $1.9-2.6$ at $\Delta =0.3$ GeV and $0.8$ GeV $ < M< 0.9$ GeV. For a given species the ratios depend here quite significantly 
on $\Delta$ and $M$ and can even reach values below one in some cases. Whenever a ratio larger than one arises one must not expand the function $f_0^{(d)}(x^2)$ 
around $d=d_0$. 

In conclusion it is advantageous not to further expand the function $f^{(d)}_0(x^2)$ around $d= d_0$. Otherwise an asymmetric treatment of the baryon octet and decuplet would arise 
necessarily. In the following we keep the function $f_0^{(d)}(x^2)$ as it is but expand the coefficient function in $f^{(d)}_{n>0}(x^2)$ in powers of $d- d_0$. More 
specifically we note the decomposition
\begin{eqnarray}
&&(x^2-d^2)\,f_2^{(d)}(x^2) = d\,\tilde \gamma_1+ \sum_{n =1 }^\infty  \frac{\tilde \gamma_{2\,n}}{d^{2\,n-1}}\,x^{2\,n}\,, \qquad \tilde \gamma_3 = -2\,d\,,
\label{def-tilde-gamma}
\end{eqnarray}
characterized by suitable coefficients $\tilde \gamma_n$ that depend on $d$ only. 
Given such a strategy  we do not see any stringent reason to expect a non-converging chiral expansion 
of the baryon masses already at the physical quark masses. However, from our analysis it also follows that a conventional chiral expansion of the baryon masses 
that is not formulated in terms of physical meson and baryon masses can not be convergent at physical quark masses. 

\subsection{Chiral expansion of the bubble-loop: third order}
\vskip0.3cm

We derive the N$^2$LO and N$^3$LO chiral correction term of the baryon masses applying the correlated expansion strategy of the previous 
section. For the baryon octet and decuplet states we expand the loop functions  (\ref{result-loop-8}) and (\ref{result-loop-10}) in powers of 
the meson masses at fixed ratios $m_Q/\Delta$ and $\Delta/M$. 

Consider first the contributions with either $B,R\in[8]$ or $B,R\in[10]$ in (\ref{result-loop-8}, \ref{result-loop-10}). 
At third order $Q^3$ there are  terms only from the scalar bubble-loop function $\bar I_{QR}$  that are relevant. If expanded according 
to (\ref{IQR-x}, \ref{fn-exp}) we obtain 
\begin{eqnarray}
&& \Sigma^{(3-\chi)}_{B \in [8]} \,= \sum_{Q\in [8], R\in [8]} \left(\frac{1}{4\,\pi\,f}\,G_{QR}^{(B)} \right)^2 \Big\{ 
\frac{m_Q^2}{2\,M_B}\,\Big( 1-\log\frac{m_Q}{M_R} \Big)
\nonumber\\
&& \qquad \qquad \qquad -\,\frac{\pi}{2}\,m_Q \Big\}\,\Big( m_Q^2- (M_R-M_B)^2\Big) \,,
\nonumber\\
&&  \Sigma^{(3-\chi)}_{B \in [10]} \!= \sum_{Q\in [8], R\in [10]} \left(\frac{1}{4\,\pi\,f}\,G_{QR}^{(B)} \right)^2 \frac{5}{9}\,
\Big\{ 
 \frac{m_Q^2}{2\,M_B}\,\Big( 1-\log\frac{m_Q}{M_R} \Big) 
 \nonumber\\
&& \qquad \qquad \qquad -\,\frac{\pi}{2}\,m_Q\Big\}\,\Big( m_Q^2- (M_R-M_B)^2\Big) \,,
\label{res-Q3}
\end{eqnarray}
where all terms  are implied by (\ref{IQR-x}) with all $f_{n} (x^2)$ truncated at leading order. In addition we kept the $\log (1+d)$
structure in (\ref{IQR-x-delta}) unexpanded and protected the property of the  
phase-space factor in front of $\bar I_{QR}$ that it vanishes at the threshold and pseudo-threshold conditions with $x=\pm\, d$.

We recall the significant cancellation amongst the $m^3_Q$ and $m_Q^4$ terms in (\ref{res-Q3}). As worked 
out in the previous section this cancellation does not determine the convergence domain of the chiral expansion. Therefore it is justified to slightly 
reorganize the chiral expansion insisting on a specific correlation as suggested by the general decompositions (\ref{IQR-x}) and (\ref{IQR-x-delta})
of the bubble loop. 

There are further terms of third chiral order that arise from the loop contributions involving terms with either $B\in[8], R\in[10]$ or 
$B\in[10], R\in [8]$. The corresponding scalar bubble-loop  function was analyzed in (\ref{IQR-x-delta}, \ref{fndelta-exp}). 
An appropriate correlated expansion of the one-loop expressions (\ref{result-loop-8}, \ref{result-loop-10})  leads to the following 
expressions
\allowdisplaybreaks[1]
\begin{eqnarray}
&&\bar \Sigma^{{\rm bubble}-3}_{B \in [8]} =\Sigma^{(3-\chi)}_{B \in [8]} +\!\!\!\sum_{Q\in [8], R\in [10]}
\left(\frac{1}{4\,\pi\,f}\,G_{QR}^{(B)} \right)^2 \,\frac{\alpha_1}{3}\, \Bigg\{ \hat \gamma_2\, \Delta\,m_Q^2
\nonumber\\
&& \qquad \quad + \, \Big[\gamma_1\, \Delta_B -\tilde \gamma_1\,\big( M_R-M_B\big)\Big]\,\Delta_Q^2 + \hat \gamma_1 \,\Delta^2\,\Big( M_R-M_B -\Delta_B \Big)
\nonumber\\
&& \qquad\quad - \,\frac{2\,M+ \Delta}{2\,M}
\Bigg[ \Big( \Delta_Q^2\,-  \frac{1}{2}\,m_Q^2\Big)\,\big( M_R-M_B\big)\, \log  \frac{m_Q^2}{M_R^2} 
\nonumber\\
&& \qquad  \qquad \quad  +\,  \Delta_Q^3\,\Big( \log \big( M_R-M_B + \Delta_Q \big) - \log \big( M_R-M_B - \Delta_Q\big)\Big) \Bigg]
\nonumber\\
&&\qquad \quad +\,     \frac{m_Q^2}{\Delta_B}\, \Big[ -\tilde  \gamma_2 \,\Delta_Q^2+ \tilde \gamma_3\,m_Q^2\,\log \frac{m_Q^2}{M_R^2} \Big]\Bigg\}\,,
\nonumber\\ \nonumber\\
&&  \Delta_Q = \Big[ (M_B -M_R)^2- m_Q^2\Big]^{1/2} \,,  \qquad \qquad  \Delta_B = \Delta \,M_B \,\lim_{m_{u,d,s}\to \,0}\frac{1}{M_B}\,, 
\nonumber\\
&& \hat \gamma_1 =\frac{2\,M+\Delta}{2\,M} \frac{\partial }{\partial \Delta}\,\frac{2\,\Delta\,M}{2\,M+\Delta}\,\big(\gamma_1-\tilde \gamma_1\big) + \tilde \gamma_1\,, \quad \quad 
\nonumber\\
&& \hat \gamma_2 = \gamma_2  + \frac{1}{2}\,(\gamma_1 - \tilde \gamma_1)\,\frac{\Delta^2}{(2\,M+\Delta)^2} \,,
\label{loop-HB-3}
\end{eqnarray}
where the coefficients $\alpha_n$ and $\gamma_n, \tilde \gamma_n$ are detailed 
at the beginning of Appendix A. All such coefficients are dimension less and depend on the ratio $\Delta/M$ only.  
The various terms in (\ref{loop-HB-3}) are a consequence of the  expansion strategy illustrated in  (\ref{IQR-x-delta}, \ref{fndelta-exp}). 
At the given order we use in particular $d=d_0$ in the truncated functions $f^{(d)}_{n\neq 0 }(x^2)$ with
\begin{eqnarray}
 d_0=\frac{\Delta}{M} \,,\qquad \qquad \qquad \frac{x}{d_0} = \frac{m_Q}{\Delta_B}\,.
\end{eqnarray}

We recall that $\alpha_n$  characterize the 
chiral expansion of the coefficients in front of $\bar I_{QR}$ and $\bar I_Q$ 
in  (\ref{result-loop-8}). In the limit $d_0=\Delta/M \to 0 $ it holds $\alpha_n \to 1$. The coefficients $\gamma_n$  follow from a chiral 
expansion of $\bar I_{QR}$  at $m_Q < \Delta$ (see (\ref{def-gammas})). They are supplemented by $\tilde \gamma_n$ 
which encode the chiral moments of the functions $f^{(d)}_n(x^2)$ considered in (\ref{IQR-x-delta}, \ref{def-tilde-gamma}). 
This implies in particular that all coefficients  $\tilde \gamma_n$  are analytic functions at $d_0 =0$. 
In contrast the coefficients $\gamma_{1,2,4}$  have a  branch point  of the type $\log d_0$.  

All contributions in (\ref{loop-HB-3}) 
originate from an expansion of terms 
proportional to the scalar bubble integral $\bar I_{QR}$,  
an anomalous contribution proportional to  $\Delta \,\alpha_4\,\bar I_Q $ 
and the subtraction terms introduced in (\ref{result-loop-8},\ref{eliminate-mu}, \ref{def-alphaBR}). 
Like in (\ref{res-Q3}) we keep the $\log (1+d)$
structure in (\ref{IQR-x-delta}) unexpanded and the kinematical constraint  
that the phase-space factor in front of $\bar I_{QR}$ vanishes at the thresholds $x=\pm\, d$. 
The result (\ref{loop-HB-3}) generalizes (\ref{res-Q3}) which follows from the leading order expansion of $\bar I_{QR}$ only, with  $\Delta \to 0$ 
in this case (see (\ref{IQR-x}, \ref{fn-exp})). 

A few more comments on (\ref{loop-HB-3}) may be useful for the reader. We recall that the  $m_Q^2\,\log \mu^2$ dependence in $ (M_R-M_B)\,\bar I_Q \to (M_R-M_B)\,I^R_Q$  is eaten up by subtraction terms discussed 
at (\ref{eliminate-mu}). The  troublesome $(M_R-M_B)\,I_Q^R$ term is  canceled identically by a corresponding contribution from the scalar bubble $\bar I_{QR}$. 
This is reflected in a particular relation, $ \alpha_4= 2\,\alpha_1\,\gamma_3$,  amongst various coefficients we introduced. Owing to the  coefficients $\gamma^R_{B}$ and $\alpha_{QR}$ in (\ref{result-loop-8})
the loop contributions neither renormalize the chiral limit mass of the octet states  nor any of the counter terms $b_0, b_D, b_F$. These conditions are indeed respected  
by our result (\ref{loop-HB-3}), where we point at the specific role played by the terms proportional to $\gamma_1-\tilde\gamma_1$ and $\hat \gamma_2$. 
Moreover, the baryon wave function derived from (\ref{loop-HB-3}) turns one in the chiral limit. 
This is a consequence of (\ref{def-alphaBR}) also and at the given order of the term proportional to $\hat \gamma_1$ in (\ref{loop-HB-3}).

We return to the loop contributions for the decuplet masses. Like in the octet case at the given order we 
use  $d=d_0$ in the truncated functions $f^{(d)}_{1,2}(x^2)$ with
\begin{eqnarray}
 d_0=-\frac{\Delta}{M+ \Delta} \,,\qquad \qquad \frac{x}{d_0} = -\frac{m_Q}{\Delta_B}\,, 
\end{eqnarray}
In contrast to the octet states, for which the expansion (\ref{f0-exp}) can be justified, here it is crucial not to expand the function $f^{(d)}_0(x^2)$ around 
$d=d_0$. Following this strategy we obtain 
\begin{eqnarray}
&&\bar \Sigma^{{\rm bubble}-3}_{B \in [10]} =\Sigma^{(3-\chi)}_{B \in [10]} +\!\!\! \sum_{Q\in [8], R\in [8]}
\left(\frac{1}{4\,\pi\,f}\,G_{QR}^{(B)} \right)^2 \, \frac{\beta_1}{6}\,\Bigg\{ \hat \delta_2\,\Delta\,m_Q^2
\nonumber\\
&& \qquad \;\quad + \,\Big[\delta_1\,\Delta_B - \tilde \delta_1\,\big( M_B-M_R\big) \Big]\,\Delta_Q^2 -   \hat \delta_1\,\Delta^2\, \Big( M_R-M_B + \Delta_B \Big)
\nonumber\\
&& \qquad  \;\quad + \, \frac{(2\,M+ \Delta)\,M}{2\,(M+ \Delta)^2} \,
\Bigg[ \Big( \Delta_Q^2\,-  \frac{1}{2}\,m_Q^2\Big)\, \big(M_B-M_R \big)\, \log  \frac{m_Q^2}{M_R^2} 
\nonumber\\
&&  \qquad  \; \qquad \quad  +\,  \Delta_Q^3\,\Big( \log \big( M_R-M_B - \Delta_Q \big) - \log \big( M_R-M_B + \Delta_Q\big)\Big)
\Bigg]\,
\nonumber\\
&&  \qquad  \;\quad +\, \frac{m_Q^2}{\Delta_B}\, \Big( - \tilde \delta_2\,\Delta_Q^2 +\tilde \delta_3\,m_Q^2\, \log \frac{m_Q^2}{M_R^2} \Big)
\Bigg\} \,,
\nonumber\\ \nonumber\\
&& \hat \delta_1 = \frac{(2\,M+ \Delta)}{2\,M} \,\frac{\partial }{\partial \Delta }\,\frac{2\,(M+ \Delta)}{2\,M+ \Delta}\,\Delta \,( \delta_1 - \tilde \delta_1) 
+\tilde \delta_1\,, \qquad \quad 
\nonumber\\
&& \hat \delta_2 = \delta_2 + \frac{1}{2}\,(\delta_1 - \tilde \delta_1)\,\frac{\Delta^2}{(2\,M+\Delta)^2} \,,
\label{loop-HB-3-B}
\end{eqnarray}
where the decuplet coefficients $\beta_n$ and $ \delta_n ,\tilde \delta_n$ play the role of the octet coefficients $\alpha_n$ and $\gamma_n, \tilde \gamma_n$
respectively. The latter are documented at the beginning of  Appendix B.

We claim the functional form of $\bar \Sigma^{{\rm bubble}-3}_B$ on  $M_B, M_R$ and $\Delta/M$ to be model independent. They are 
a consequence of relativistic kinematics for the meson-baryon vertex and can not be altered by higher order terms in the 
chiral expansion. Clearly, a further expansion of the terms (\ref{loop-HB-3}, \ref{loop-HB-3-B}) in powers of $m_Q$ has 
a convergence domain strictly bounded by $m_Q <\Delta$. 

The leading order results of the heavy-baryon formulation 
\cite{Bernard:1993nj,Banerjee:1994bk,Banerjee:1995wz,Lehnhart2004,Semke2005}
are recovered from (\ref{res-Q3}, \ref{loop-HB-3}) by an additional formal expansion: the baryon masses $M_B$ and $M_R$ 
in (\ref{res-Q3}, \ref{loop-HB-3}) have to be replaced by their chiral limit values and an expansion in 
powers of $\Delta/M $ is to be applied.  

\subsection{Chiral expansion of the bubble-loop: fourth order}
\vskip0.3cm

Finally there are the 4th order contributions from the one-loop expressions (\ref{result-loop-8}, \ref{result-loop-10}), following the expansion scheme  
illustrated with  (\ref{IQR-x-delta}, \ref{fndelta-exp}, \ref{f0-exp}). 
While it is straight forward to extract the fourth order terms from the loop contributions with $B,R\in [8]$ or $B,R\in [10]$ this may not 
be so immediate for some of the remaining terms. All contributions from $\bar I_{QR}$ can be deduced by an appropriate expansion of the 
functions $f^{(d)}_n(x^2)$ in (\ref{IQR-x-delta}). The latter functions characterize the scalar bubble loop at $M_B \neq M_R$ with $d = M_R/M_B -1$ 
and $x = m_Q/M_B$. We recall that the third order terms in (\ref{loop-HB-3}) are implied by a leading order truncation of 
\begin{eqnarray}
f^{(d)}_0(x^2) \to  f^{(d)}_0(x^2) \,,\qquad  \qquad 
f^{(d)}_1(x^2) \to f^{(d=d_0)}_1(x^2=0)\,,
\end{eqnarray}
in (\ref{fndelta-exp}). The next term in $f^{(d)}_1(x^2)$ proportional to $x^2\sim Q^2$ is two orders suppressed as compared to the leading term 
and therefore does not contribute to our fourth order terms. In contrast, the expansion $f^{(d)}_{n\neq 0}(x^2)$  around $d=d_0$ does lead to terms 
of chiral order four.  Note that the diagonal cases with $B,R\in [8]$ or $B,R\in [10]$  is covered with $d_0 =0$ in the latter expansion.
An expansion of $f^{(d)}_{2}(x^2)$  around $d=d_0$ generates powers of 
\begin{eqnarray}
\Bigg(\frac{M_R-M_B }{M_B} \Bigg)^n \sim Q^{2\,n} \qquad {\rm or} \qquad \Bigg(\frac{M_R-M_B \pm \Delta_B}{M_B} \Bigg)^n \sim Q^{2\,n} 
\label{def-mass-differences:A}
\end{eqnarray}
depending on the species $B$ and $R$. While we convinced ourselves that such an expansion is rapidly convergent, this may not 
necessarily be the case once  mass differences $M_R-M_B$ are decomposed further into chiral moments. Recall that despite 
the convergence of the  chiral expansion there are significant cancellations amongst various chiral terms. 
Therefore we avoid the approximate treatment of the mass differences in (\ref{def-mass-differences:A}) and work out  
the correlated expansion of the baryon self energies in powers of $x$ and $d-d_0$. 

We collect the contributions from (\ref{result-loop-8}, \ref{result-loop-10}) as implied by our chiral decomposition. 
While this is straight forward for the diagonal cases with $B,R\in[8]$ or  $B,R\in[10]$ the results for the off-diagonal cases are most efficiently 
derived from the third order result $\bar \Sigma^{{\rm bubble}-3}_{B  \in [8]}$ in (\ref{loop-HB-3}) by taking an appropriate derivative of it with respect 
to $d_0$. After some algebra the following  expressions are obtained
\begin{eqnarray}
&&\bar \Sigma^{{\rm bubble}-4}_{B \in [8]} =  \!\! \sum_{Q\in [8], R\in [8]}
\left(\frac{1}{4\,\pi\,f}\,G_{QR}^{(B)} \right)^2 \Big( M_R-M_B\Big) \,\Bigg\{ \Big( 1+ \log \frac{m_Q}{M_R } 
\nonumber\\
&& \qquad \qquad  -\, \frac{3\,\pi}{4} \,\frac{m_Q}{M_B} \Big)\,\Big(m_Q^2 - (M_R-M_B)^2 \Big) + \frac{1}{4}\,m_Q^2\,\log \frac{m_Q^2}{M_R^2} \Bigg\}
\nonumber\\
&& \quad  +\sum_{Q\in [8], R\in [10]}
\left(\frac{1}{4\,\pi\,f}\,G_{QR}^{(B)} \right)^2 \,\Bigg\{ \frac{\alpha_1}{3}\,\Delta^2\,\Big( \frac{\partial}{\partial \Delta}\,\Delta\,  \gamma_1 
- \hat \gamma_1 \Big) +\frac{\tilde \alpha_4}{3}\,\Delta_Q^2 
\nonumber\\
&& \quad \quad  -\, \frac{\tilde \alpha_5}{3}\,\frac{ M_R-M_B}{\Delta_B}\,\Delta_Q^2  
- \frac{\tilde \alpha_1}{3\,\Delta_B}\,  \Bigg[ \Big(  \Delta^2_Q- \frac{1}{2}\,m_Q^2\Big) \, \big( M_R-M_B\big)\,\log \frac{m_Q^2}{M_R^2} 
\nonumber\\
&& \qquad  \qquad + \, \Delta_Q^3\,\Big(\log ( M_R-M_B + \Delta_Q ) -\log (M_R-M_B - \Delta_Q )\Big)
   \Bigg]
\nonumber\\ 
&& \quad \quad  +\,\frac{1}{3}\,\frac{m_Q^2}{\Delta^2_B}\,\Big( -\tilde \alpha_2\,\Delta_Q^2+ \tilde \alpha_3\,m_Q^2\,\log \frac{m_Q^2}{M_R^2}\Big)
\Bigg\} \,\Big( M_R - M_B -  \Delta_B\Big)\,,
\label{loop-HB-4}
\end{eqnarray}
with $\Delta_Q$ and $\Delta_B$ already introduced in (\ref{loop-HB-3}).
Our fourth order term (\ref{loop-HB-4}) considers contributions that are formally one power suppressed. 
The extra structures are either proportional to $\pi\,m_Q/M_B$ or to $\tilde \alpha_{1,2,3} $.  
Such a rearrangement is required since there are significant cancellations in (\ref{loop-HB-4}). For instance 
the $1 +\log (m_Q/M_R) $ term in (\ref{loop-HB-4}) may be largely canceled  by the associated  $\pi\,m_Q/M_B$ term. 
This resembles the cancellation mechanism within our third order terms (\ref{loop-HB-3}). We emphasize 
that our reordering of terms is not caused by a poor convergence of the chiral expansion, rather it is suggested by 
specific correlation properties of the chiral moments. 

We turn to the decuplet sector. The derivation of our results is analogous to the octet sector. Again the role of $\alpha_n, \tilde \alpha_n$ and $\gamma_n,\tilde \gamma_n$
is taken over by $\beta_n, \tilde \beta_n$ and $\delta_n, \tilde \delta_n$. Altogether we find:
\begin{eqnarray}
&& \bar \Sigma^{{\rm bubble}-4}_{B \in [10]} =  \!
\sum_{Q\in [8], R\in [10]}\!
\left(\frac{1}{4\,\pi\,f}\,G_{QR}^{(B)} \right)^2 \!\frac{5}{9}\,\Big( M_R-M_B\Big) \, \Bigg\{\Big( 1  +\log \frac{m_Q}{M_R }
\nonumber\\
&& \qquad  \qquad  - \, \frac{3\,\pi}{4} \,\frac{m_Q}{M_B}\Big)\,\Big(m_Q^2 - (M_R-M_B)^2 \Big) + \frac{1}{4}\,m_Q^2\,\log \frac{m_Q^2}{M_R^2} \Bigg\}
\nonumber\\
&& \quad + \sum_{Q\in [8], R\in [8]}\!
\left(\frac{1}{4\,\pi\,f}\,G_{QR}^{(B)} \right)^2 \Bigg\{ 
\frac{\beta_1}{6}\,\Delta^2\,\Big(   \hat \delta_1 -\frac{M+ \Delta}{M}\,\frac{\partial}{\partial \Delta}\,\Delta \, \delta_1\Big)   -\frac{\tilde \beta_4}{6}\,\Delta_Q^2
\nonumber\\
&& \quad \quad   -\,\frac{\tilde \beta_5}{6}\, \frac{ M_R-M_B}{\Delta_B}\,\Delta_Q^2
 - \frac{\tilde \beta_1}{6\,\Delta_B}\,\Bigg[  
\Big( \Delta_Q^2\,-\frac{1}{2}\,m_Q^2\Big) \,\big( M_B-M_R\big)\, \log \frac{m_Q^2}{M_R^2}
\nonumber\\
&& \qquad \qquad   + \,
\Delta_Q^3\, 
\Big( \log ( M_R-M_B - \Delta_Q ) -\log ( M_R-M_B + \Delta_Q ) \Big) 
\Bigg] 
\nonumber\\
&& \quad  \quad -\, \frac{1}{6} \,\frac{m_Q^2}{\Delta^2_B}\,\Big(- \tilde \beta_2\,\Delta_Q^2 + \tilde \beta_3\,m_Q^2\,\log \frac{m_Q^2}{M_R^2}\Big)
\Bigg\} \,\Big( M_R - M_B +  \Delta_B\Big)\,.
\label{loop-HB-4-B}
\end{eqnarray}
Like our third order terms (\ref{loop-HB-3}, \ref{loop-HB-3-B}) the fourth order loop contributions (\ref{loop-HB-4}, \ref{loop-HB-4-B}) neither renormalize the 
chiral limit mass of the baryon states  nor any of the counter terms $b_0, b_D, b_F$ and $ d_0, d_D$. Moreover, the baryon wave functions derived from (\ref{loop-HB-4}, \ref{loop-HB-4-B}) remain 
one in the chiral limit. 

\subsection{Convergence at physical quark masses}
\vskip0.3cm

In this section we scrutinize the chiral decomposition of the one-loop contribution into its chiral moments as developed in the 
previous section by providing numerical values. Since in our approach the use of physical meson and baryon masses inside the loop functions is a crucial element such 
a convergence test requires the solution of a set of coupled and nonlinear equations at each given order of the truncation. 

At this stage we can perform such studies meaningfully at $c=e=0$ and $\bar g = \bar h =0$ only.  The size of the set of those low-energy parameters 
is not well established yet. Ultimately they have to be extracted from QCD lattice simulation data, which will be the target of the next chapter. 
Any ad-hoc choice thereof may disguise the expected convergence pattern. We will return to this issue after a determination of such parameters from lattice data. 
In this section we identify the nth moment of the baryon self energy 
\begin{eqnarray}
 \Sigma^{(n)}_B = \bar \Sigma_B^{{\rm bubble}-n}\,.
\end{eqnarray}
with the nth moment of the one-loop expressions.

\begin{table}[t]
\setlength{\tabcolsep}{3.5mm}
\renewcommand{\arraystretch}{1.2}
\begin{center}
\begin{tabular}{c|rr||rrr }\hline
$B$         & $\Sigma_B$\,  & $\Sigma^{(3+4+5)}_B$      & $\Sigma^{(3)}_B$    &  $\Sigma^{(4)}_B$ & $\Sigma^{(5)}_B$   \\ \hline \hline
$N$         & -247.3        &  -250.4                   & -185.2              &  -87.4     &   22.2  \\
$\Lambda$   & -340.6        &  -319.1                   & -434.7              &  75.3      &   40.3  \\
$\Sigma$    & -513.4        &  -515.3                   & -453.6              &  -87.9     &   26.1  \\
$\Xi$       & -571.6        &  -554.1                   & -714.8              &   126.9    &   33.8  \\ \hline

$\Delta$    &  -198.3       &  -198.7                   &  -174.5             &  -39.1     &   15.0  \\
$\Sigma^*$  &  -262.3       &  -256.4                   &  -270.2             &  -10.3     &   24.1  \\
$\Xi^*$     &  -335.4       &  -327.6                   &  -383.4             &   28.8     &   27.1  \\
$\Omega$    &  -425.6       &  -418.4                   &  -511.6             &   67.8     &   25.3  \\ \hline
\end{tabular}
\caption{The self energies $\Sigma^{}_B$ in units of MeV are evaluated at the  
physical isospin averaged meson and baryon masses. The axial coupling constants are $F=0.45$ and $D =0.80$ together with $H = 9\,F-3\,D$  and $C=0$.  }
\label{tab:3}
\end{center}
\end{table}
A first result, that avoids an explicit solution of the set of coupled and nonlinear Dyson equations, can be obtained at the physical quark masses.  This is so since typically the mass splittings of the 
physical baryon masses can be reproduced quite accurately in terms of the three parameters $b_D,b_F, d_D$ only (see e.g. Tab. \ref{tab:FitParameters:N2LO}). Given such 
a parameter set it is justified to analyze the loop function at the physical masses directly.

We begin with a discussion of results for the 'diagonal' sector implied by the particular choice $C=0$. 
The baryon octet and decuplet self energies are computed 
according to (\ref{result-loop-8}, \ref{result-loop-10}) with (\ref{eliminate-mu}). The values for the baryon octet and decuplet self energies are listed in the 2nd column 
of Tab. \ref{tab:3}. Those numbers are to be compared with the various chiral moments $\Sigma^{(3)}_B$ of (\ref{loop-HB-3}, \ref{loop-HB-3-B}), $\Sigma^{(4)}_B$ of (\ref{loop-HB-4}, \ref{loop-HB-4-B}) 
and $\Sigma^{(5)}_B$ of (\ref{loop-HB-5}, \ref{loop-HB-5-B}) for the octet and decuplet states respectively. We make two encouraging observations. First 
we see a clear hierarchy in the successive orders in the self energies for all octet and decuplet states. The fifth order terms are significantly smaller than the third order terms. 
Second, the expansion truncated at fifth order is characterized by a mean deviation from the exact expressions of 
about 8 MeV only. 

We continue with a discussion of the 'offdiagonal' sector implied by the particular choice $F=D=H=0$. The results of this case study are collected in Tab. \ref{tab:4}.
We confirm the pattern observed before for the diagonal case. A clear hierarchy in the successive orders in the self energies is observed.  
The expansion truncated at fifth order is characterized by a mean deviation of about 2 MeV only. 
\begin{table}[t]
\setlength{\tabcolsep}{3.5mm}
\renewcommand{\arraystretch}{1.2}
\begin{center}
\begin{tabular}{c|rr||rrr }\hline
$B$         & $\Sigma_B$\, & $\Sigma^{(3+4+5)}_B$      & $\Sigma^{(3)}_B$    &  $\Sigma^{(4)}_B$ & $\Sigma^{(5)}_B$   \\ \hline \hline
$N$         & -56.6        &   -56.8                   &  -46.5              &   -8.3     &   -2.0  \\
$\Lambda$   & -118.3       &  -118.2                   & -127.9              &    9.2     &    0.5  \\
$\Sigma$    & -140.2       &  -135.4                   & -423.8              &   260.1    &   28.3  \\
$\Xi$       & -193.2       &  -190.2                   & -380.0              &   174.5    &   15.2  \\ \hline

$\Delta$    &  -115.4      &  -118.6                   &   -50.9             &   -61.2    &  -6.5  \\
$\Sigma^*$  &  -62.7       &   -64.7                   &   -32.0             &  -36.0     &   3.3  \\
$\Xi^*$     &   -4.7       &    -5.9                   &     1.5             &   -13.5    &   6.2  \\
$\Omega$    &   54.2       &    53.8                   &    43.5             &     4.5    &   5.8  \\ \hline
\end{tabular}
\caption{The self energies $\Sigma^{}_B$ in units of MeV are evaluated at the  
physical isospin averaged meson and baryon masses. The axial coupling constants are $F=D=H=0$  and $C=1.6$. }
\label{tab:4}
\end{center}
\end{table}

\begin{table}[t]
\setlength{\tabcolsep}{2.4mm}
\renewcommand{\arraystretch}{1.2}
\begin{center}
\begin{tabular}{c||r|rrr ||c|rrr}
$B$         & $Z_B$\,   & $Z^{[3]}_B$    &  $Z^{[4]}_B$ & $ Z^{[5]}_B$ & $\bar \Sigma^{\rm bubble}_B/Z_B$\,   & $\bar \Sigma^{[3]}_B$    &  $\bar \Sigma^{[4]}_B$ & $ \bar \Sigma^{[5]}_B$  \\ \hline \hline
$N$         &  1.118    & 0.463          & 1.226        & 1.167        &   -271.9  &   -500.8      &  -267.2     &  -263.3       \\
$\Lambda$   &  2.064    & 0.851          & 1.906        & 2.179        &   -222.4  &   -660.9      &  -250.9     &  -200.7      \\
$\Sigma$    &  2.507    & 0.615          & 2.433        & 2.300        &   -260.7  &   -1426.7     &  -289.9     &  -283.0       \\
$\Xi$       &  3.423    & 1.022          & 3.111        & 3.386        &   -223.4  &   -1071.7     &  -255.0     &  -219.8       \\ \hline
$\Delta$    &  1.570    & 0.757          & 1.514        & 1.615        &   -199.9  &   -297.7      &  -215.1     &  -196.4       \\
$\Sigma^*$  &  1.914    & 1.104          & 1.913        & 1.982        &   -169.8  &   -273.8      &  -182.2     &  -162.0       \\
$\Xi^*$     &  2.438    & 1.525          & 2.472        & 2.516        &   -139.5  &   -250.4      &  -148.3     &  -132.5       \\
$\Omega$    &  3.064    & 1.936          & 3.115        & 3.151        &   -121.2  &   -241.7      &  -127.1     &  -115.7       \\ 
\end{tabular}
\caption{The axial coupling constants are $F=0.45$ and $D =0.80$ together with $H = 9\,F-3\,D$  and $C=1.6$. The $\bar \Sigma^{[n]}_B$ as introduced in 
(\ref{def-Zn}) are measured in units of MeV. Physical meson and baryon masses are used for $M_B$, $M_R$ and $m_Q$.  }
\label{tab:5}
\end{center}
\end{table}

We turn to the effects of the wave-function factors $Z_B$. As was already demonstrated in Tab. \ref{tab:1}  
there is a significant deviation from the chiral limit value $Z_B\to 1$ implied by the loop contribution to the 
baryon self energy. This has a significant effect on the set of Dyson equations (\ref{gap-equation-B}). To illustrate this further 
we introduce $\bar \Sigma^{[n]}_B$ and $\bar Z^{[n]}_B$ with
\begin{eqnarray}
&& Z^{[n]}_B = 1 -\sum_{k=3}^{n}\, \frac{\partial}{\partial M_B}\,\bar \Sigma_B^{ {\rm bubble}-k} \,, \qquad 
 \bar \Sigma^{[n]}_B = \sum_{k=3}^n \,\bar \Sigma_B^{ {\rm bubble}-k} / Z^{[n]}_B \,,
 \label{def-Zn}
\end{eqnarray}
and collect numerical values thereof in Tab. \ref{tab:5}. We observe a convincing convergence pattern for all baryons. Note however, that due to the large third order 
terms for the baryon octet states significant result can be expected only at the accuracy level four and higher.

From this section we conclude that indeed a systematic decomposition of  the baryon self energies into chiral moments is feasible and 
appears well converging. However, the sizes of the fifth order terms are a bit too large so that a full control of the baryon masses 
accurate at the few MeV level may require the consideration of the 'full' fifth order contribution, which involves the computation of a class of 
two-loop diagrams.

\newpage

\section{Low-energy parameters from QCD lattice data}
\vskip0.3cm
In the previous chapter we illustrated that the power counting domain (PCD) of the chiral expansion for the baryon octet and decuplet 
masses may be surprisingly large encompassing most likely the physical quark masses. In order to demonstrate such a 
behavior it is  useful to reorganize the chiral expansion. If its various moments are expressed in terms of the physical meson and 
baryon masses the first few terms are able to reproduce the full one-loop expressions with increasing accuracy. We would like to challenge this 
picture by a realistic parameter set that includes all low-energy parameters relevant at N$^3$LO  and that is adjusted to a QCD lattice data 
set on the baryon masses at different sets of unphysical quark masses. 

In the previous analysis \cite{Lutz:2014oxa} a large set of QCD lattice data 
points were quite accurately reproduced and in part predicted. Unfortunetely, the underlying set of low-energy parameters can not be so easily 
used for our purpose. In principle one may envisage a matching of the scheme used in \cite{Lutz:2014oxa} with the one developed here.  
However, since such a matching would rely necessarily on an expansion in powers of $(\Delta/M)$, the extremely poor convergence properties 
of this expansion make a quantitative application of such a matching futile. Moreover, as was illustrated in Section 3.4 the effects of the baryon's
wave-function renormalization were not sufficently well treated in  our previous work \cite{Lutz:2014oxa}. In any case we deem the scheme proposed here 
superior to the one in \cite{Lutz:2014oxa}. While our approach is strictly consistent with chiral constraints in the chiral regime 
with $m_Q < \Delta$ these constraints are realized in \cite{Lutz:2014oxa} only at a formal level to some order in $(\Delta/M)^n$.

\subsection{The chiral extrapolation scheme}
\vskip0.3cm

We adjust the low-energy parameters to a set of QCD lattice data. While this can in principle
be done at different chiral orders, we do so using the subtracted loop expressions (\ref{result-loop-8},  \ref{result-loop-10}) in (\ref{gap-equation-B})
supplemented by (\ref{def-tadpole}, \ref{eliminate-mu}) and the finite volume corrections of the scalar loop expressions as worked out previously 
in  \cite{Lutz:2014oxa}.  The good convergence properties of our chiral expansion as formulated in terms of the physical meson and baryon masses 
we take as a reasonable justification of this strategy. 

Moreover, despite the rapid convergence properties of the reordered chiral expansion we found that the 
fifth order contributions from the bubble-loop can still be sizable of the order of 20 MeV. Therefore it is a matter of convenience to perform  
our fits using the one-loop functions as detailed in Chapter 3. Therewith the finite volume corrections specific to the various chiral moments, 
whose derivation would require further tedious algebra, are not required. It should also be mentioned that the complete fifth order expression (N$^4$LO) would 
receive further contributions from a set of two-loop diagrams as explored for instance in \cite{Schindler:2007dr}. In a complete study such two-loop diagrams, 
however, also including the decuplet degrees of freedom, should be considered. This is beyond the scope of the present work and not yet available in the literature.

\begin{table}[t]
\setlength{\tabcolsep}{4.5mm}
\renewcommand{\arraystretch}{1.3}
\begin{center}
\begin{tabular}{l|rrr}
                                                & Fit 1      &  Fit 2       & Fit 3       \\   \hline

$10^3\,(L_4 - 2\,L_6)\, $                        & -0.0462    &  -0.0405     & -0.0488     \\
$10^3\,(L_5 - 2\,L_8)\, $                        & -0.0892    &  -0.1084     & -0.1103     \\
$10^3\,(L_8 + 3\,L_7)\, $                        & -0.4808    &  -0.4828     & -0.4872     

\end{tabular}
\caption{Low-energy parameters  from a fit to the baryon octet and decuplet masses
of the PACS, LHPC,  HSC, NPLQCD, QCDSF-UKQCD and ETMC groups as described in the text. All parameters  $L_i$ are scale dependent 
given at $\mu =$ 770 MeV together with $f =92.4$ MeV (see (\ref{meson-masses-q4})). }
\label{tab:FitParametersA}
\end{center}
\end{table}

It is pointed out that the values of the low-energy constants $L_4 -2\,L_6, L_5 -2\,L_8$ have a crucial 
impact on the description of the baryon masses from lattice QCD simulations \cite{Lutz:2014oxa}. In our approach  we use the published pion and kaon masses from 
a given lattice ensemble. With (\ref{meson-masses-q4}) the later translate into the quark masses that are used in our chiral formulae for the baryon masses. In turn the three combinations 
of Gasser and Leutwyler low-energy parameters as collected in Tab. \ref{tab:FitParametersA} are determined from a fit to the baryon masses. Note however,
that for any given choice of $f, L_4 - 2\,L_6$ and $L_5 -2\,L_8$  the value of $L_8 + 3\,L_7$ is determined by the physical 
eta-meson mass at the one-loop level (\ref{meson-masses-q4}). With Tab. \ref{tab:FitParametersA}  significant results are obtained for the low-energy constants. In contrast their latest determination  
from the phenomenology of the meson sector suffers from substantial uncertainties \cite{Bijnens:2011tb,Bijnens:2014lea}. For instance the value of the
combination $L_5 -2\,L_8$ may be positive or negative.  There is yet a further interesting issue to be discussed. From Tab. \ref{tab:FitParametersA} one may infer the quark-mass ratio
$m_s/m \simeq 26.1 \pm 0.1$, which comes close to the empirical value claimed in the PDG within a $5 \%$ uncertainty only \cite{PDG}.

\subsection{Three fit scenarios}
\vskip0.3cm
From the many distinct fits we document three typical scenarios that all rely on the large-$N_c$ sum rules (\ref{res-FDCHs}, \ref{def-combinations}, \ref{ces-subleading}, \ref{Q4-subleading}) as they 
are recalled in Chapter 2 at subleading order. This is supplemented by the renormalization scale-invariance conditions (\ref{ces-subleading-Gamma}).
While Fit 1 is characterized by insisting on the two identities
\begin{eqnarray}
 b_F + b_D = d_D/3 \,, \qquad \qquad d_0 + d_D/3 - b_0 = 2\,b_D \,,
 \label{rescall-b-Nc}
\end{eqnarray}
Fit 2 insists on only the first of the two equations in (\ref{rescall-b-Nc}). Last, in Fit 3 all low-energy parameters $b_0, b_D, b_F$ and $d_0, d_D$ are kept independent. We remind the reader 
that given the imposed large-$N_c$ relations there are altogether only 8 independent low-energy parameters that drive the terms proportional to the square of the quark masses. 
They are complemented by 7 degrees of freedom that determine the symmetry conserving two-body counter terms.

\subsection{How to fit the lattice data}
\vskip0.3cm

The applied strategy how to arrive at a realistic parameter set is described in the following. We first identify the lattice data set used in our fits. 
Only published and documented results for QCD lattice simulations with three light flavours at pion and kaon masses smaller than $600$ MeV are considered. 
That leaves data sets from  PACS, LHPC,  HSC,  NPLQCD, QCDSF-UKQCD and ETMC \cite{PACS-CS2008,LHPC2008,HSC2008,NPLQCD:2011,Bietenholz:2011qq,Alexandrou:2013joa}. 
We are aware of the recent lattice ensembles of the CLS group with 2+1 flavors based on nonperturbatively improved Wilson fermions \cite{Bruno:2014jqa,Bali:2016umi,Bruno:2016plf}. Results for 
baryon masses are not available yet. Based on the published pion and kaon masses of the various ensembles \cite{Bruno:2016plf} we will attempt to make a prediction of the baryon masses 
as they follow from our sets of low-energy parameters. 

Like in the previous 
work \cite{Lutz:2014oxa} we use the empirical values of the physical baryon octet and decuplet masses for a determination of the lattice scales. All of 
our parameter sets are tuned as to reproduce the isospin averaged baryon masses from the PDG \cite{PDG} within an error window of at most three MeV. It is not our purpose to show that QCD lattice 
simulations are consistent with the empirical baryon masses, rather we assume the latter and wish to extract from the lattice data the 
low-energy parameters of the chiral Lagrangian.

\begin{table}[t]
\setlength{\tabcolsep}{2.5mm}
\renewcommand{\arraystretch}{1.0}
\begin{center}
\begin{tabular}{l|ccc|c} 

                                                      &  Fit 1   &  Fit 2    &  Fit 3    &  lattice  \;group  \\ \hline

$a_{\rm PACS-CS}\,   \hfill \mathrm{[fm]}$            &  0.0943 & 0.0929    &  0.0925  & 0.0907(14)  \hfill \cite{PACS-CS2008}  \\ 
$a_{\rm LHPC} \,   \hfill \mathrm{[fm]}$              &  0.1318 & 0.1289    &  0.1285  & 0.1241(25)  \hfill \cite{LHPC2008} \\
$a_{\rm HSC} \,   \hfill \mathrm{[fm]}$               &  0.1229 & 0.1218    &  0.1225  & 0.1229(7)   \hfill \cite{HSC2008} \\
$a_{\rm QCDSF-UKQCD}\,  \hfill \mathrm{[fm]}$         &  0.0759 & 0.0758    &  0.0758  & 0.0765(15)  \hfill \cite{Bietenholz:2011qq} \\ 
$a^{\beta =1.90}_{\rm ETMC}\,   \hfill \mathrm{[fm]}$ &  0.1030 & 0.1019    &  0.1016  & 0.0934(37)  \hfill \cite{Alexandrou:2013joa}\\   
$a^{\beta =1.95}_{\rm ETMC}\,   \hfill \mathrm{[fm]}$ &  0.0921 & 0.0920    &  0.0920  & 0.0820(37)  \hfill \cite{Alexandrou:2013joa} \\   
$a^{\beta =2.10}_{\rm ETMC}\,   \hfill \mathrm{[fm]}$ &  0.0680 & 0.0679    &  0.0680  & 0.0644(26)  \hfill \cite{Alexandrou:2013joa}\\ 

                                                      &         &           &          &   systematic error  \\\hline

$\chi^2_{\rm PACS-CS}/N$                              & 1.7241     & 1.0507   & 0.8253    &    10 {\rm MeV}  \\
                                                      & 2.3080     & 1.4678   & 1.1703    &    5 {\rm MeV}  \\ \hline

$\chi^2_{\rm LHPC}/N$                                 & 0.9029     & 2.1253   & 1.4668    &    20 {\rm MeV}  \\
                                                      & 6.8158     & 5.6362   & 3.7141    &    10 {\rm MeV}\\
                                                      & 19.842     & 17.454   & 12.216    &    5 {\rm MeV}  \\ \hline

$\chi^2_{\rm HSC}/N$                                  & 0.7954     & 0.7631   & 0.7367    &   10 {\rm MeV}   \\
                                                      & 1.0224     & 0.9916   & 0.9453    &   5 {\rm MeV}   \\ \hline
                                                      
$\chi^2_{\rm NPLQCD}/N$                               & 0.4100     & 0.3185   & 0.4077    &   10 {\rm MeV}   \\
                                                      & 1.2859     & 0.9788   & 1.2473    &   5 {\rm MeV}   \\ \hline

$\chi^2_{\rm QCDSF-UKQCD}/N$                          & 0.6776     & 0.6741   & 0.7045    &   10 {\rm MeV}   \\
                                                      & 0.9174     & 0.9195   & 0.9353    &   5 {\rm MeV}   \\ \hline
                                                      
$\chi^2_{\rm ETMC}/N$                                 & 0.9744     & 0.9772   & 1.1405    &   10 {\rm MeV}   \\ 
                                                      & 1.2944     & 1.2673   & 1.5210    &    5 {\rm MeV}  \\

\end{tabular}
\caption{Our determination of the lattice scale for PACS-CS, LHPC, HSC, NPLQCD, QCDSF-UKQCD and ETMC with $a_{\rm HSC} = a_{\rm NPLQCD}$. 
The set of lattice data fitted is described in the text. The corresponding low-energy parameters of Fit 1-3 are given in Tab. \ref{tab:FitParametersA} - \ref{tab:FitParametersD}. }
\label{tab:lattice-scale}
\end{center}
\end{table}

We do not implement discretization effects in our chiral extrapolation approach since the majority of the data sets is  
available at a single beta value only. As a consequence a systematic error analysis is not possible yet in our present study. 
In order to cope with this deficiency we supplement the statistical errors given by the lattice groups by a systematic error in mean quadrature.  
A lower limit for such an error may be put by the isospin splittings of the baryon octet and decuplet masses, which is in the range of 1-4 MeV. Surely the budget for the systematic error 
should further increase from discretization effects as well as from the absence of chiral N$^4$LO effects. Following 
\cite{Lutz:2014oxa} we perform fits at different ad-hoc values for the systematic error. We would argue that once this error is sufficiently large the $\chi^2$ per data point should be close to one. 
By this method we may estimate the net size of the systematic error in our chiral extrapolation approach.

We found considerable tension amongst the world lattice data set which prohibits a significant fit with a common systematic error estimate. A natural solution to this misery arises if 
a larger error budget is assigned to the data of LHPC. We note that this is the case even if we ignore the already previously in \cite{Lutz:2014oxa} questioned data on the baryon 
decuplet masses of this group. We would expect that discretization errors are largest for the baryon data of LHPC. Their studies are the only ones that are based on a mixed action framework 
with domain-wall valence quarks but staggered sea-quark ensembles generated by  MILC \cite{Orginos:1999cr,Orginos:1998ue,MILC2001,MILC2004}. 
In anticipation of the details of our fit results we find the systematic error to be about 10 MeV for all lattice groups but LHPC, for which our estimate comes at about 20 MeV.

To actually perform such fits is quite a computational challenge. For any set of the low-energy parameters eight coupled non-linear equations are to be solved for each lattice configuration 
considered. We chose to apply the evolutionary algorithm of GENEVA 1.9.0-GSI \cite{Geneva}. Any of our attempts to use a gradient 
approach as offered for instance by MINUIT did not lead to any competitive results. We submitted typical GENEVA runs with a population size of 8000 on 700 parallel CPU cores. The runtime of fits last 
to about a week and more.

In Tab. \ref{tab:lattice-scale} we show the consequence of three different parameter sets for the spatial lattice scales.
The values from the three fits  are consistent with the corresponding values determined from the different lattice groups given their uncertainties. The spread implied 
by the different fits is much smaller than the uncertainty of the lattice scales given by the lattice groups with the exception of the LHPC data. As was emphasized 
in \cite{Lutz:2012mq,Alexandrou:2013joa,Lutz:2014oxa} the physical baryon masses pose a strong constraint on the precise value of the 
QCD lattice scale. The table illustrates the quality of the various fits by providing the chi-square ($\chi^2$) per number of fitted data point ($N$). 
Two values are provided in the table for all collaborations with the exception of LHPC, for which three cases are provided. 
While the upper value is computed with  respect to a systematic error of 10 MeV, the lower one with respect to 5 MeV. For the LHPC the upper value corresponds to the additional 20 MeV.
All $\chi^2/N$ values are close  to one at the advocated estimates for the systematic error of 10 MeV for the PACS-CS, HSC, QCDSF-UKQCD, NPLQCD, ETMC and 20 MeV for the data of LHPC.

\subsection{Baryon masses on QCD lattice ensembles}
\vskip0.3cm

\begin{figure}[t]
\center{
\includegraphics[keepaspectratio,width=0.97\textwidth]{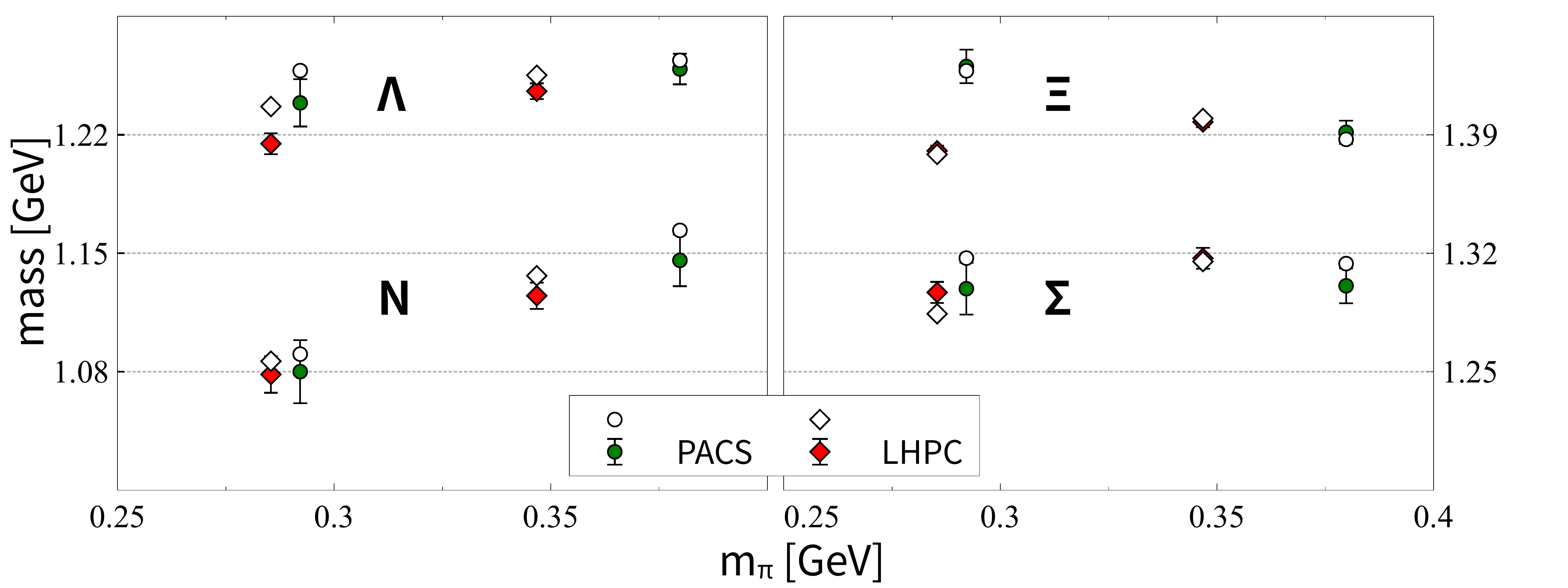} }
\vskip-0.2cm
\caption{\label{fig:lattice-1} Baryon octet masses from Fit 3  compared to results from  PACS-CS and LHPC \cite{PACS-CS2008,LHPC2008}.  }
\end{figure}

\begin{figure}[t]
\center{
\includegraphics[keepaspectratio,width=0.97\textwidth]{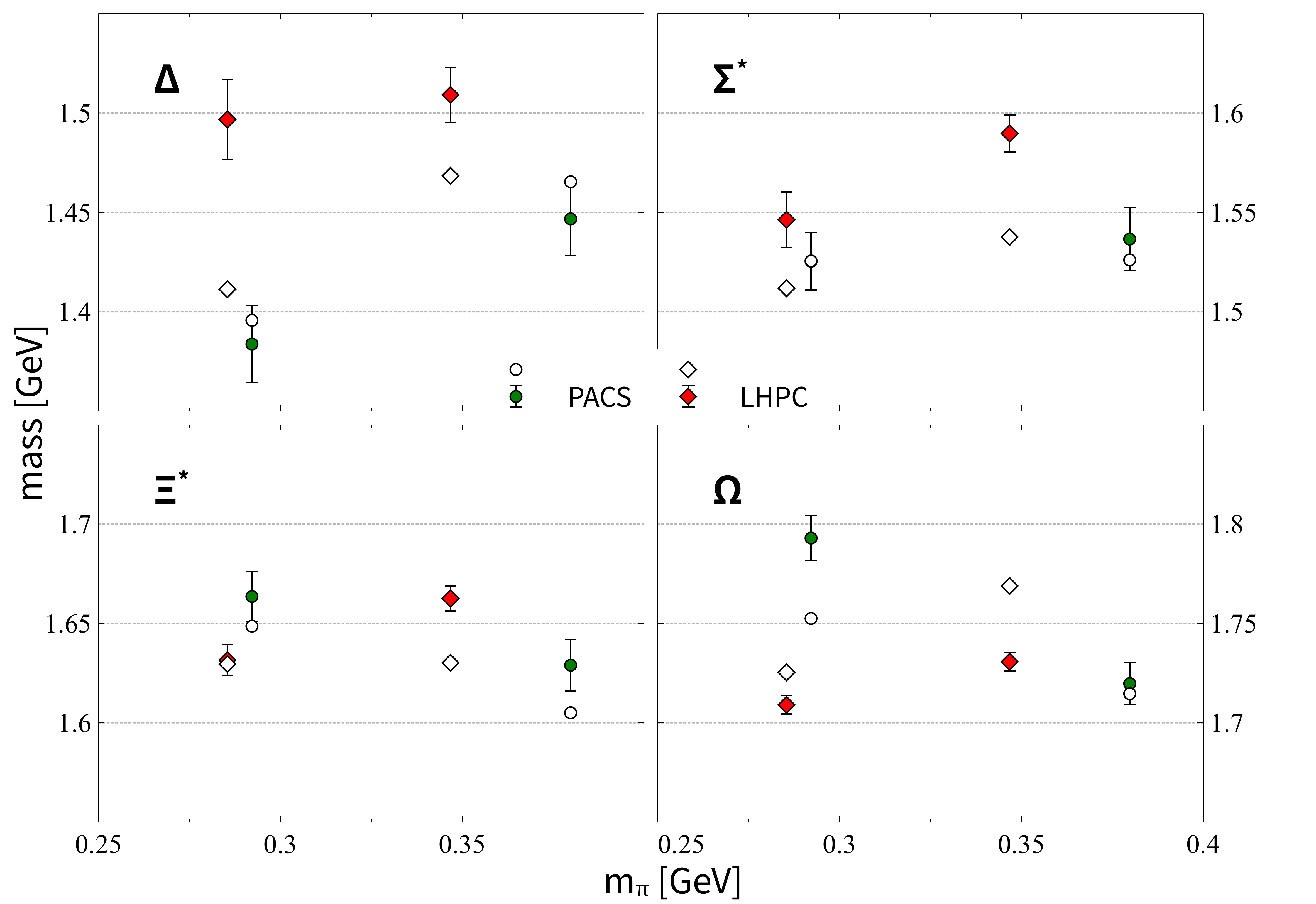} }
\vskip-0.2cm
\caption{\label{fig:lattice-2} Baryon decuplet masses from Fit 3 compared to results from PACS-CS and LHPC \cite{PACS-CS2008,LHPC2008}.  }
\end{figure}

Let us illustrate our results with a direct comparison to the lattice data. The following series of figures is based on Fit 3 and well illustrates our conclusions. 
In all figures the lattice data are shown only with their statistical errors as provided by the collaborations. Our results are presented with open symbols always that lie on top of the data points. 
First the PACS-CS and LHPC data are scrutinized with Fig. \ref{fig:lattice-1} and Fig \ref{fig:lattice-2}, where the baryon octet and decuplet masses are compared with the results of Fit 3. Note that 
while the PACS-CS data are based on 32$^3$ lattices, the LHPC data on 20$^3$ ensembles. There are three comments we wish to make here. 
As was anticipated the description of the PACS-CS data is significantly better than the LHPC data. 
Note that we did not consider an additional set of masses from PACS-CS at $m_\pi \simeq 150$ MeV. The evaluation of finite volume effects is difficult to fully control, since the product $ m_\pi\, L \simeq 2 $
is unfavorably small. The considerable tension amongst the two data sets is most clearly illustrated by the results for the decuplet masses. For the $\Delta$, $\Sigma^*$ and $\Xi^*$ masses of LHPC one may 
speculate that the temporal lattice separation was not sufficiently large so that the asymptotic exponential mass was not reached yet in their mass determination. This possible explanation is less likely for the 
$\Omega$ mass, where we expect a significant underestimated mass from LHPC. Here a comparison with the results of PACS-CS may be instructive. 
For instance, the LHPC ensemble at the pion mass of about 347 MeV has an associated kaon mass of about 587 MeV. This is to be compared to the PACS-CS ensemble at pion mass 292 MeV, which comes with a kaon 
mass of about 586 MeV. It is striking to see that the $\Omega$ baryon masses of the two ensembles are split nevertheless by almost 100 MeV. We take this as a hint that there indeed must be significant 
discretization effects in the LHPC ensembles. We note that the chisquare values for the LHPC ensembles in Tab. \ref{tab:FitParametersA} include the four baryon octet masses together with the $\Omega$ mass 
only. Also in addition the $\Delta, \Sigma^*$ and $\Xi^*$ masses were not considered in the chisquare of the previous study \cite{Lutz:2014oxa}.

\begin{figure}[t]
\center{
\includegraphics[keepaspectratio,width=0.97\textwidth]{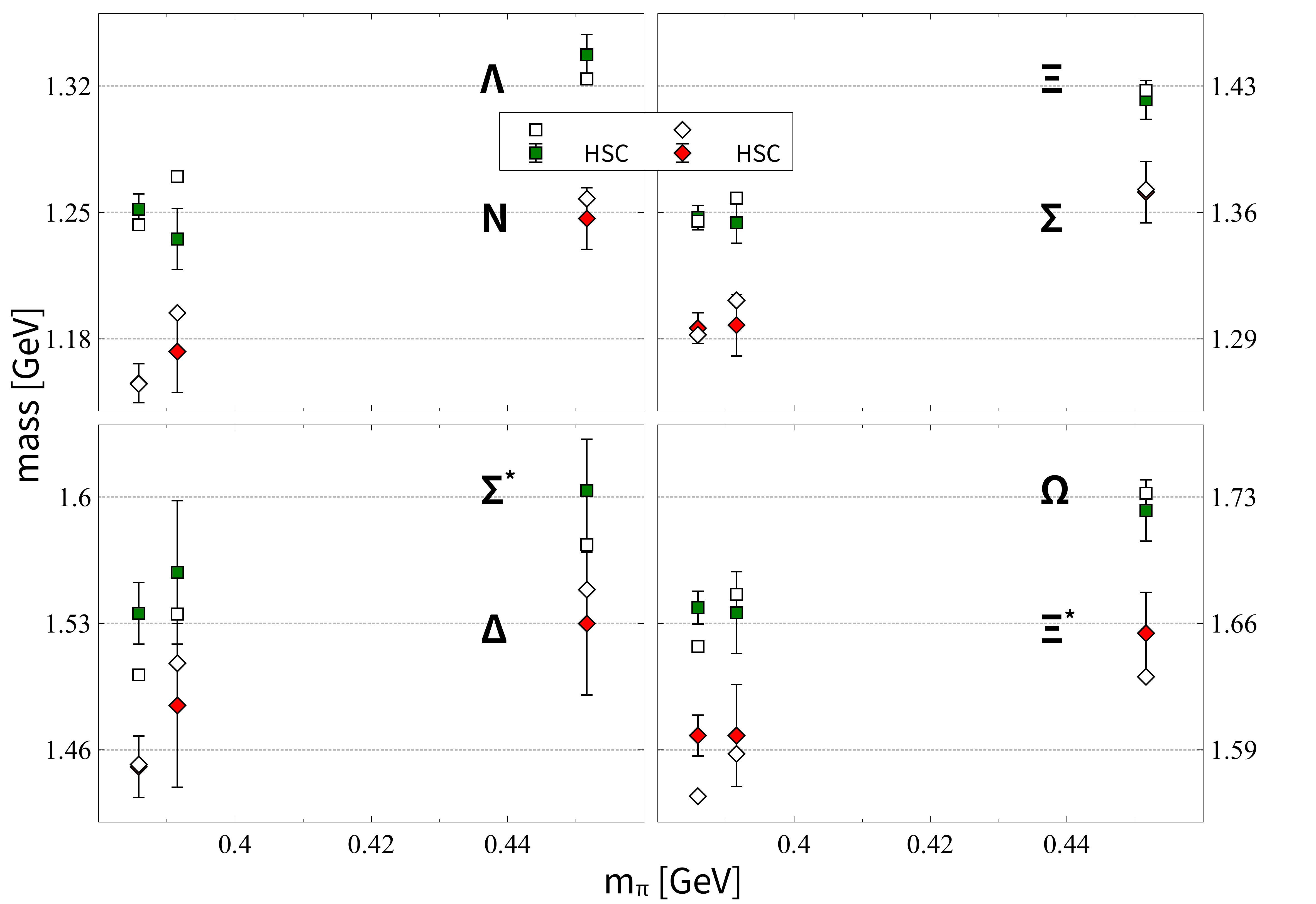} }
\vskip-0.2cm
\caption{\label{fig:lattice-3} Baryon masses from Fit 3 compared to results from HSC \cite{HSC2008}.  }
\end{figure}

\begin{figure}[t]
\center{
\includegraphics[keepaspectratio,width=0.97\textwidth]{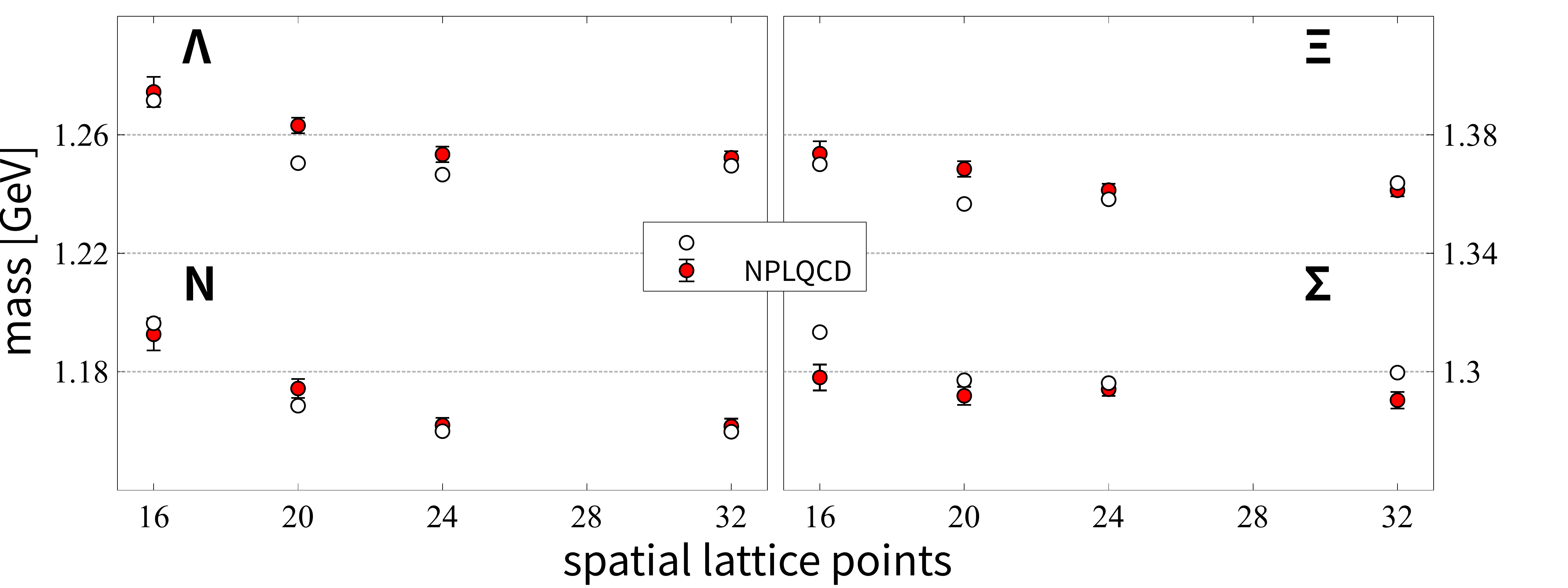}}
\vskip-0.2cm
\caption{\label{fig:lattice-4} Baryon octet masses from Fit 3  compared to results from NPLQCD \cite{NPLQCD:2011}.  }
\end{figure}

We continue with Fig. \ref{fig:lattice-3}, where we present our results for the HSC ensembles on 16$^3$ and 24$^3$ lattices. For all ensembles presented a convincing reproduction of the lattice 
data is achieved. Additional studies on the ensemble at pion mass 390 MeV were generated by NPLQCD, for which we offer Fig. \ref{fig:lattice-4}. Such data are at four different lattice volumes, however only 
for the baryon octet masses. The lattice data are reproduced well for all states with possibly some reservation for the $\Sigma$. Incidentally, for the latter state the plateau signals as shown 
in Fig. 9 of \cite{NPLQCD:2011} suffer from sizable fluctuations in particular on the 32$^3$ ensemble. If compared to the signals from the four octet states on the 24$^3$ lattices the plateau is 
much less pronounced and stable. See for instance the $\Lambda $ in Fig. 7 of \cite{NPLQCD:2011}. We note that the qualitative pattern of Fig. \ref{fig:lattice-3} remains through all of our three fit scenarios.

\begin{figure}[t]
\center{
\includegraphics[keepaspectratio,width=0.85\textwidth]{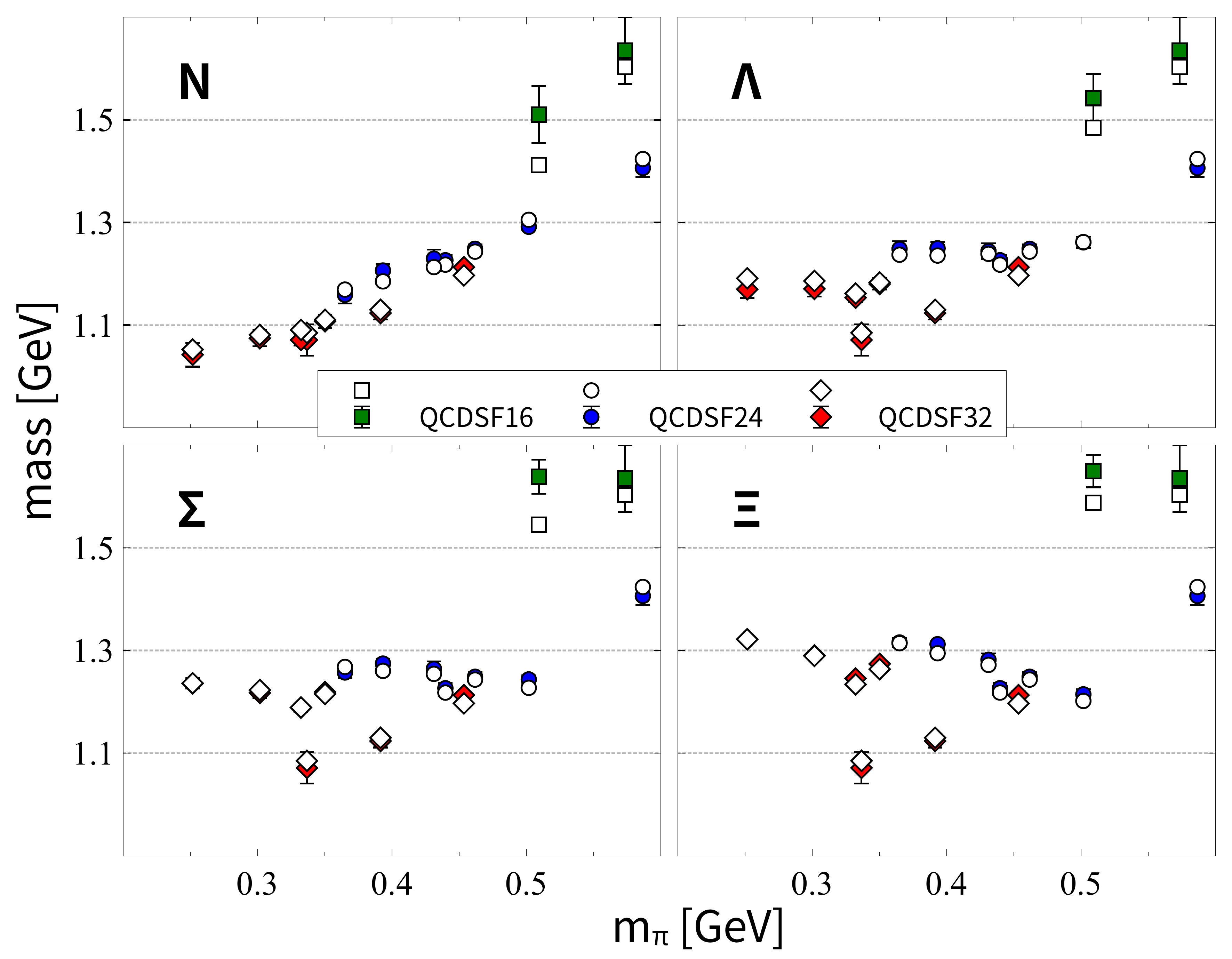}
\includegraphics[keepaspectratio,width=0.85\textwidth]{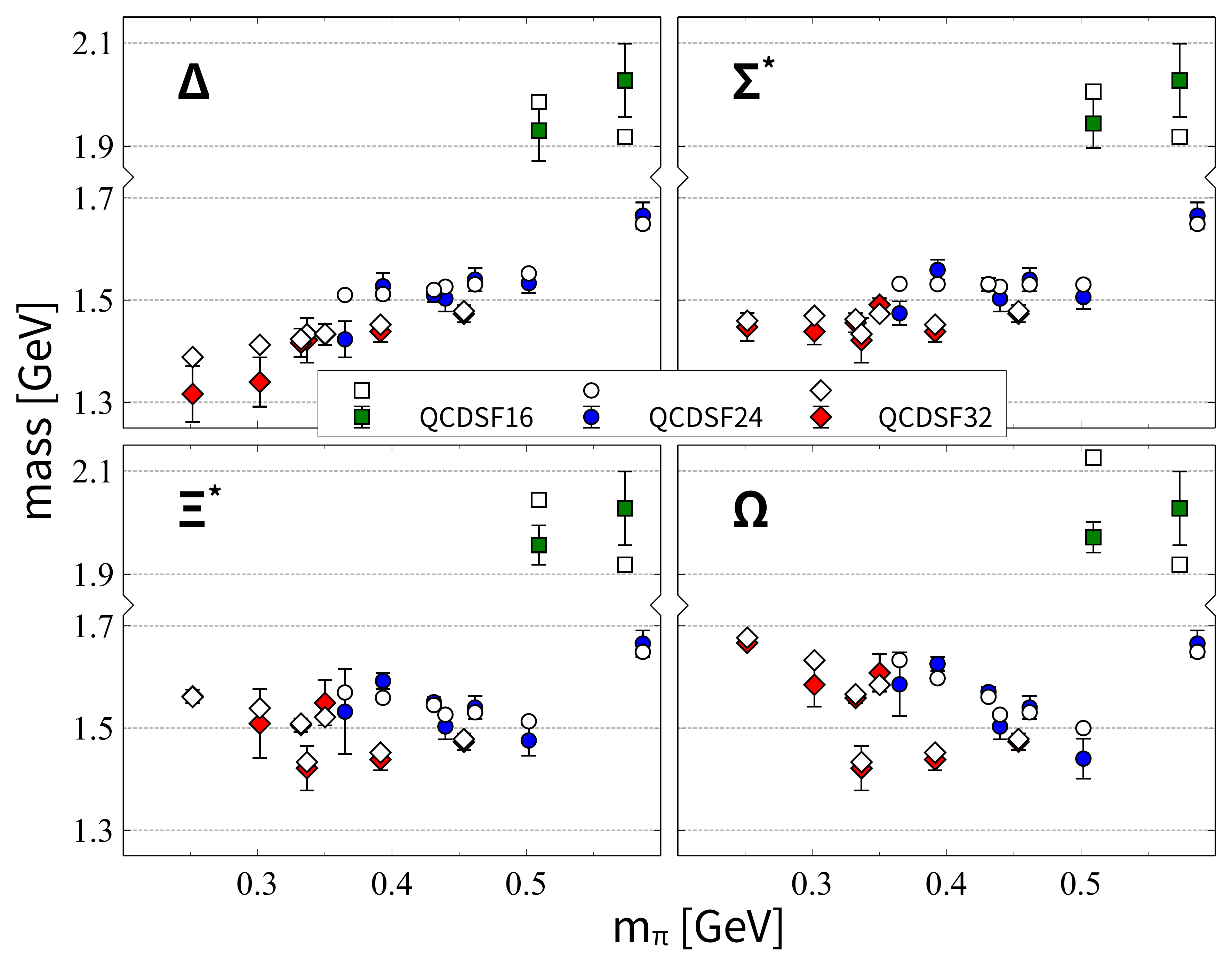} }
\vskip-0.2cm
\caption{\label{fig:lattice-5} Baryon masses from Fit 3  compared to results from QCDSF-UKQCD \cite{Bietenholz:2011qq}.  }
\end{figure}

\begin{figure}[t]
\center{
\includegraphics[keepaspectratio,width=0.85\textwidth]{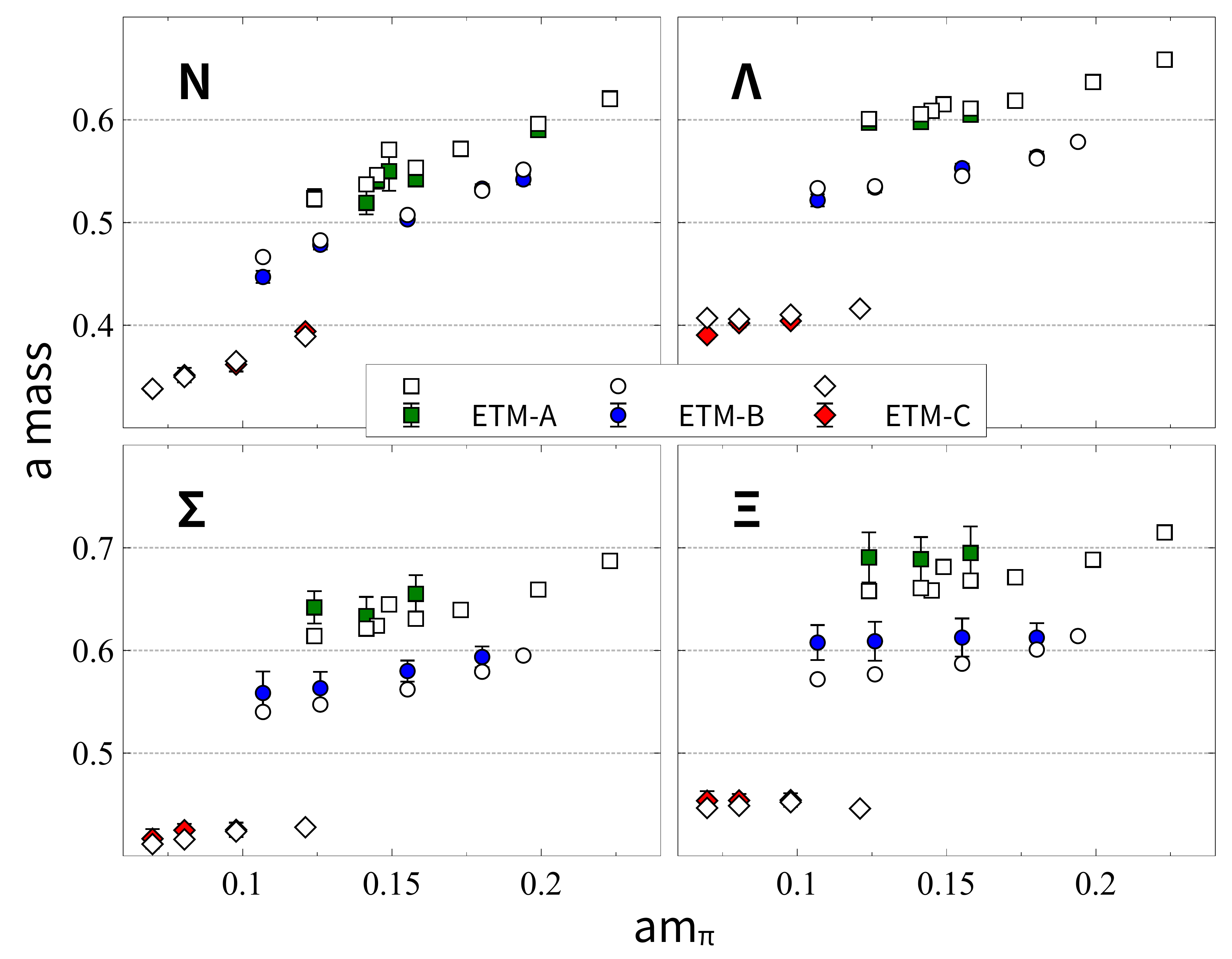}
\includegraphics[keepaspectratio,width=0.85\textwidth]{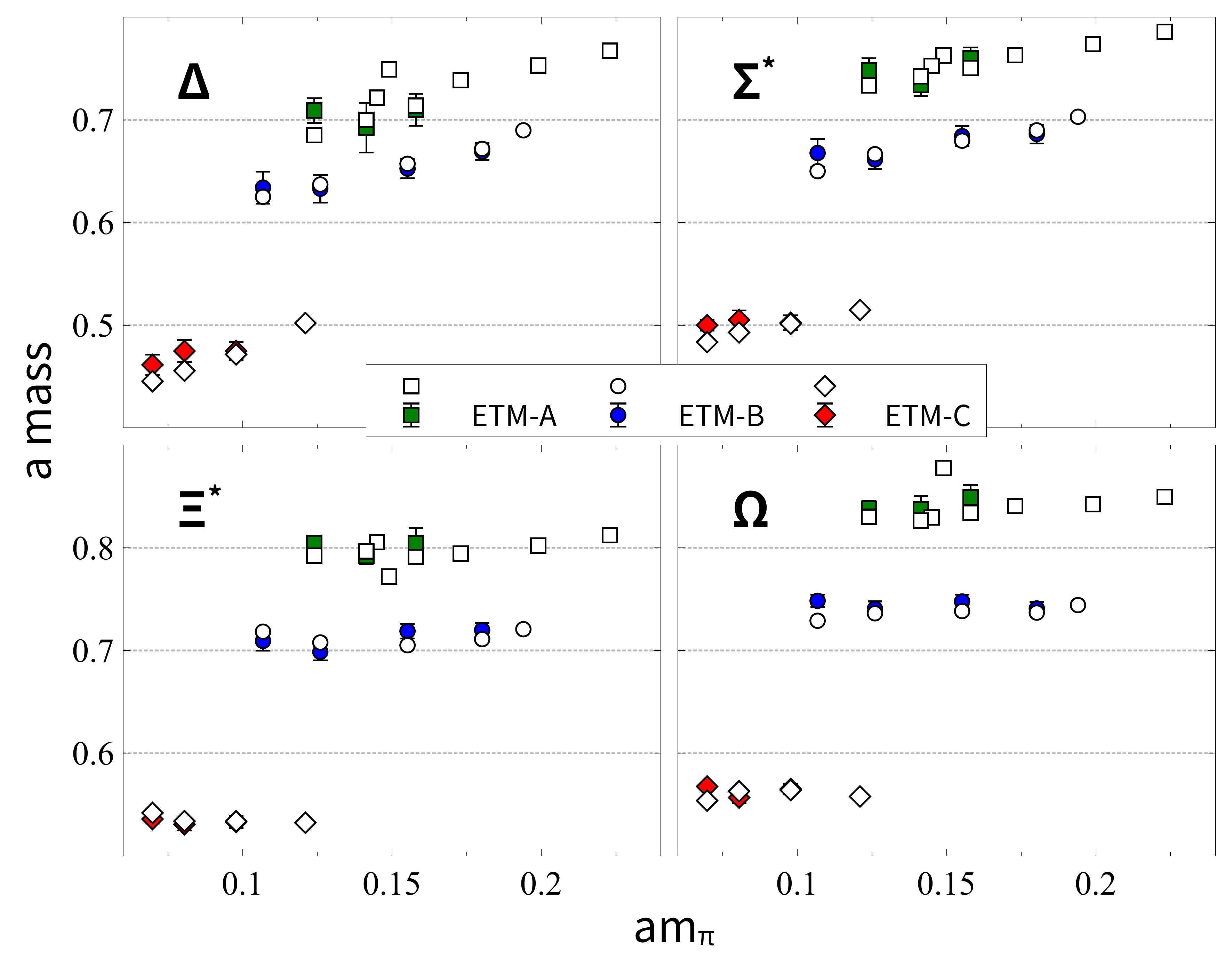} }
\vskip-0.2cm
\caption{\label{fig:lattice-7} Baryon  masses from Fit 3  compared to results from ETMC \cite{Alexandrou:2014sha}.  }
\end{figure}

\clearpage

Consider the ensembles of QCDSF-UKQCD, that are generated on 16$^3$, 24$^3$ and 32$^3$ lattices. In Fig. \ref{fig:lattice-5}  the baryon masses are shown in 
application of the lattice scales taken from Fit 3 as given in Tab. \ref{tab:FitParametersA}. An excellent description of all lattice points is achieved. For more discussions, in particular on some decuplet masses, we refer 
to the detailed previous study \cite{Lutz:2014oxa}. 

It remains to illustrate our results for the ensembles of the ETMC, which generated baryon masses at different lattice scales and volumes as summarized in Fig. \ref{fig:lattice-7}. 
In the previous study \cite{Lutz:2014oxa} their baryon masses were predicted almost quantitatively based on the knowledge of the pion and kaon masses of the ensembles only. This came as a 
great surprise to the community and deserves further studies. With our current fits based on a further improved chiral extrapolation framework a full quantitative description is achieved, given, 
however,  the assumption that the LHPC ensembles indeed suffer from significant discretization effects. In the two figures results for all ensembles of ETMC are included, also for those for which their baryon masses 
are so far not derived. With our figures such masses are to be taken as predictions of our chiral extrapolation studies which await confirmation or rejection.

\subsection{Estimates for low-energy parameters}
\vskip0.3cm

We turn to the low-energy parameters determined by our three fits. They fall into two classes. In 
Tab. \ref{tab:FitParametersB} and Tab. \ref{tab:FitParametersC} we collect parameters that are scale invariant and in Tab. \ref{tab:FitParametersD} those that 
run on the renormalization scale $\mu$. Given our particular self consistent summation scheme all baryon masses arise as $\mu$ independent quantities.
It is emphasized that our results are heavily relying on the large-$N_c$ sum rules presented in Chapter 2 in great detail. Even at subleading order in the $1/N_c$ expansion the 
number of free fit parameters is reduced  to a great extent and only therefore significant results from the lattice data set can be deduced. 
Our prime interest are the parameters listed in Tab. \ref{tab:FitParametersB} and Tab. \ref{tab:FitParametersC}  since they are 
of central importance for coupled-channel computations of low-energy meson-baryon scattering based on the chiral Lagrangian. 

We refrain from providing statistical errors in the parameters being 
not relevant. The number of considered lattice data points of about 300 makes the statistical error negligible. Any uncertainty is driven by systematic effects that are not fully under control.  
The most reasonable we can do is to offer a set of distinct scenarios as we do with our Fit 1-3.

\begin{table}[t]
\setlength{\tabcolsep}{4.5mm}
\renewcommand{\arraystretch}{1.1}
\begin{center}
\begin{tabular}{l|rrr}
                                            & Fit 1      &  Fit 2     & Fit 3       \\   \hline

$ M\;\;$ \hfill [GeV]                       &  0.8796    & 0.8745     & 0.8692    \\
$\Delta $\hfill [GeV]                       &  0.3510    & 0.3724     & 0.3673       \\ \\ \hline

$ b_0\, \hfill \mathrm{[GeV^{-1}]}$         & -0.6784    &  -0.6805   & -0.6911     \\
$ b_D\, \hfill \mathrm{[GeV^{-1}]}$         &  0.0788    &   0.0820   &  0.0761     \\
$ b_F\, \hfill \mathrm{[GeV^{-1}]}$         & -0.3281    &  -0.3409   & -0.3281     \\
$ d_0\, \hfill \mathrm{[GeV^{-1}]}$         & -0.2714    &  -0.2331   & -0.2595     \\
$ d_D\, \hfill \mathrm{[GeV^{-1}]}$         & -0.7479    &  -0.7768   & -0.7533     \\  \\  \hline

$F$                                         & 0.4893     &  0.4849    & 0.4858         \\
$D$                                         & 0.7408     &  0.7453    & 0.7442         \\
$C$                                         & 1.4815     &  1.4905    & 1.4884         \\
$H$                                         & 2.1809     &  2.1285    & 2.1395         \\    

\end{tabular}
\caption{The leading order low-energy parameters  from a fit to the baryon  masses of the lattice data set as described in the text. All parameters are scale independent.  
}
\label{tab:FitParametersB}
\end{center}
\end{table}

\begin{table}[t]
\setlength{\tabcolsep}{4.5mm}
\renewcommand{\arraystretch}{1.2}
\begin{center}
\begin{tabular}{l|rrr}
                                                  & Fit 1      &  Fit 2    & Fit 3       \\   \hline
 
$\bar g^{(S)}_0\,\hfill\mathrm{[GeV^{-1}]}$       & -7.9775    & -7.9905   & -8.0106             \\
$\bar g^{(S)}_1\,\hfill\mathrm{[GeV^{-1}]}$       &  0.2632    &  0.7972   &  0.7894      \\
$\bar g^{(S)}_D\,\hfill\mathrm{[GeV^{-1}]}$       & -1.0381    & -1.2850   & -1.3066      \\
$\bar g^{(S)}_F\,\hfill\mathrm{[GeV^{-1}]}$       & -4.4782    & -5.2778   & -4.6855     \\  \hline
$\bar h^{(S)}_1\,\hfill\mathrm{[GeV^{-1}]}$       & -3.3318    & -2.5412   & -3.2253       \\
$\bar h^{(S)}_2\,\hfill\mathrm{[GeV^{-1}]}$       &  0.0000    &  0.0000   &  0.0000      \\
$\bar h^{(S)}_3\,\hfill\mathrm{[GeV^{-1}]}$       & -8.1001    & -9.8693   & -9.0970     \\
$\bar h^{(S)}_4\,\hfill\mathrm{[GeV^{-1}]}$       & -3.9791    & -3.0050   & -2.9407       \\
$\bar h^{(S)}_5\,\hfill\mathrm{[GeV^{-1}]}$       &  0.2204    &  1.2208   &  1.2929      \\
$\bar h^{(S)}_6\,\hfill\mathrm{[GeV^{-1}]}$       &  3.9791    &  3.0050   &  2.9407       \\    \\

$\bar g^{(V)}_0\,\hfill\mathrm{[GeV^{-2}]}$       & -0.0735    & -0.1456   & -0.3197         \\
$\bar g^{(V)}_1\,\hfill\mathrm{[GeV^{-2}]}$       & -6.7576    & -6.9246   & -6.7135     \\ 
$\bar g^{(V)}_D\,\hfill\mathrm{[GeV^{-2}]}$       &  8.8166    &  9.0872   &  9.0061        \\   
$\bar g^{(V)}_F\,\hfill\mathrm{[GeV^{-2}]}$       & -2.0591    & -2.1625   & -2.1445      \\  \hline
$\bar h^{(V)}_1\,\hfill\mathrm{[GeV^{-2}]}$       &  3.9697    &  3.9756   &  3.9915       \\
$\bar h^{(V)}_2\,\hfill\mathrm{[GeV^{-2}]}$       &  9.9114    &  9.7755   &  9.9144      \\
$\bar h^{(V)}_3\,\hfill\mathrm{[GeV^{-2}]}$       & -9.9114    & -9.7755   & -9.6922       \\

\end{tabular}
\caption{Renormalization-scale independent low-energy parameters from our fits to the lattice data as explained in the text. 
Owing to the imposed large-$N_c$ sum rules there are  7 independent parameters only for 
any of the three fits.
}
\label{tab:FitParametersC}
\end{center}
\end{table}

Consider first Tab. \ref{tab:FitParametersB} where the low-energy parameters that enter the baryon masses at N$^2$LO are listed.  Conceptually the most crucial low-energy parameter 
are the chiral limit values of the baryon masses at $m=m_s = 0$. All of our three fits predict masses of around 875 MeV and 1238 MeV for octet and decuplet states respectively. Such values are somewhat 
larger than those obtained in the previous most comprehensive analysis \cite{Lutz:2014oxa}. This was to be expected since as the pion and kaon masses start to get smaller than $\Delta$ the approach 
used in \cite{Lutz:2014oxa} looses control and therefore the values of $M$ and $\Delta$ can not be trusted fully. Moreover, one should note that there is very little direct constraints on those 
limit values since lattice ensembles at small pion {\it and} kaon masses are not available so far.

It is interesting to compare the five $Q^2$ parameters in Tab. \ref{tab:FitParametersB} with the values collected in Tab. \ref{tab:FitParameters:N2LO}, where the latter were determined from a N$^2$LO 
analysis of the physical baryon masses only. We recall that the values from Fit 1 are subject to the 
large-$N_c$ relations as recalled with
\begin{eqnarray}
 d_D = 3\,b_D + 3\,b_F \,, \qquad \qquad d_0 = b_0 + b_D - b_F  \,.
 \label{rescall-b-Nc-B}
\end{eqnarray}
Only in Fit 3 all 5 parameters were adjusted in an independent manner. We would conclude that both tables present low-energy parameters 
reasonably close to each other. Also the spread of the low-energy parameters amongst the three different fits is comfortably small.  

In Tab. \ref{tab:FitParametersB} we also collected the axial coupling constants of the baryons. In all three fits they are subject to the large-$N_c$ constraint equations 
\begin{eqnarray}
C = 2\,D \,,\qquad \qquad H = 9\,F -3\,D\,.
\label{reacll-FDHC}
\end{eqnarray}
With our fits we explored the impact of variations around the values $F = 0.45$ and $D = 0.80$ used in our N$^2$LO studies. 
Again we find a small spread amongst the three different fits. Our values are within a reasonable range for $F$ and $D$ that are required
to reproduce the empirical axial-vector coupling constants of the baryon octet states \cite{Okun}.

We continue with the set of symmetry preserving low-energy parameters that drive the leading chiral correction in the two-body meson baryon interaction. 
They are collected in Tab. \ref{tab:FitParametersC}. For all three fit scenarios the 17 counter terms are parameterized by 7 operators only, which are detailed 
at length in Chapter 2. The reader may be reminded that the sum rules do depend on the values of the $b_0, b_D, b_F$ and $d_0, d_D$ counter terms. 
While Fit 1 is based on the large-$N_c$ scenario II of Chapter 2, Fit 2 and 3 on scenario III and IV respectively.  
It is interesting to confront our results to the leading order large-$N_c$ scenario I of Chapter 2.  At leading order an additional 4 sum rules appear
that take the form
\begin{eqnarray}
&&\bar  h_5^{(S)} = 3\,\bar g_1^{(S } + \bar g_D^{(S)}\,, \qquad\bar  h_6^{(S)} = -3\,\bar g_D^{(S)} -\frac{9}{2}\,\bar g_1^{(S)} \,,\qquad 
\nonumber\\
&& \bar  g_D^{(V)} = - \frac{3}{2}\,\bar  g_1^{(V)}\,, \qquad \quad \;\bar  h_3^{(V)} = \frac{3}{2}\,\bar g_1^{(V)}\,.
\label{recall-gh-sum}
\end{eqnarray}
We find it reassuring that all four relations are approximately realized by our numerical estimates displayed in Tab. \ref{tab:FitParametersC}. 
Using the values for $\bar g_1^{(S)}, \bar g_D^{(S)}$ and $\bar g_1^{(V)}$ we roughly recover the numerical values for $\bar h_5^{(S)}, \bar h_6^{(S)}$ and $ \bar g_D^{(V)}, \bar h_3^{(V)}  $.
In turn we may state that the overall pattern of the symmetry conserving two-body counter terms is governed by 3 degrees of freedom only. This is 
an amazing success showing the huge predictive power of large-$N_c$ QCD in hadron physics.

\begin{table}[t]
\setlength{\tabcolsep}{3.5mm}
\renewcommand{\arraystretch}{1.2}
\begin{center}
\begin{tabular}{l|cccc} 
                                                &  Fit 1     &  Fit 2      &  Fit 3      \\ \hline
$c_0\,\hfill\mathrm{[GeV^{-3}]}$                & -0.0776    & -0.0837     & -0.0725     \\ 
$c_1\,\hfill\mathrm{[GeV^{-3}]}$                & -0.1133    & -0.1236     & -0.1336        \\ 
$c_2\,\hfill\mathrm{[GeV^{-3}]}$                &  0.1229    &  0.1408     &  0.1470    \\     
$c_3\,\hfill\mathrm{[GeV^{-3}]}$                &  0.4464    &  0.4761     &  0.4216       \\     
$c_4\,\hfill\mathrm{[GeV^{-3}]}$                & -0.1818    & -0.2158     & -0.2312        \\     
$c_5\,\hfill\mathrm{[GeV^{-3}]}$                &  0.0929    &  0.0601     &  0.1037     \\     
$c_6\,\hfill\mathrm{[GeV^{-3}]}$                &  0.1414    &  0.1695     &  0.1507        \\  

$\zeta_0\,\hfill\mathrm{[GeV^{-2}]}$            &  0.0114    & -0.0018     &  0.0222       \\ 
$\zeta_D\,\hfill\mathrm{[GeV^{-2}]}$            &  0.1119    &  0.1139     &  0.1497       \\ 
$\zeta_F\,\hfill\mathrm{[GeV^{-2}]}$            & -0.1103    & -0.1076     & -0.1412     \\  \\  \hline

$e_0\,\hfill\mathrm{[GeV^{-3}]}$                & -0.6170    & -0.6386     & -0.5801       \\     
$e_1\,\hfill\mathrm{[GeV^{-3}]}$                &  0.2762    &  0.2055     &  0.1955     \\     
$e_2\,\hfill\mathrm{[GeV^{-3}]}$                &  1.0919    &  1.2743     &  1.1096     \\  
$e_3\,\hfill\mathrm{[GeV^{-3}]}$                & -0.2667    & -0.4670     & -0.3828         \\ 
$e_4\,\hfill\mathrm{[GeV^{-3}]}$                &  0.0827    &  0.1133     &  0.0487     \\ 

$\xi_0\,\hfill\mathrm{[GeV^{-2}]}$              &  0.2336    &  0.2197     &  0.3131       \\ 
$\xi_D\,\hfill\mathrm{[GeV^{-2}]}$              &  0.0046    &  0.0189     &  0.0256       \\

\end{tabular}
\caption{The renormalization-scale dependent parameters from a fit to the baryon octet and decuplet masses
at  $\mu = 770$ MeV.  Owing to the imposed large-$N_c$ sum rules there are  8 independent parameters that correlate the set of low-energy parameters $c_n$ and $e_n$ only for 
any of the three fits. }
\label{tab:FitParametersD}
\end{center}
\end{table}

We note that a meaningful comparison of Tab. \ref{tab:FitParametersC} with parameters from available coupled-channel studies like \cite{Lutz:2001yb}, is not possible in a straight forward manner, since at present such computations 
do not yet consider the baryon decuplet degrees of freedom explicitly.  Also a direct comparison with the results of \cite{Lutz:2014oxa} can only be expected at a qualitative level. 
It is not straight forward how to match the two distinct schemes in view of the extremely slow convergence of the $\Delta/M$ expansion. Nevertheless, it is assuring to see that 
similar patterns for the scalar and vector coupling constants emerge. Again the spread in the low-energy parameters as seen in Tab. \ref{tab:FitParametersC} amongst the three different fits is comfortably small. 

Finally we turn to Tab. \ref{tab:FitParametersD}, in which we collect all symmetry breaking low-energy constants that derive the quadratic quark mass terms in the baryon self energies. In all three fit scenarios the 
12 low-energy constants $c_n$ and $e_n$ are parameterized by eight operators only. Again it is instructive to consider the three additional constraints that arise at the leading order large-$N_c$ scenario I as derived in Chapter 2. 
We recall the three sum rules
\begin{eqnarray}
c_2 = - \frac{1}{2}\,e_1 \,,\qquad c_3 = \frac{1}{2}\,e_1 + \frac{1}{3}\,e_2  \,,\qquad c_4 = \frac{1}{3}\,e_1\,,
\label{recall-c-e}
\end{eqnarray}
and find that the numerical estimates collected in Tab.  \ref{tab:FitParametersD} are qualitatively in line with (\ref{recall-c-e}). Using the estimates for $e_1$ and $e_2$ one roughly recovers the 
values for $c_3$ but only the small magnitude of $c_2$ and $c_4$. Note that such a comparison is scale invariant only if the symmetry conserving counter terms 
in Tab. \ref{tab:FitParametersC} would follow the additional sum rules (\ref{recall-gh-sum}), which is not the case for any of the three fit scenarios. 
Nevertheless we may state that the overall pattern of the symmetry conserving two-body counter terms is governed by 5 degrees of freedom only. 

It is left to discuss the wave-function renormalization terms $\zeta_0, \zeta_D,\zeta_F$ and $\xi_0, \xi_D$ as summarized in Tab. \ref{tab:FitParametersD}. In all our three fit scenarios 
the two sum rules 
\begin{eqnarray}
\xi_D = 3\,\zeta_F + 3\,\zeta_D  \,, \qquad \qquad \xi_0 = \zeta_0 + \zeta_D - \zeta_F \,,
\label{rescall-zeta-xi-Nc}
\end{eqnarray}
are imposed. As compared to the previous study where the size of such counter terms were estimated from the baryon masses, we observed a significant reduction in the size of such terms. 
This is expected since in \cite{Lutz:2014oxa} the way such parameters are used was quite distinct as compared to the present  more reliable approach.  

We conclude that the spread in the low-energy parameters as advocated in  Tab. \ref{tab:FitParametersB}-\ref{tab:FitParametersD} amongst the three different fits is comfortably small. 
Given the number of about 300 described lattice data points the number of fit parameters is reasonably small and significant results are established that characterize important low-energy properties of QCD.

\begin{table}[t]
\setlength{\tabcolsep}{2.5mm}
\renewcommand{\arraystretch}{0.9}
\begin{center}
\begin{tabular}{l||c|c|rrr}\hline
                               &  \cite{Durr:2011mp}          &  \cite{Horsley:2011wr}         & Fit 1        &  Fit 2   &  Fit 3   \\ \hline \hline
$\sigma_{\pi N}$               & $39(4) ^{+ 18}_{- 7}  $      & $ 31 (3)(4)$                   & 48.81        & 48.38    & 47.56        \\
$\sigma_{\pi \Lambda}$         & $29(3)  ^{+11}_{-5} $        & 24(3)(4)                       & 52.40        & 58.49    & 57.12        \\
$\sigma_{\pi \Sigma}$          & $ 28(3) ^{+19}_{-3}$         & 21(3)(3)                       & 20.55        & 20.37    & 21.21      \\
$\sigma_{\pi  \Xi}$            & $16(2)^{+8}_{-3}  $          &  16(3)(4)                      & 20.02        & 23.79    & 22.87            \\
\\
$\sigma_{s N}$                 & $\;\;34(14)^{+28}_{-24}  $   & $ 71(34)(59)$                  & 42.59        & 52.46    & 22.91         \\
$\sigma_{s \Lambda}$           & $\;\;90(13)  ^{+24}_{-38} $  & 247(34)(69)                    & 452.26       & 441.84   & 435.88        \\
$\sigma_{s \Sigma}$            & $ 122(15) ^{+25}_{-36}$      & 336(34)(69)                    & 279.29       & 285.94   & 291.73         \\
$\sigma_{s  \Xi}$              & $156(16)^{+36}_{-38}  $      & 468(35)(59)                    & 640.47       & 638.52   & 657.13        \\
\hline
\\
                               & \cite{MartinCamalich:2010fp} &  \cite{Ren:2013oaa}            &  Fit 1       &  Fit 2   &  Fit 3   \\ \hline \hline
$\sigma_{\pi \Delta}$          & $55(4)(18) $                 & $28(1)(8) $                    & 45.16        & 40.21    & 40.98         \\
$\sigma_{\pi \Sigma^*}$        & $39(3)(13) $                 & $22(2)(9) $                    & 22.68        & 22.99    & 22.47         \\
$\sigma_{\pi \Xi^*}$           & $22(3)(7)  $                 & $11(2)(6)  $                   & -2.35        & 2.15     & 1.86         \\
$\sigma_{\pi  \Omega}$         & $\;\;5(2)(1) $               & $\;\;5(2)(2) $                 & 2.96         & 0.74     & 0.88        \\
\\                                                            
$\sigma_{s \Delta}$            & $\;\;56(24)(1) $             & $\;\;88(22)(3) $               & -313.38      &-285.82   & -307.68         \\
$\sigma_{s \Sigma^*}$          & $160(28)(7) $                & $243(24)(31) $                 & 93.85        & 67.06    &  69.87         \\
$\sigma_{s \Xi^*}$             & $ 274(32)(9)$                & $391(24)(67)$                  & 468.32       & 427.30   &  450.41       \\
$\sigma_{s  \Omega}$           & $360(34)(26)  $              & $528(26)(101)  $               & 125.13       & 171.31   &  151.58          \\ \hline
\end{tabular}
\caption{Pion- and strangeness sigma terms of the baryon octet and decuplet states in units of MeV. A comparison
with some previous works \cite{Durr:2011mp,Horsley:2011wr} and \cite{MartinCamalich:2010fp,Ren:2013oaa} is provided. }
\label{tab:sigmaterms}
\end{center}
\end{table}

\subsection{Sigma terms from a chiral extrapolation}
\vskip0.3cm

The pion-nucleon sigma term $\sigma_{\pi N}$ and the strangeness sigma term $\sigma_{sN}$ are defined as follows
 \begin{eqnarray}
&&\sigma_{\pi N} =  m\,\frac{\partial}{\partial m} m_N\,, \qquad \qquad 
 \sigma_{s N} =  m_s\,\frac{\partial}{\partial m_s} m_N\,.
\label{def-sigmapiN}
\end{eqnarray} 
From the knowledge of the quark-mass dependence of the baryon masses such sigma terms can readily be determined \cite{Durr:2011mp,Horsley:2011wr,Ren:2013dzt,MartinCamalich:2010fp,Ren:2013oaa}. 
On the other hand the size of the pion-nucleon sigma term can be extracted  from the pion-nucleon scattering data set. 
Recently the seminal value of $\sigma_{\pi N}= 45(8)$ MeV  established long ago by Gasser, Leutwyler and Sainio in  \cite{Gasser:1990ce} was questioned by an 
analysis of an updated data set using the Roy-Steiner equations in \cite{Hoferichter:2016ocj,RuizdeElvira:2017stg}. A significantly larger value of  
$\sigma_{\pi N}= 58(5)$ MeV was obtained. This new result triggered further studies from the lattice community which provided a series of direct computations of the 
sigma terms close to the physical point \cite{Durr:2015dna,Bali:2016lvx,Abdel-Rehim:2016won,Yang:2015uis}. All of them claim values 
that appear more consistent with the seminal result of Gasser, Leutwyler and Sainio in  \cite{Gasser:1990ce} rather than the {\it large} sigma term scenario of \cite{RuizdeElvira:2017stg}. 
It is noteworthy to recall also the previous analysis \cite{Alvarez-Ruso:2013fza} that obtained 
$\sigma_{\pi N} =52(3)(8)$ MeV based on a flavour SU(2) extrapolation of a large selection of lattice data for the nucleon mass \cite{Procura:2006bj}.

We consider our previous results on the sigma terms \cite{Semke:2012gs,Lutz:2014oxa} outdated. In \cite{Semke:2012gs} the finite volume effects were not considered yet and only a partial 
set of lattice data were considered. More critically is the fact that both works are not fully consistent in the chiral regime with $m_\pi < \Delta $. While this does not appear to prohibit 
a reproduction of the lattice data set, the extrapolation down to the physical pion mass does suffer from significant uncertainties as is illustrated by our current results for the sigma terms. 
It should be noted also, that as long as the systematic error in the baryon masses on the various lattice ensembles is not available, a full quantitative control on the final error budget of our results 
is basically impossible.

In Tab. \ref{tab:sigmaterms} we present our predictions for the pion- and strangeness sigma terms of the baryon octet and decuplet states based on our parameter sets Fit 1-3. 
The sigma terms of the baryon octet and decuplet states are defined by analogy with the definition of the sigma terms of the nucleon in (\ref{def-sigmapiN}).
Since a full estimate of the systematic uncertainties is beyond the scope of this work, we again illustrate uncertainties by providing sigma terms in three different fit scenarios.

Our values for the baryon octet states in Tab. \ref{tab:sigmaterms} are compared with two lattice determinations \cite{Durr:2011mp,Horsley:2011wr} that rely on a sizable extrapolation of the lattice data 
down to the physical quark masses. Such extrapolations are quite a challenge and may lead to significant uncertainties.  
Nevertheless, we note that our values for the non-strange sigma terms are in reasonable range of the lattice results. 
For the strangeness sigma terms those lattice studies come typically with a large error budget. Still,  
there appears to be a striking conflict amongst the values obtained by the BMW and QCDSF-UKQCD 
groups \cite{Durr:2011mp,Horsley:2011wr}. Our results are neither close to the values of any of the two groups. A more significant observation is that 
a recent direct evaluation of the $\chi$QCD group with $\sigma_{\pi N} = 45.9(7.4)(2.8) $ MeV and $\sigma_{s N} = 40.2(11.7)(3.5) $ MeV in  \cite{Yang:2015uis}
predicted values in the range of our results. Note that the size of the nucleon's strangeness content  is compatible also with the lattice average $\sigma_{sN}= 40^{+10}_{-10}$ MeV advocated 
previously in \cite{Junnarkar:2013ac}

The sigma terms for the baryon decuplet states are 
compared with two previous extrapolation studies \cite{MartinCamalich:2010fp,Ren:2013oaa}. While our results are in range of  
\cite{MartinCamalich:2010fp,Ren:2013oaa} for the pion sigma terms, this is clearly not the case for the strangeness sigma terms. 
In particular we find striking the large and negative values for the isobar we predict in our study. This hints at a strong and non-linear 
dependence of this state on the strange quark mass.

\begin{table}[t]
\setlength{\tabcolsep}{2.5mm}
\renewcommand{\arraystretch}{0.9}
\begin{center}
\begin{tabular}{lc|cc|r}
$\beta $   & $a_{\rm CLS}$ [fm]  &  $m_\pi$ [MeV]         &  $m_K$ [MeV]              & $N_s$      \\ \hline
3.40       & 0.08636(98)(40)           &  420                   & 420                       & 32         \\
           &                           &  350                   & 440                       & 32         \\
           &                           &  280                   & 460                       & 32         \\
           &                           &  220                   & 470                       & 48         \\ \\ \hline
3.55       & 0.06426(74)(17)           &  420                   & 420                       & 32         \\
           &                           &  280                   & 460                       & 48         \\
           &                           &  200                   & 480                       & 64         \\ \\ \hline 
3.70       & 0.04981(56)(10)           &  420                   & 420                       & 48         \\
           &                           &  260                   & 470                       & 64         \\

\end{tabular}
\caption{Pion and kaon masses on various CLS ensembles taken from \cite{Bruno:2016plf}.}
\label{tab:CLS}
\end{center}
\end{table}

\subsection{Predictions for baryon masses in CLS ensembles}
\vskip0.3cm

We consider 9 ensembles of the CLS collaboration at three different $\beta$  values \cite{Bruno:2016plf}. 
In Tab. \ref{tab:CLS} the characteristics thereof are recalled. Given our parameter sets we compute all baryon masses in the finite box 
as specified by $N_s$ and the lattice scale estimates of \cite{Bruno:2016plf}.

\begin{figure}[t]
\center{
\includegraphics[keepaspectratio,width=0.8\textwidth]{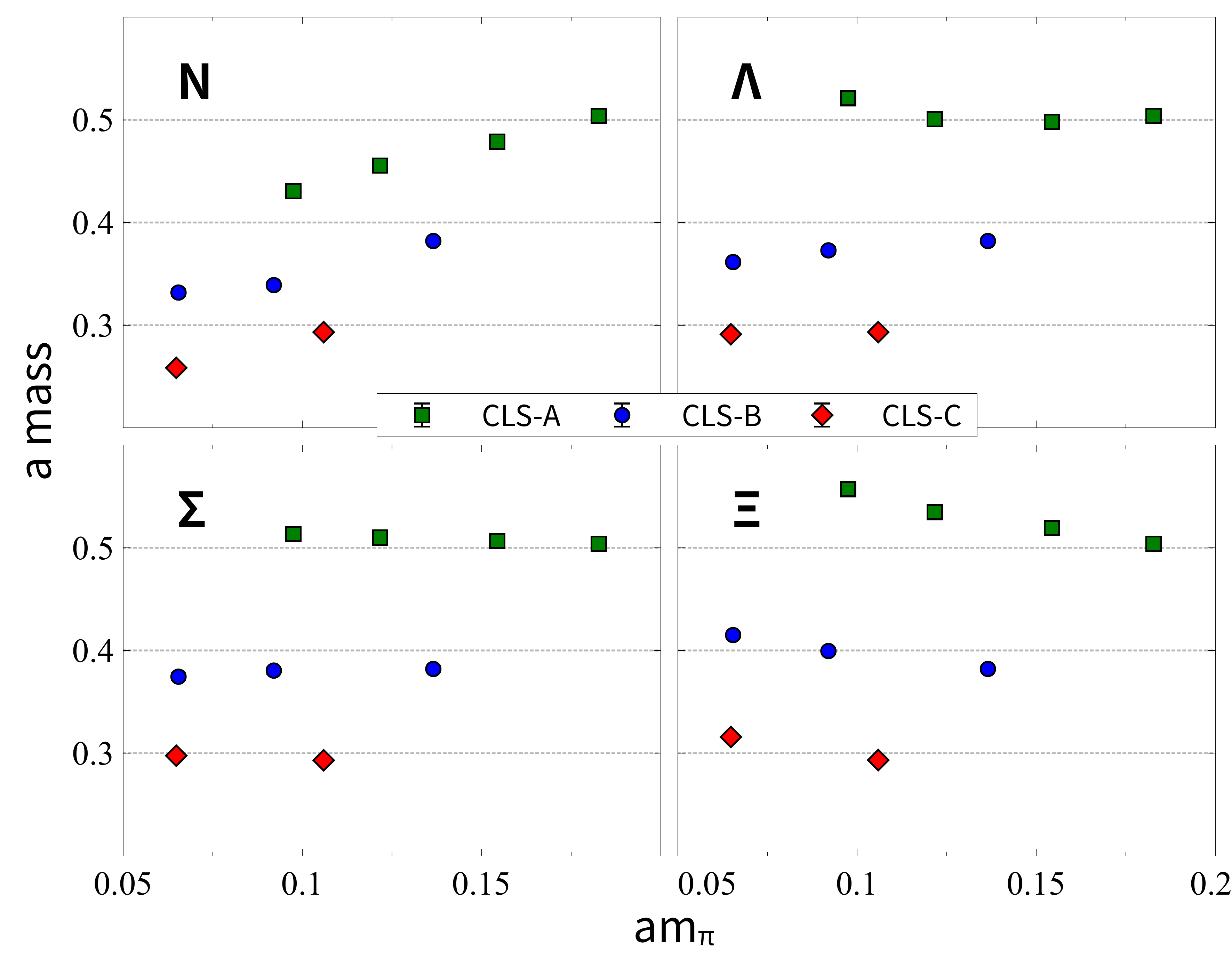} 
\includegraphics[keepaspectratio,width=0.8\textwidth]{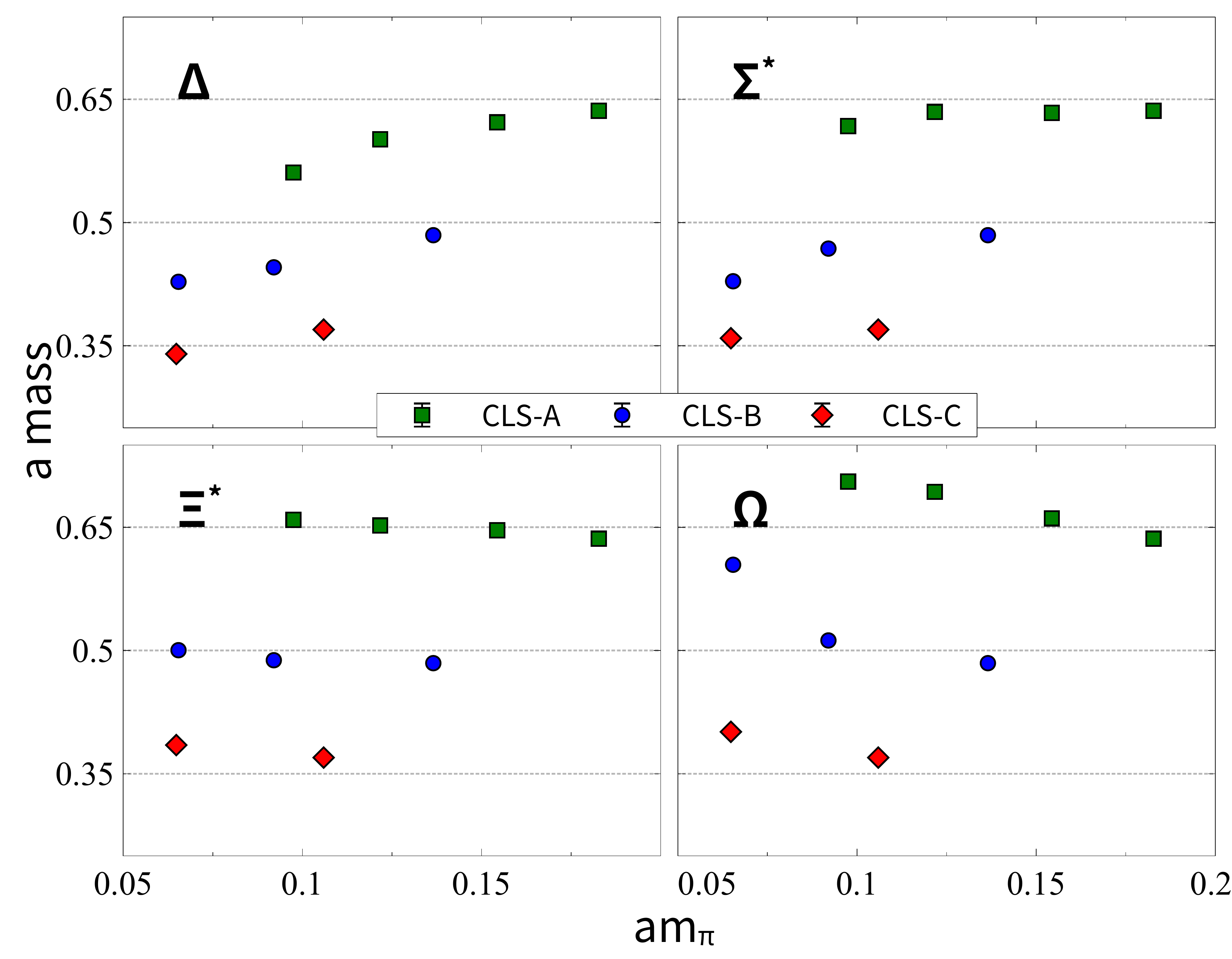} }
\vskip-0.2cm
\caption{\label{fig:lattice-9} Baryon masses from Fit 3  in the CLS ensembles of Tab. \ref{tab:CLS}.  }
\end{figure}

Our results are presented with Fig. \ref{fig:lattice-9}. The baryon masses are shown based the scenario Fit 3. In order to minimize any possible 
uncertainty in the lattice scale determination of \cite{Bruno:2016plf}, we show the baryon masses in units of 
the lattice scale. Here we assume the central value of the lattice scales as recalled in Tab. \ref{tab:CLS}. 
As compared to the Fit 1 and Fit 2 no significant deviations from the  masses of Fit 3 are observed in the octet masses. For the decuplet masses our solutions, in particular Fit 1
predicts baryon masses off by at most 5$\%$ from the masses of the Fit 3 scenario. We conclude that 
we obtained stable and significant results for the baryon masses on the CLS ensembles which await a critical evaluation of the CLS collaboration. 
The results of CLS may be of  importance to unravel possible further systematic uncertainties in the analyzed QCD lattice data set. 

Once the CLS collaboration make their results on the baryon masses available we will include their results in our global fit and study the impact of such data 
on the low-energy constants of QCD.

\clearpage

\section{Summary and outlook}
\label{sec:summary}

We reconsidered the chiral extrapolation of baryon masses based on the three-flavour chiral Lagrangian 
formulated with the baryon octet and decuplet fields. The main achievements of our work  are summarized:

\begin{itemize}
\item[-]
At N$^3$LO the number of relevant low-energy parameters is quite large. A significant parameter reduction was 
obtained by the application of large-$N_c$ sum rules. We reviewed all sum rules that are relevant for the chiral extrapolation of 
the baryon masses. So far unknown subleading terms in the $1/N_c$ expansion were established for the first time.
\item[-]
A subtraction scheme for the one-loop contributions to the baryon self energies was devised that ensures consistent results 
in a chiral expansion with either $m_Q \sim M_\Delta -M_N$ or $m_Q \ll M_\Delta -M_N$ with $m_Q$ denoting a meson mass. 
The significant role played by effects from the baryon wave-function renormalization was emphasized. It was argued that such effects 
should be considered already at the N$^3$LO level.
\item[-]
A reordering of terms in the chiral expansion of the baryon masses was suggested that makes the convergence properties more effective.
The various moments are expressed in terms of physical meson and baryon masses. 
It was illustrated that the one-loop contributions to the baryon self energy  if expanded in powers 
of $x=m_Q/M_B$ with $M_B$ and $m_Q$ a physical baryon and meson mass respectively is rapidly converging. 
Based on the analytic properties of the loop functions such an expansion was proven to converge up to  $x=m_Q/M_B < 2$. Flavour breaking effects are considered by an expansion 
of $d=M_R/M_B -1$ around its chiral limit value $d_0$, where two different types of baryons $B,R$ are involved. The convergence condition 
$|d-d_0|< |x+d_0|$ was derived and shown to be satisfied by the physical meson and baryon masses in the chiral expansion of the baryon octet masses. 
For the baryon decuplet states such an expansion is not convergent and a further partial summation scheme is required. 
Explicit expressions up to chiral order five were derived for all baryon masses at the one-loop level. The novel chiral ordering was successfully  
tested for physical meson and baryon masses. This suggests a rather large PCD which includes the physical masses of the up, down and strange quarks.
\item[-]
Results are obtained that are invariant under changes of the renormalization scale. Given the need to formulate the chiral expansion in terms of 
physical meson and baryon masses this is achieved with a particular rewrite of the contributions form the counter terms using physical masses. 
\item[-]
The set of low-energy parameters was adjusted to QCD lattice data at N$^3$LO, where the low-energy parameters are systematically correlated 
by large-$N_c$ sum rules. Results from PACS-CS, LHPC, HSC, NPLQCD, QCDSF-UKQCD and ETMC are considered. We observe  considerable tension between the PACS-CS and the LHPC data set. 
The most natural resolution requires 
a reduction of the LHPC weight in our global fit indicating significant discretization effects. Predictions for baryon masses on ensembles from CLS as well
as all low-energy constants that enter the baryon masses at N$^3$LO are made. For the nucleon we then obtain a sigma 
term of $\sigma_{\pi N} = 48(1)$ MeV and a  strangeness content of $\sigma_{sN} = 38(15)$ MeV. 
\end{itemize}

To further substantiate the claimed chiral low-energy parameters it is necessary to take further data on a QCD lattice in particular 
at unphysical quark masses. In order to consolidate the PCD the chiral extrapolation formulae should be extended to the two-loop level, 
where the effects of the baryon decuplet degrees of freedom should be considered explicitly. The low-energy parameters determined in our study 
should be scrutinized in future coupled-channel computations of low-energy meson baryon scattering that consider the baryon octet and decuplet fields 
simultaneously.

\section*{Acknowledgments}

M.F.M. Lutz thanks Kilian Schwarz and Jan Knedlik for significant support on distributed computing issues. 
Particular thanks go to Walter Sch\"on who is operating the HPC cluster at GSI with his department in an outstanding manner. Without the ongoing 
and constructive interactions with his group this work would not have been possible. 
We are grateful to R\"udiger Berlich of Gemfony scientific UG for help with their optimization library Geneva. 
John Bulava and Ulrich Sauerwein are acknowledged for critically reading the manuscript.
\newpage

\section*{Appendix A}

We collect all kinematic coefficients that enter the chiral decomposition of the one-loop contributions to the baryon octet self 
energies. All such terms depend on the ratio $\Delta/M$ only.
\begin{eqnarray}
&&\alpha_1 = \frac{(2\,M+\Delta)^4}{16\,M^2\,(M+\Delta)^2}\,, \qquad \alpha_3 =\frac{8\,M\,(M +\Delta) +3\,\Delta^2 }{8\,(M+\Delta)^2}\,,
\nonumber\\
&& \alpha_2  =\big(4\,M\,( M+\Delta) +3\,\Delta^2 \big)\,\frac{(2\,M+\Delta)^2}{16\,M^2\,(M+\Delta)^2}\,,\qquad  
 \alpha_4 = \frac{(2\,M+\Delta )^3}{8\,M\,(M+\Delta)^2}\,,
\nonumber\\
&& \alpha_5 = \frac{4\,M^2+\Delta\,M -\Delta^2}{4\,(M+\Delta)^2}\,,\qquad 
 \alpha_6 = \frac{M^2}{(M+\Delta)^2}\,,
\nonumber\\ \nonumber\\ \nonumber\\
&& \gamma_1 = \frac{2\,M+ \Delta}{M}\,\log \frac{\Delta\,(2\,M + \Delta)}{(M+ \Delta)^2} \,, \qquad \qquad 
\nonumber\\
&&  \gamma_2 = - \frac{2\,M^2+ 2\,\Delta\,M+ \Delta^2}{M\,(2\,M+ \Delta)}\,\log \frac{\Delta\,(2\,M + \Delta)}{(M+\Delta)^2} 
-\frac{M}{2\,M+\Delta} \,,
\nonumber\\
&&   \gamma_3 = \frac{M}{2\,M+ \Delta}\,,\qquad \qquad \gamma_5 = \frac{M\,(M+ \Delta)^2}{(2\,M+ \Delta)^3}\,,
\nonumber\\
&&  \gamma_4 = \frac{M^3}{2\,(2\,M+ \Delta)^3} -2\,\frac{M\,(M+ \Delta)^2}{(2\,M+ \Delta)^3}\,\log \frac{\Delta\,(2\,M + \Delta)}{(M+\Delta)^2}\,,
\nonumber\\ \nonumber\\ \nonumber\\
&& \tilde \gamma_1 = \gamma_1- \frac{2\,M+ \Delta}{M}\,\log \frac{2\,\Delta}{M+\Delta} \,,
\nonumber\\
&& \tilde \gamma_2 = \gamma_2 + \frac{2\,M^2 + 2\,\Delta\,M + \Delta^2}{(2\,M+\Delta)\,M}\,\log \frac{2\,\Delta}{M} + \frac{2\,M + \Delta}{4\,M}
\nonumber\\
&& \qquad \qquad+\,2\,\big(\tilde \gamma_3 -\gamma_3\big)\, \log \frac{M+\Delta}{M}\,,
\nonumber\\
&& \tilde \gamma_3 = \gamma_3  - \frac{2\,M^2 + 2\,\Delta\,M + \Delta^2}{2\,M\,(2\,M + \Delta)}\,, \qquad \qquad 
\tilde \gamma_5 = 0\,,
\nonumber\\
&& \tilde \gamma_4 = \gamma_4 + \frac{2\,M\,(M + \Delta)^2}{(2\,M+\Delta)^3}\,\log\frac{2\,\Delta}{M}
\nonumber\\
&& \qquad-\, \frac{4\,M^2 + \Delta\,(4\,M + 5\,\Delta)}{32\,M\,(2\,M + \Delta)}+ 2\,\big(\tilde \gamma_5 -\gamma_5\big)\, \log \frac{M+\Delta}{M}\,,
\nonumber\\ \nonumber\\
&& \tilde \alpha_1 = \Delta \,\frac{ \partial }{\partial \Delta} \, \alpha_1 \,\frac{2\,M+ \Delta}{2\,M}  \,,
\nonumber\\
&& \tilde \alpha_2 = \Delta^2\,\frac{\partial }{\partial \Delta} \, \frac{\alpha_1 \,\tilde \gamma_2 }{\Delta}\,,\qquad  \qquad\;\,
\tilde \alpha_3 =  \Delta^2\,\frac{\partial }{\partial \Delta} \,\frac{\alpha_1 \,\tilde \gamma_3 }{\Delta}\, ,
\nonumber\\
&& \tilde \alpha_4 =\Delta\,\gamma_1\,\frac{\partial}{\partial \Delta}\, \alpha_1\,, \qquad \qquad \quad \tilde \alpha_5 = \Delta\,\frac{\partial}{\partial \Delta}\, \alpha_1\,\tilde \gamma_1\,,
\nonumber\\
&& \tilde \alpha_6 = \frac{\Delta^2\, \partial^2 }{\partial \Delta \,\partial \Delta} \, 
\Bigg(\alpha_1 \,\frac{2\,M+ \Delta}{2\,M} \Bigg) \,, \qquad \qquad  \tilde \alpha_7 =  \Delta\,\frac{\Delta^2\,\partial^2 }{\partial \Delta\,\partial \Delta} \, \frac{\alpha_1 \,\tilde \gamma_2 }{\Delta}\,,
\nonumber\\
&& \tilde \alpha_8 =  \Delta\,\frac{\Delta^2\,\partial^2 }{\partial \Delta \,\partial \Delta} \,\frac{\alpha_1 \,\tilde \gamma_3 }{\Delta}\, ,\qquad \qquad \qquad \qquad 
\nonumber\\ \nonumber\\
&& \tilde \alpha_9 =  \gamma_1  \,\frac{\Delta^2\,\partial ^2}{\partial \Delta \,\partial \Delta}\,\alpha_1\,, \qquad \qquad
 \tilde \alpha_{10} =  \frac{\Delta^2\,\partial^2}{\partial \Delta \,\partial \Delta}\,\alpha_1\,\tilde \gamma_1\,,
\nonumber\\
&& \tilde \alpha_{11} = -\frac{1}{4}\,\alpha_1\,\frac{M}{2\,M+\Delta} + \Big(\alpha_1-\alpha_2 \Big)\,\frac{(2\,M+ \Delta)\,M}{2\,\Delta^2}\,,
\end{eqnarray}
While the $\alpha_i$ characterize the chiral expansion of the coefficients in front of $\bar I_{QR}$ and $\bar I_Q$ in  (\ref{result-loop-8}), the 
$\gamma_i$ follow from a chiral expansion of $\bar I_{QR}$ with $M_B = M$ and $M_R = M+ \Delta$ and $m_Q < \Delta$. The $\tilde \gamma_n$ follow 
from the corresponding $\gamma_n$ after subtractions from a chiral expansion of the anomalous term $f^{(d)}_0(x^2)\,f^{(d)}_1(x^2)$ in (\ref{IQR-x-delta}). 
Finally the coefficients $\tilde \alpha_n$ characterize the central results (\ref{loop-HB-4}) and \ref{loop-HB-5}).

The symmetry breaking term in (\ref{def-tadpole}) takes the form

\allowdisplaybreaks[1]
\begin{eqnarray}
\Sigma^{(4-\chi)}_N \!\!\!\!&=&\!\!\! - 4\,B_0 \left( b^{\rm eff}_0\, (2\,m+m_s) + b^{\rm eff}_D\, (m+m_s) + b^{\rm eff}_F\, (m-m_s) \right)
\nonumber \\
\!\!\!\!&\phantom =& \!\!\!-4\,B_0^2\, \Big( c_0\, (2\,m^2 + m_s^2) + c_2\, (m^2+m_s^2) + c_3\, (m^2-m_s^2) \Big)
\nonumber \\
\!\!\!\!&\phantom =& \!\!\! - 2\,B_0\, \Big( \zeta_0\, (2\,m+m_s) + \zeta_D\, (m+m_s)
+ \zeta_F\, (m-m_s) \Big) \Big(M_N - M\Big) \,,
\nonumber \\
\Sigma^{(4-\chi)}_\Lambda \!\!\!\!&=&\!\!\! - 4\,B_0\, \Big(b^{\rm eff}_0\, (2\,m+m_s) + \frac{2}{3}\,b^{\rm eff}_D\, (m+2\,m_s) \Big)
\nonumber \\
\!\!\!\!&\phantom =& \!\!\!- 4\,B_0^2\, \Big( c_0\, (2\,m^2 + m_s^2) + \frac 23 \,c_1\, (m - m_s)^2  + \frac 23 \,c_2\, (m^2 + 2\,m_s^2) \Big)
\nonumber \\
\!\!\!\!&\phantom =& \!\!\!- 2\,B_0\, \Big(\zeta_0\, (2\,m+m_s) + \frac{2}{3}\,\zeta_D\, (m+2\,m_s) \Big)  \Big(M_\Lambda - M\Big)\,,
\nonumber \\
\Sigma^{(4-\chi)}_\Sigma\!\!\!\!&=&\!\!\! - 4\,B_0\, \Big(b^{\rm eff}_0\, (2\,m+m_s) + 2\,b^{\rm eff}_D\,m \Big)
\nonumber \\
\!\!\!\!&\phantom =& \!\!\! - 4\,B_0^2\, \Big(c_0\, (2\,m^2+m_s^2) + 2\,c_2\, m^2 \Big)
\nonumber \\
\!\!\!\!&\phantom =& \!\!\!- 2\,B_0\, \Big(\zeta_0\, (2\,m+m_s) + 2\,\zeta_D\,m
\Big)  \Big(M_\Sigma - M\Big) \,,
\nonumber \\
\Sigma^{(4-\chi)}_\Xi \!\!\!\!&=&\!\!\! - 4\,B_0\, \Big( b^{\rm eff}_0\, (2\,m+m_s)  + b^{\rm eff}_D\, (m+m_s)  - b^{\rm eff}_F\, (m - m_s) \Big)
\nonumber\\
\!\!\!\!&\phantom =& \!\!\!- 4\,B_0^2\left( c_0\, (2\,m^2+m_s^2) + c_2\,(m^2 + m_s^2) - c_3\,(m^2 - m_s^2) \right)
\nonumber\\
\!\!\!\!&\phantom =& \!\!\!- 2\,B_0\, \Big( \zeta_0\, (2\,m+m_s)  + \zeta_D\,
(m+m_s)  - \zeta_F\, (m - m_s) \Big)  \Big(M_\Xi - M\Big) \,,
\nonumber\\ \nonumber\\
&& \!\!\!\!\! \!\!\! \!\!\!\!\! \!\!\! b^{\rm eff}_0 = c_6\, B_0\,(2\,m+m_s)\,, \qquad \qquad 
 b^{\rm eff}_D = c_4\, B_0\,(2\,m+m_s)\,, \qquad   
\nonumber\\
&&  \!\!\!\!\! \!\!\! \!\!\!\!\! \!\!\! b^{\rm eff}_F = c_5\, B_0\,(2\,m+m_s)\,,
\label{result-counter-terms-octet}
\end{eqnarray}
with
 the low-energy parameters  $c_i$. The renormalization scale dependence 
of the $c_i$ as implied by (\ref{def-tadpole}) is
\begin{eqnarray}
&& \mu^2\,\frac{d }{d \,\mu^2} \,c_i = -\frac{1}{4}\,\frac{ \Gamma_{c_i}}{(4\,\pi\,f )^2}\,,
\end{eqnarray}
 with
\begin{eqnarray}
&& \Gamma_{c_0} =   \frac{20}{3}\,b_0+ 4\,b_D  -
\frac{1}{36}\, \Big( 30 \,\bar g_0^{(S)}+9 \,\bar g_1^{(S)}+26 \,\bar g_D^{(S)}\Big) 
\nonumber\\
&& \qquad \qquad -\, \frac{M}{144}\, \Big(30\,
   \bar g_0^{(V)}+9 \,\bar g_1^{(V)}+26 \,\bar g_D^{(V)}\Big) \,,
\nonumber\\
&& \Gamma_{c_1} = -
\frac{1}{24} \,\Big(4 \,\bar g_1^{(S)}+\bar g_1^{(V)}\, M \Big)\,,
\nonumber\\
&& \Gamma_{c_2} =  \frac{2}{3}\,b_D + \frac{1}{16} \,\Big(4\, (\bar g_1^{(S)}+\bar g_D^{(S)})+M\, (\bar g_1^{(V)}+\bar g_D^{(V)})\Big) \,,
\nonumber\\
&& \Gamma_{c_3} = \frac{2}{3}\,b_F  +\frac{1}{16} \,\Big(4\, \bar g_F^{(S)}+\bar g_F^{(V)}\, M\Big ) \,,
\nonumber\\
&& \Gamma_{c_4} = \frac{44}{9}\,b_D-\frac{1}{72} \,\Big( 36\, \bar g_1^{(S)}+52 \,\bar g_D^{(S)} + M\,( 9\, \bar g_1^{(V)} + 13\, \bar g_D^{(V)})
  \Big)\,,
\nonumber\\
&& \Gamma_{c_5} = \frac{44}{9}\,b_F-\frac{13}{72} \,\Big(4\, \bar g_F^{(S)}+\bar g_F^{(V)}\, M\Big) \,,
\nonumber\\
&& \Gamma_{c_6} = \frac{44}{9}\,b_0+ \frac{1}{432}\,\Big(-264 \,\bar g_0^{(S)}+108\, \bar g_1^{(S)} +32\, \bar g_D^{(S)} 
\nonumber\\
&& \qquad \qquad 
+ \, M\,\big(-66 \,\bar g_0^{(V)}+27\, \bar g_1^{(V)}  +8\, \bar g_D^{(V)} \big)\Big)\,.
\label{res-Gamma-ci}
\end{eqnarray}
Large-$N_c$ sum rules for the low-energy parameters $c_n, e_n$ are detailed in (\ref{ces-subleading}). Corresponding relations for the $\Gamma_{c_n}, \Gamma_{e_n}$ follow 
from (\ref{ces-subleading}) by the substitution $c_n\to \Gamma_{c_n} $ and $e_n\to \Gamma_{e_n}$.

We derive the form of the anomalous scaling term, $c^{\rm ano}_i$, introduced in (\ref{def-c-e-ano}, \ref{c-e-running}). The prescription (\ref{eliminate-mu}) applied to the 
$(M_R -M_B)\,\bar I_Q$ terms in (\ref{result-loop-8}) implies a renormalization of the 
symmetry breaking counter terms
\begin{eqnarray}
&& c^{\rm ano}_i =   \frac{1}{4}\,\frac{1}{(4\,\pi\,f )^2}\, \Gamma^{(1)}_{c_i}\,\gamma^{(1)}_c
+ \frac{1}{4}\,\frac{C^2}{(4\,\pi\,f )^2}\, \Big( \Gamma^{(2)}_{c_i}\gamma^{(2)}_c + \Gamma^{(3)}_{c_i}\gamma^{(3)}_c \Big) \,,
\qquad 
\nonumber\\
&&  \gamma^{(1)}_c = - 4 - \log \frac{M^2}{\mu^2} \,, \qquad \qquad 
 \gamma^{(2)}_c =  -\frac{2}{3}\left(\frac{\partial \Delta\, \alpha_4}{\partial \Delta}\right) \log\frac{\mu^2}{(M+\Delta)^2} \,,
\nonumber\\
&& \gamma^{(3)}_c = -\,\frac{2}{3}\left(\frac{\partial \Delta \,\alpha_4}{\partial M}\right) \log \frac{\mu^2}{(M+\Delta)^2}
- \gamma^{(2)}_c\,.
\nonumber\\
&& \mu^2\,\frac{d }{d \,\mu^2} \,c^{\rm ano}_i = \frac{1}{4}\,\frac{-1}{(4\,\pi\,f )^2}\, \Gamma^{(1)}_{c_i}   
+\frac{1}{6}\,\frac{C^2}{(4\,\pi\,f )^2}\,\Bigg\{  \Gamma^{(2)}_{c_i} \,\Big[ 1+ \Delta\,\frac{\partial}{\partial \Delta} \Big]
\nonumber\\
&& \qquad \qquad \; -\,  \Gamma^{(3)}_{c_i} \,\Big[ 1 + \Big(1+\frac{\Delta}{M} \Big)\,\Delta\,\frac{\partial}{\partial \Delta }\Big] 
\Bigg\}\,\alpha_4\,,  
\label{res-c-ano}
\end{eqnarray}
with
\allowdisplaybreaks[1]
\begin{eqnarray}
&& \Gamma^{(1)}_{c_0} = 
- \frac{4}{3} \, b_D \left(D^2+3 \,F^2\right)-8\, b_F\, D\, F \,, \qquad 
 \Gamma^{(1)}_{c_1} =  -\frac{64}{9} \,b_D \,D^2\,,
\nonumber\\
&& \Gamma^{(1)}_{c_2} =  \frac{2}{3} \, b_D \left(5 \,D^2+9 \,F^2\right)+12\, b_F\, D\, F  \,, \qquad 
\nonumber\\
&& \Gamma^{(1)}_{c_3} = \frac{20}{3} \,b_D \,D \,F +\frac{2}{3}\, b_F \left(5\, D^2+9 \,F^2\right)\,, \qquad 
 \Gamma^{(1)}_{c_4} =  \frac{32}{9} \,b_D \,D^2\,, \qquad 
\nonumber\\
&& \Gamma^{(1)}_{c_5} = 0 \,, \qquad 
 \Gamma^{(1)}_{c_6} = -\frac{8}{3}  \,b_D \,D^2\,,
 \nonumber\\ \nonumber\\
&& \Gamma^{(2)}_{c_0} = - \frac{11}{9}\,d_D \,, \qquad 
\Gamma^{(2)}_{c_1} = -\frac{10}{9}\,d_D\,,  \qquad 
 \Gamma^{(2)}_{c_2} =  \frac{8}{3}\,d_D\,, \qquad  
\nonumber\\
&&\Gamma^{(2)}_{c_3} = -\frac{2}{3}\,d_D\,,\qquad   \Gamma^{(2)}_{c_4} = -\frac{2}{9}\,\Big(  9\,d_0 + 8\,d_D\Big)\,,  \quad 
\nonumber\\
&&  \Gamma^{(2)}_{c_5} = \frac{1}{9}\,\Big( 15\,d_0 + 19\,d_D\Big)\,, \qquad  \Gamma^{(2)}_{c_6} = \frac{1}{9}\,\Big( 42\,d_0 + 19\,d_D\Big)\,,
\nonumber\\ \nonumber\\
&& \Gamma^{(3)}_{c_0} =-2\,b_D - \frac{5}{3}\,b_F  \,, \qquad 
\Gamma^{(3)}_{c_1} = -\frac{2}{9}\,\Big(6\,b_D + 15\,b_F \Big)\,,  \qquad
\nonumber\\
&& \Gamma^{(3)}_{c_2} = 2\,b_D + 5\,b_F \,,  \qquad \Gamma^{(3)}_{c_3} = \frac{1}{3}\,\Big(5\,b_D - 6\,b_F \Big)\,, \qquad
\nonumber\\
&&\Gamma^{(3)}_{c_4} = -\frac{2}{9}\,\Big(9\,b_0 - 3\,b_D + 15\,b_F \Big)\,, \qquad  \Gamma^{(3)}_{c_5} = \frac{1}{9}\,\Big(15\,b_0 + 42\,b_F \Big)\,,\qquad 
\nonumber\\
&&  \Gamma^{(3)}_{c_6} = \frac{1}{9}\,\Big(42\,b_0 + 18\,b_D + 15\,b_F \Big)\,.
\end{eqnarray}
We close this Appendix with explicit expressions of the fifth moment of the octet self energy. With  $m_{QR}^2 = m_Q^2-(M_R-M_B)^2$ we find
\allowdisplaybreaks[1]
\begin{eqnarray}
&&\bar \Sigma^{{\rm bubble}-5}_{B \in [8]} =  \!\! \sum_{Q\in [8], R\in [8]}
\left(\frac{m_{QR}}{4\,\pi\,f}\,G_{QR}^{(B)} \right)^2 \Bigg\{ \frac{\pi}{16}\,\frac{m^3_Q}{M_B^2}   
-   \frac{m^4_Q}{24\,M^3_B} 
 \nonumber\\
&& \qquad  \quad +\,\Big(M_R-M_B\Big)^2\,\Bigg[  \frac{\pi}{4\,m_Q} + \frac{1}{2\,M_B}\,\Big( 2 + 3\,\log \frac{m_Q}{M_R} \Big)\Bigg] \Bigg\}
\nonumber\\
&& \; +\sum_{Q\in [8], R\in [10]}
\left(\frac{m_Q}{4\,\pi\,f}\,G_{QR}^{(B)} \right)^2 \, \Bigg\{ \frac{\tilde \alpha_{11}}{3}\,\frac{\Delta^3}{M^2}\,\log \frac{4\,\Delta^2}{(M+\Delta)^2}
   \nonumber\\
&& \qquad \quad + \, \frac{1}{3}\,  \Big( \alpha_1- \alpha_2\Big )\,\Bigg( \frac{2\,M+ \Delta}{2\,M}\,\frac{M_R-M_B}{\Delta_B^2}\, m_Q^2\,\log \frac{m_Q^2}{M_R^2}
 \nonumber\\
&& \qquad \qquad \quad  +\,\big(\gamma_1-\tilde \gamma_1 \big)\,\frac{\Delta_Q^2-\Delta\,\Delta_B }{\Delta_B}-  \tilde \gamma_1\, \frac{\Delta_Q^2}{\Delta_B^2}\,\Big( M_R-M_B-\Delta_B\Big) \Bigg)
\nonumber\\
&& \qquad  \quad -\,\frac{\tilde \alpha_{11}}{3\,M_B^2}\,\Bigg[  \big(M_R-M_B \big)^3\,\log \frac{m_Q^2}{M_R^2} 
+ \Delta_Q^3\,\Big( \log \big( M_R-M_B +  \Delta_Q \big) 
\nonumber\\
&& \qquad  \qquad \quad -\, \log \big(M_R-M_B -  \Delta_Q\big) \Big)\Bigg]
\nonumber\\
&& \qquad  \quad +\,\frac{m_Q^2\, \Delta_Q^2}{3\,\Delta^3_B} \,\Bigg[\Big(\alpha_2-\alpha_1 \Big)\, \Big(  \tilde \gamma_2 +\tilde \gamma_3\,\log \frac{m_Q^2}{M_R^2} \Big)
-  \alpha_1\,\Big( \tilde  \gamma_4 + \tilde \gamma_5\,\log \frac{m_Q^2}{M_R^2} \Big)  \Bigg]
 \nonumber\\
&& \qquad \quad + \, \Big(M_R - M_B-\Delta_B \Big)^2\Bigg(\frac{\tilde \alpha_9}{6}\,\frac{\Delta_Q^2}{m_Q^2\,\Delta_B}
 -\frac{\tilde \alpha_{10}}{6}\,\frac{\Delta_Q^2}{m_Q^2\,\Delta^2_B}\,\big(M_R-M_B \big)
\nonumber\\
 && \qquad \qquad \quad 
- \frac{\tilde \alpha_6}{6\,\Delta_B^2\,m_Q^2}\,\Bigg[ 
\Big( \Delta_Q^2-\frac{1}{2}\,m_Q^2\Big)\,\big(M_R-M_B\big)\,\log \frac{m_Q^2}{M_R^2} 
  \nonumber\\
&& \qquad \qquad \quad
 +\, \Delta_Q^3\,\Big( \log \big(M_R-M_B + \Delta_Q \big)-\log \big(M_R-M_B - \Delta_Q\big)  \Big)
\Bigg]  
\nonumber\\
&& \qquad\qquad \quad    +\, \frac{1}{6\,\Delta^3_B}\,\Big(-\tilde \alpha_7\,\Delta_Q^2 + \tilde \alpha_8\,m_Q^2\,\log \frac{m_Q^2}{M_R^2}\Big) 
\Bigg)
 \Bigg\}\,.
\label{loop-HB-5} 
\end{eqnarray}

\newpage

\section*{Appendix B}

We collect all kinematic coefficients that enter the chiral decomposition of the one-loop contributions to the baryon decuplet self 
energies. All such terms depend on the ratio $\Delta/M$ only.
\begin{eqnarray}
&&\beta_1 = \frac{(2\,M+\Delta)^4}{16\,M\,(M+\Delta)^3}\,, \qquad \qquad \beta_3 =M\,\frac{8\,M\,(M +\Delta) +3\,\Delta^2 }{8\,(M+\Delta)^3}\,,
\nonumber\\
&& \beta_2  =\big(4\,M\,( M+\Delta) +3\,\Delta^2 \big)\,\frac{(2\,M+\Delta)^2}{16\,M\,(M+\Delta)^3}\,, \qquad 
 \beta_4 = \frac{(2\,M+\Delta )^3}{8\,(M+\Delta)^3}\,,
\nonumber\\
&& \beta_5 = \frac{4\,M^3+ 5\,\Delta\,M^2 +2\,\Delta^2\,M }{4\,(M+\Delta)^3}\,, \qquad 
 \beta_6 = \frac{M}{M+\Delta}\,,
\nonumber\\ \nonumber\\ \nonumber\\
&& \delta_1 = -\frac{M\,(2\,M+ \Delta)}{(M+\Delta)^2}\,\log \frac{\Delta\,(2\,M+\Delta)}{M^2}\,, \qquad  
\nonumber\\
&& \delta_2 = \frac{M}{2\,M+ \Delta } +M\,\frac{2\,M^2 + 2\,\Delta\,M+\Delta^2}{(2\,M + \Delta )\,(M+ \Delta )^2}\,
\log \frac{\Delta\,(2\,M+ \Delta)}{M^2}  \,,
\nonumber\\
&&  \delta_3 = - \frac{M}{2\,M+\Delta } \,, \qquad \qquad \qquad \delta_5 = -\frac{M^3}{(2\,M+\Delta )^3}\,,
\nonumber\\
&& \delta_4 =- \frac{M\,(M+\Delta)^2}{2\,(2\,M+ \Delta)^3 } 
+ \frac{2\,M^3 }{(2\,M + \Delta )^3}\,\log \frac{\Delta\,(2\,M+ \Delta)}{M^2}  \,,
\nonumber\\ \nonumber\\ \nonumber\\
&& \tilde \delta_1 = \delta_1 +\frac{M\,(2\,M+ \Delta)}{(M+\Delta)^2}\, \log \frac{2\,\Delta}{M} \,,
\nonumber\\
&& \tilde \delta_2 = \delta_2 - \frac{M\,(2\,M^2 + 2\,\Delta\,M + \Delta^2)}{(2\,M+\Delta)\,(M+\Delta)^2}\,\log \frac{2\,\Delta}{M+\Delta} 
\nonumber\\
&& \qquad -\,  \frac{M\,(2\,M + \Delta)}{4\,(M +\Delta)^2}- 2\,\big(\tilde \delta_3 -\delta_3\big)\, \log \frac{M+\Delta}{M}\,,
\nonumber\\
&& \tilde \delta_3 = \delta_3  + \frac{M\,(2\,M^2 + 2\,\Delta\,M + \Delta^2)}{2\,(M +\Delta)^2\,(2\,M + \Delta)}\,, \qquad \qquad 
\tilde \delta_5 = 0\,,
\nonumber\\
&& \tilde \delta_4 = \delta_4 - \frac{2\,M^3}{(2\,M+\Delta)^3}\,\log\frac{2\,\Delta}{M+\Delta}
\nonumber\\
&& \qquad +\, \frac{M\,(4\,M^2+4\,\Delta\,M+5\,\Delta^2)}{32\,(M+\Delta)^2\,(2\,M + \Delta)}
- 2\,\big(\tilde \delta_5-\delta_5\big)\, \log \frac{M+\Delta}{M}\,,
\nonumber\\ \nonumber\\
&& \tilde \beta_1 =\frac{M+\Delta}{M}\,\Delta\,\frac{\partial }{\partial \Delta} \,\beta_1 \,\frac{(2\,M+ \Delta)\,M}{2\,(M+\Delta)^2}  \,,
\label{def-hat-alpha} \\
&& \tilde \beta_2 = \Delta^2\,\frac{\partial }{\partial \Delta} \, \frac{M+\Delta}{M}\,\frac{\beta_1 \,\tilde \delta_2 }{\Delta}\,, \qquad \qquad \quad \; 
 \tilde \beta_3 = \Delta^2\,\,\frac{\partial }{\partial \Delta} \, \frac{M+\Delta}{M}\,\frac{\beta_1 \,\tilde \delta_3 }{\Delta}\, ,
\nonumber\\
&& \tilde \beta_4 = \frac{\Delta}{M}\,\frac{(M+\Delta)^2}{M}\,\delta_1\,\frac{\partial}{\partial \Delta}\, \frac{M\,\beta_1}{M+\Delta}\,, \qquad 
\tilde \beta_5 = \Big( 1+ \frac{\Delta}{M}\Big)\,\Delta\,\frac{\partial}{\partial \Delta}\,\beta_1\,\tilde \delta_1 \,,
\nonumber\\ 
&& \tilde \beta_6 =  D_{\Delta \Delta}\,\frac{(2\,M+ \Delta)\,M}{2\,(M+\Delta)^2}\,\beta_1 \,, \qquad \qquad 
\tilde \beta_7 = \frac{\Delta}{M+\Delta}\,D_{\Delta \Delta} \, 
\frac{M+\Delta }{\Delta}\,\beta_1 \,\tilde \delta_2 \,,\qquad  \quad
\nonumber\\
&& \tilde \beta_8 =   \frac{\Delta}{M+\Delta}\,D_{\Delta \Delta} \, 
\frac{M+\Delta }{\Delta}\,\beta_1 \,\tilde \delta_3\,,
\nonumber\\
&& \tilde \beta_9 = \delta_1\,\frac{M+\Delta}{M}\, D_{\Delta \Delta} \,\frac{M\,\beta_1}{M+\Delta} \,, \qquad \qquad 
 \tilde \beta_{10} =D_{\Delta \Delta} \,\beta_1\,\tilde \delta_1 \,,
\nonumber\\
&& \tilde \beta_{11} = -\frac{1}{4}\,\beta_1 \,\frac{M}{2\,M + \Delta}+ \Big( \beta_1- \beta_2\Big )\,\frac{(2\,M+ \Delta)\,M}{2\,\Delta^2} \,,
\nonumber\\ \nonumber\\
&& \qquad \qquad D_{\Delta \Delta} = \frac{(M+\Delta)^2}{M^2}\,\Big( \frac{\Delta^2\, \partial^2 }{\partial \Delta \,\partial \Delta} 
+\frac{2\,\Delta}{M+\Delta}\,\frac{\Delta \,\partial }{\partial \Delta} \Big)\,,
\end{eqnarray}
While the $\beta_i$ characterize the chiral expansion of the coefficients in front of $\bar I_{QR}$ and $\bar I_Q$ in  (\ref{result-loop-10}), the 
$\delta_i$ follow from a chiral expansion of $\bar I_{QR}$ with $M_B = M+ \Delta$ and $M_R = M$ and  $m_Q < \Delta$. 
Again the $\tilde \delta_n$ follow from the corresponding $\delta_n$ after subtractions from a chiral expansion of 
the anomalous term $f^{(d)}_0(x^2)\,f^{(d)}_1(x^2)$ in (\ref{IQR-x-delta}). Finally the coefficients $\tilde \alpha_n$ characterize the central results (\ref{loop-HB-4-B}) and \ref{loop-HB-5-B}).

Consider the symmetry breaking counter terms
\allowdisplaybreaks[1]
\begin{eqnarray}
\Sigma^{(4-\chi)}_\Delta \!\!\!\!&=&\!\!\! - 4\,B_0\, \Big( d^{\rm eff}_0\, (2\,m+m_s)  + d^{\rm eff}_D\, m\Big)
 \nonumber\\
\!\!\!\!&\phantom =& \!\!\! - 4\,B_0^2\, \Big( e_0\,(2\,m^2 + m_s^2) + e_2\, m^2 \Big)
 \nonumber\\
\!\!\!\!&\phantom =& \!\!\! - 2\,B_0\, \Big( \xi_0\, (2\,m+m_s)  +
\xi_D\, m\Big)  \Big(M_{\Delta } - (M+\Delta)\Big) \,,
\nonumber \\
\Sigma^{(4-\chi)}_{\Sigma^*} \!\!\!\!&=&\!\!\! - 4\,B_0\, \Big( d^{\rm eff}_0\, (2\,m+m_s)  + \frac{1}{3} \,d^{\rm eff}_D\, (2\,m+m_s) \Big)
\nonumber \\
\!\!\!\!&\phantom =& \!\!\!- 4\,B_0^2\, \Big( e_0\, (2 \,m^2 + m_s^2) + \frac 13 \,e_1\, (m - m_s)^2 + \frac 13\, e_2\,(2\,m^2 + m_s^2) \Big)
\nonumber \\
\!\!\!\!&\phantom =& \!\!\!- 2\,B_0\, \Big( \xi_0\,
(2\,m+m_s)  + \frac{1}{3}\, \xi_D\, (2\,m+m_s) \Big)
\Big(M_{\Sigma^*} - (M+\Delta)\Big)\,,
\nonumber \\
\Sigma^{(4-\chi)}_{\Xi^*} \!\!\!\!&=&\!\!\! - 4\,B_0\, \Big( d^{\rm eff}_0\, (2\,m+m_s)  + \frac{1}{3} \,d^{\rm eff}_D\, (m+2\,m_s) \Big)
\nonumber \\
\!\!\!\!&\phantom =& \!\!\!-4\,B_0^2\, \Big( e_0\, (2\,m^2 + m_s^2)  + \frac 13\, e_1\, (m - m_s)^2 + \frac 13\, e_2\, (m^2 + 2\,m_s^2) \Big)\,,
\nonumber \\
\!\!\!\!&\phantom =& \!\!\!- 2\,B_0\, \Big( \xi_0\, (2\,m+m_s)
+ \frac{1}{3} \,\xi_D\, (m+2\,m_s) \Big)  \Big(M_{\Xi^*} - (M+\Delta)\Big)\,,
\nonumber \\
\Sigma^{(4-\chi)}_\Omega \!\!\!\!&=&\!\!\! - 4\,B_0\, \Big( d^{\rm eff}_0\, (2\,m+m_s) + d^{\rm eff}_D\, m_s \Big)
\nonumber \\
\!\!\!\!&\phantom =& \!\!\! - 4\, B_0\, \Big( e_0\, (2\,m^2+m_s^2) + e_2\, m_s^2 \Big)
\nonumber \\
\!\!\!\!&\phantom =& \!\!\!  - 2\,B_0\, \Big( \xi_0\, (2\,m+m_s) +
\xi_D\, m_s \Big) \,\Big(M_\Omega - (M+\Delta)\Big)\,,
\nonumber\\
&& \!\!\!\!\! \!\!\! \!\!\!\!\! \!\!\! d^{\rm eff}_0 = e_4 \,B_0\,(2\,m+m_s),\qquad \qquad d^{\rm eff}_D = e_3\, B_0\,(2\,m+m_s)\,,
\label{result-counter-terms-decuplet}
\end{eqnarray}
with
\begin{eqnarray}
&& \mu^2 \,\frac{d}{d\,\mu^2}\,e_i = -\frac{1}{4}\,\frac{\Gamma_{e_i}}{(4\,\pi\,f )^2} \,, 
\nonumber\\
&& \Gamma_{e_0} = \frac{20}{3}\,d_0+2\,d_D - \frac{1}{18}\,\Big(15\, \tilde h_1^{(S)}+13 \,\tilde h_2^{(S)}+9\, \tilde h_3^{(S) }
\Big)
\nonumber\\
&& \qquad -\, \frac{1}{72}\,( M+\Delta ) \,\Big( 15 \,\bar h_1^{(V)}+13\,\bar  h_2^{(V)} +9\,  \bar h_3^{(V)} \Big)  \,,
\nonumber\\
&& \Gamma_{e _1} = -\frac{1}{3} \,\tilde h_3^{(S)} - \frac{1}{12} \,( M+\Delta ) \,\bar h_3^{(V)}  \,,
\nonumber\\
&& \Gamma_{e_2} =  \frac{2}{3}\,d_D +  \frac{1}{2}\,\Big(\tilde h_2^{(S)}+\tilde h_3^{(S)}\Big)+
\frac{1}{8}\,( M+\Delta )\, \Big( \bar h_2^{(V)}+ \bar h_3^{(V)} \Big)\,,
\nonumber\\
&& \Gamma_{e_3} =   \frac{44 }{9}\,d_D -\frac{1}{9} \,\Big( 13 \,\tilde h_2^{(S)}+ 9\, \tilde h_3^{(S)}\Big) 
- \frac{1}{36} \,( M+\Delta )\,\Big( 13\, \bar h_2^{(V)} +9\,   \bar h_3^{(V)} \Big)\,,
\nonumber\\
&& \Gamma_{e_4} = \frac{44 }{9}\,d_0 +\frac{1}{54}\,\Big(-33 \,\tilde h_1^{(S)}+4 \,\tilde h_2^{(S)} \Big)
\nonumber\\
&& \qquad \qquad -\,\frac{1}{216}\,( M+\Delta )\,\Big( 33 \,\bar h_1^{(V)} -4\,\bar h_2^{(V)} \Big)\,.
\label{res-Gamma-ei}
\end{eqnarray}
We derive the form of the anomalous scaling term, $e^{\rm ano}_i$,  introduced in (\ref{def-c-e-ano}, \ref{c-e-running}). The prescription (\ref{eliminate-mu}) applied to the 
$(M_R -M_B)\,\bar I_Q$ terms in (\ref{result-loop-10}) implies a renormalization of the 
symmetry breaking counter terms
\begin{eqnarray}
&& e^{\rm ano}_i =  \frac{1}{4}\,\frac{1}{(4\,\pi\,f )^2}\, \Gamma^{(1)}_{e_i}\,\delta^{(1)}_{e}
+ \frac{1}{4}\,\,\frac{C^2}{(4\,\pi\,f )^2}\,\Big( \Gamma^{(2)}_{e_i}\, \delta^{(2)}_{e} + \Gamma^{(3)}_{e_i}\, \delta^{(3)}_{e} \Big) \,,
\nonumber\\
&&  \delta^{(1)}_e= -4 - \log \frac{(\Delta + M)^2}{\mu^2} \,, \qquad \qquad 
 \delta^{(2)}_e = \frac{1}{3}\left(\frac{\partial \Delta\, \beta_4}{\partial \Delta}\right)\log\frac{\mu^2}{M^2}   \,,
\nonumber\\
&& \delta^{(3)}_e =  \frac 13\left(\frac{\partial \Delta \,\beta_4 }{\partial M}\right)\log\frac{\mu^2}{M^2}
 - \delta^{(2)}_e \,,
\nonumber\\
&& \mu^2\,\frac{d }{d \,\mu^2} \,e^{\rm ano}_i = \frac{1}{4}\,\frac{-1}{(4\,\pi\,f )^2}\, \Gamma^{(1)}_{e_i}   
-\frac{1}{12}\,\frac{C^2}{(4\,\pi\,f )^2}\,\Bigg\{  \Gamma^{(2)}_{e_i} \,\Big[ 1+ \Delta\,\frac{\partial}{\partial \Delta} \Big]
\nonumber\\
&& \qquad \qquad \; -\,  \Gamma^{(3)}_{e_i} \,\Big[ 1 + \Big(1+\frac{\Delta}{M} \Big)\,\Delta\,\frac{\partial}{\partial \Delta }\Big] 
\Bigg\}\,\beta_4\,,  
\label{res-e-ano}
\end{eqnarray}
with
\begin{eqnarray}
&& \Gamma^{(1)}_{e_0} = -\frac{10}{81} \, d_D\,H^2\,, \qquad 
 \Gamma^{(1)}_{e_2} = \frac{10}{27} \, d_D\,H^2\,, \qquad 
\Gamma^{(1)}_{e _1} = \Gamma^{(1)}_{e_3} =  \Gamma^{(1)}_{e_4} = 0\,,
\nonumber\\ \nonumber\\ 
&& \Gamma^{(2)}_{e_0} = 0\,, \qquad 
\Gamma^{(2)}_{e_1} = - \frac{4}{3}\,d_D \,, \qquad \Gamma^{(2)}_{e_2} = 2\,d_D \,,   
\nonumber\\
&& \Gamma^{(2)}_{e_3} =  2\,( d_0 + d_D )\,, \qquad
\Gamma^{(2)}_{e_4} = 2\, d_0\,,
\nonumber\\ \nonumber\\   
&& \Gamma^{(3)}_{e_0} = 2\,(b_F - b_D)\,, \qquad 
\Gamma^{(3)}_{e_1} = -4\,b_F  \,, \qquad \Gamma^{(3)}_{e_2} =   4\,b_D\,,   
\nonumber\\
&& \Gamma^{(3)}_{e_3} = 2\,(b_0 + 2\,b_F) \,, \qquad
\Gamma^{(3)}_{e_4} = 2\,\big(b_0 + b_D - b_F \big)\,.
\end{eqnarray} 

We close this Appendix with explicit expressions for the fifth moment of the decuplet self energy:
\allowdisplaybreaks[1]
\begin{eqnarray}
&& \bar \Sigma^{{\rm bubble}-5}_{B \in [10]} =  \!
\sum_{Q\in [8], R\in [10]}
\left(\frac{m_{QR}}{4\,\pi\,f}\,G_{QR}^{(B)} \right)^2  \frac{5}{18}\,\Bigg\{ 
-  \frac{m^4_Q}{M^3_B} \,\Big( \frac{17}{60} - \frac{1}{5}\,\log\frac{m_Q}{M_R}\Big)
\nonumber\\
&& \qquad  \quad +\, \frac{13\,\pi}{40}\,\frac{m^3_Q}{M_B^2}   
  +\,\Big(M_R-M_B\Big)^2\,\Bigg[\frac{\pi}{2\,m_Q} + \frac{1}{M_B}\,\Big( 2 + 3\,\log \frac{m_Q}{M_R} \Big)\Bigg] \Bigg\}
\nonumber\\
&& \; + \,\sum_{Q\in [8], R\in [8]}
\left(\frac{m_Q}{4\,\pi\,f}\,G_{QR}^{(B)} \right)^2 \,\Bigg\{- \frac{\tilde \beta_{11}}{6}\,\frac{ \Delta^3}{(M+\Delta)^2}\,\log \frac{4\,\Delta^2}{M^2}
  \nonumber\\
&& \qquad \quad - \, \frac{1}{6}\,  \Big( \beta_1- \beta_2\Big )\,\Bigg( \frac{(2\,M+ \Delta)\,M}{2\,(M+ \Delta)^2} \,\frac{M_B-M_R}{\Delta_B^2}\, m_Q^2\,\log \frac{m_Q^2}{M_R^2}
\nonumber\\
&& \qquad  \qquad \quad  -\, \big(\delta_1-\tilde \delta_1 \big) \,\frac{\Delta_Q^2- \Delta\,\Delta_B}{\Delta_B}-  \tilde \delta_1\, \frac{\Delta_Q^2}{\Delta_B^2}\,\Big( M_R-M_B+\Delta_B\Big) \Bigg)
\nonumber\\
&& \qquad \quad  +\,\frac{\tilde \beta_{11}}{6\,M_B^2}\, \Bigg[  \big(M_B-M_R \big)^3\,\log \frac{m_Q^2}{M_R^2} 
+ \, \Delta_Q^3 \,\Big( \log \big(M_R-M_B - \Delta_Q \big) 
\nonumber\\
&& \qquad  \qquad \quad -\, \log \big(M_R-M_B + \Delta_Q\big) \Big)\Bigg]
 \nonumber\\
&& \qquad  \quad +\,\frac{m_Q^2\, \Delta_Q^2}{6\,\Delta^3_B} \,\Bigg[\Big(\beta_2-\beta_1 \Big)\, \Big(  \tilde \delta_2 +\tilde \delta_3\,\log \frac{m_Q^2}{M_R^2} \Big)
-  \beta_1\,\Big( \tilde  \delta_4 + \tilde \delta_5\,\log \frac{m_Q^2}{M_R^2} \Big)  \Bigg]
\nonumber\\
&& \qquad \quad +\,\Big(M_R - M_B +\Delta_B \Big)^2 \Bigg( 
\frac{\tilde \beta_9}{12}\,\frac{\Delta_Q^2}{m_Q^2\,\Delta_B} + \frac{\tilde \beta_{10}}{12}\,\frac{\Delta_Q^2}{m_Q^2\,\Delta^2_B}\,\big( M_R-M_B\big)
\nonumber\\
&&\qquad \qquad + \,\frac{\tilde \beta_6}{12\,\Delta^2_B\,m_Q^2}\,\Bigg[ 
\Big( \Delta_Q^2 - \frac{1}{2}\,m_Q^2\Big)\,\big( M_B-M_R\big)\, \log \frac{m_Q^2}{M_R^2}
\nonumber\\
&&\qquad \qquad \quad +\, \Delta_Q^3 \,\Big( 
\log (M_R-M_B - \Delta_Q ) -\log (M_R-M_B + \Delta_Q )  \Big)\Bigg]
\nonumber\\
&& \qquad\qquad  +\, \frac{1}{12\,\Delta^3_B}\,\Big( -\Delta_Q^2\,\tilde \beta_7 
+ \tilde \beta_8\,m_Q^2\,\log \frac{m_Q^2}{M_R^2}\Big) 
\Bigg)
\Bigg\} \,, 
\label{loop-HB-5-B}
\end{eqnarray}
with $m_{QR}^2 = m_Q^2-(M_R-M_B)^2$.

\clearpage

\bibliographystyle{elsarticle-num}
\bibliography{literature}

\end{document}